\documentclass[apj]{emulateapj}
\usepackage{amsmath}
\usepackage[colorlinks=true,citecolor=blue,linkcolor=blue]{hyperref}

\def\roma{1}
\def\icra{2}
\def\losalamos{3}
\def\rio{4}

\shorttitle{On the IGC scenario of GRBs associated with SNe}
\shortauthors{Becerra et al.}

\begin{document}

\title{On the induced gravitational collapse scenario of gamma-ray bursts associated with supernovae}

\author{L.~Becerra\altaffilmark{\roma,\icra},
				C.~L.~Bianco\altaffilmark{\roma,\icra},
				C.~L.~Fryer\altaffilmark{\losalamos}, 
				J.~A.~Rueda\altaffilmark{\roma,\icra,\rio}, 
				R.~Ruffini\altaffilmark{\roma,\icra,\rio}
				}

\altaffiltext{\roma}{Dipartimento di Fisica and ICRA, 
                     Sapienza Universit\`a di Roma, 
                     P.le Aldo Moro 5, 
                     I--00185 Rome, 
                     Italy}
                     
\altaffiltext{\icra}{ICRANet, 
                     P.zza della Repubblica 10, 
                     I--65122 Pescara, 
                     Italy} 
																	
\altaffiltext{\losalamos}{CCS-2, Los Alamos National Laboratory, Los Alamos, NM
87545}

\altaffiltext{\rio}{ICRANet-Rio, 
                     Centro Brasileiro de Pesquisas F\'isicas, 
                     Rua Dr. Xavier Sigaud 150, 
                     22290--180 Rio de Janeiro, 
                     Brazil}

\begin{abstract}
Following the induced gravitational collapse (IGC) paradigm of gamma-ray bursts (GRBs) associated with type Ib/c supernovae, we present numerical simulations of the explosion of a carbon-oxygen (CO) core in a binary system with a neutron-star (NS) companion. The supernova ejecta trigger a \emph{hypercritical} accretion process onto the NS thanks to a copious neutrino emission and the trapping of photons within the accretion flow. We show that temperatures 1--10~MeV develop near the NS surface, hence electron-positron annihilation into neutrinos becomes the main cooling channel leading to accretion rates $10^{-9}$--$10^{-1}~M_\odot$~s$^{-1}$ and neutrino luminosities $10^{43}$--$10^{52}$~erg~s$^{-1}$ (the shorter the orbital period the higher the accretion rate). We estimate the maximum orbital period, $P_{\rm max}$, as a function of the NS initial mass, up to which the NS companion can reach by hypercritical accretion the critical mass for gravitational collapse leading to black-hole (BH) formation. We then estimate the effects of the accreting and orbiting NS companion onto a novel geometry of the supernova ejecta density profile. We present the results of a $1.4\times 10^7$~particle simulation which show that the NS induces accentuated asymmetries in the ejecta density around the orbital plane. We elaborate on the observables associated with the above features of the IGC process. We apply this framework to specific GRBs: we find that X-ray flashes (XRFs) and binary-driven hypernovae (BdHNe) are produced in binaries with $P>P_{\rm max}$ and $P < P_{\rm max}$, respectively. We analyze in detail the case of XRF 060218.
\end{abstract}

\maketitle

\section{Introduction}\label{sec:1}

Recently, \citet{2016arXiv160202732R} proposed a binary nature for the progenitors of both long and short GRBs. In this work we focus on long GRBs associated with supernovae. For such systems the \emph{induced gravitational collapse} (IGC) paradigm \citep[see, e.g.,][]{2006tmgm.meet..369R,2008mgm..conf..368R,2012A&A...548L...5I,2012ApJ...758L...7R,2014ApJ...793L..36F} indicates as progenitor a binary system composed of a CO core and a neutron-star in a tight orbit. 
Such a binary system emerged first as a necessity for the explanation of a set of observational features of long GRBs associated with type Ic supernovae \citep{2012ApJ...758L...7R}. Besides, it also appears in the final stages of a well defined evolutionary path which includes the presence of interacting binaries responsible for the formation of stripped-envelope stars such as CO cores leading to type Ic supernovae \citep{2012ApJ...758L...7R,2015ApJ...812..100B,2015PhRvL.115w1102F}.

\begin{figure}
\includegraphics[width=\hsize,clip]{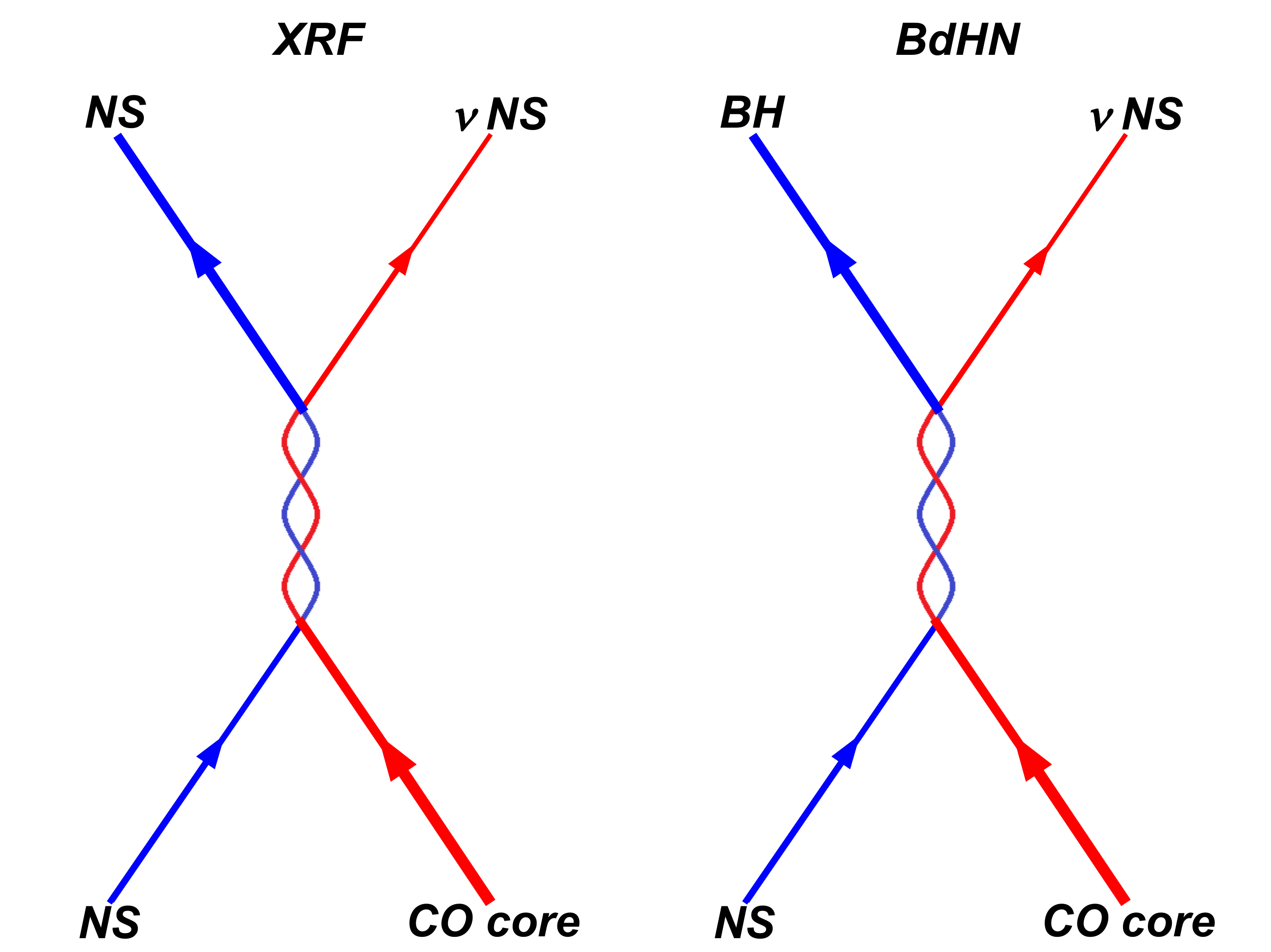}
\caption{Cosmic-matrix of XRFs and BdHNe as introduced in \citet{2015mgm..conf..242R,2015ApJ...798...10R,2015ARep...59..591R}. See text for details.}\label{fig:cmatrix}
\end{figure}

The core-collapse of the CO star produces a supernova explosion ejecting material that triggers an accretion process onto the binary neutron-star companion; hereafter indicated as NS. It was advanced in \citet{2016arXiv160202732R} the existence of two classes of long GRBs depending on whether or not a black-hole (BH) is formed in the hypercritical accretion process onto the NS:
\begin{itemize}
\item
First, there is the subclass of binary-driven hypernovae (BdHNe), long GRBs with isotropic energy $E_{\rm iso}\gtrsim 10^{52}$~erg and rest-frame spectral peak energy $0.2\lesssim E_{p,i}\lesssim 2$~MeV. Their prompt emission lasts up to $\sim 100$~s and it is at times preceded by an X-ray emission in the 0.3--10~keV band lasting up to 50~s and characterized by a thermal and a power-law component (i.e.~Episode 1 in GRB 090618 in \citealp{2012A&A...548L...5I}). They have a long lasting X-ray afterglow generally composed by a spike, a plateau, followed by a common late power-law behavior when measured in the common source rest-frame~\citep{2013A&A...552L...5P}. For all BdHNe at $z\lesssim1$, an optical supernova with luminosity similar to the one of supernova 1998bw \citep{1998Natur.395..670G} has been observed after 10--15~days in the cosmological rest-frame \citep[see, e.g.,][]{2014A&A...567A..29M}. It has been proposed that this class of GRBs occurs when the NS reaches its critical mass through the above accretion process and forms a BH (see Fig.~\ref{fig:cmatrix}). Under these conditions, the GeV emission becomes observable and it has been proposed to originate from the newly formed BH~\citep{2016arXiv160202732R}. This GRB subclass occurs in compact binaries with orbital periods as short as $P\sim 5$~min or binary separations $a\lesssim 10^{11}$~cm~\citep{2014ApJ...793L..36F}.
\item
Second, there is the subclass of X-ray flashes (XRFs), long GRBs with isotropic energies in the range $E_{\rm iso}\approx 10^{47}$--$10^{52}$~erg; spectral peak energies $E_{\rm p,i}\approx 4$--200~keV \citep{2013IJMPD..2230028A,2015ApJ...798...10R,2016arXiv160202732R}. Their prompt emission phase lasts $\sim10^2$--$10^4$~s and it is generally characterized by a spectrum composed by a thermal component (with radii $10^{10}$--$10^{12}$~cm and temperatures $0.1$--$2$~keV, \citealp{2006Natur.442.1008C}) and power-law component. They have long lasting X-ray afterglows without the characteristic common late power-law behavior encountered in the BdHNe \citep{2013A&A...552L...5P}, nor the characteristic X-ray spike. For all XRFs at $z\lesssim1$, an optical supernova with luminosity similar to the one of supernova 2010bh \citep{2012ApJ...753...67B}, has been observed after $10$--$15$ days in the cosmological rest-frame. These sources have been associated within the IGC paradigm to binaries of a CO core and an NS in which there is no BH formation (see Fig.~\ref{fig:cmatrix}): when the accretion is not sufficient to bring the NS to reach the critical mass. This occurs in  binaries with orbital periods longer than $P\sim 5$~min or binary separations $a\gtrsim 10^{11}$~cm~\citep{2016arXiv160202732R,2015ApJ...812..100B}.
\end{itemize}

The complexity of the above processes leading to two possible outcomes can be summarized schematically within the concept of \emph{Cosmic-Matrix (C-matrix)} as first introduced in \citet{2015mgm..conf..242R,2015ApJ...798...10R,2015ARep...59..591R}. The C-matrix describes these systems as a four-body problem in analogy to the case of particle physics (see Fig.~\ref{fig:cmatrix}). The \emph{in-state} is represented by the CO core and the NS companion. The interaction between these two objects given by the hypercritical accretion process triggered by the supernova explosion onto the NS companion, and which is examined in this work, lead to two possible \emph{out-states}: in the case of a BdHN it is formed by the $\nu$NS, i.e. the neutron star left by the supernova explosion of the CO core, and a BH formed from the gravitational collapse of the NS companion of the CO core in the in-state. As we have mentioned in XRFs the accretion is not enough to lead to the gravitational collapse of the NS then the out-state is a $\nu$NS and another NS (of course more massive than the initial one present in the in-state).

It is clear that the observational properties of the IGC binaries are sensitive to the binary parameters which can change the fate of the system. The first estimates of the accretion rate and the possible fate of the accreting NS in an IGC binary were presented in \citet{2012ApJ...758L...7R}. To obtain an analytic expression of the accretion rate, such first simple model assumed: 

1) a pre-supernova homogenous density profile; 

2) an homologous expansion of the density; 

3) constant mass of the NS ($\approx 1.4~M_\odot$) and the supernova ejecta ($\approx 4$--$8~M_\odot$). The first application of this model was presented in \citet{2012A&A...548L...5I} for the explanation of the Episode 1 of GRB 090618.

In \citet{2014ApJ...793L..36F} there were presented 1D numerical simulations which improved the above model. Specific density and ejection velocity profiles were adopted from numerical simulations of CO core-collapses producing type Ic supernovae. The hydrodynamic evolution of the material entering the NS accretion region was computed on the basis of models of hypercritical accretion in supernova fallback. The integration was followed up to the point where it is finally accreted by the NS on its surface.

In \citet{2015PhRvL.115w1102F} the evolution of the gravitational binding energy of the BdHNe was analyzed and it was shown that most of these systems would remain bound after the occurrence of the supernova explosion. This conclusion is in contradiction with the traditional assumption of instantaneous mass ejection which leads to the well-known limit that the system becomes unbound when the binary looses more than half of the total mass. This novel result was obtained by the accurate consideration of two ingredients: the accretion process which changes the mass and the momentum of the binary, and the orbital period which can be as short as the explosion time scale.

In \citet{2015ApJ...812..100B} the amount of angular momentum transported by the ejecta entering the Bondi-Hoyle region was estimated and how much of it can be transferred to the NS when it is finally accreted. The ejecta density profile was adopted as a power-law in radius and its evolution with time homologous. It is important to recall some of the conclusions obtained in that work: 

1) the angular momentum the ejecta inside the accretion region is such that it circularizes around the NS forming a disk-like structure;

2) the time scale of the disk angular momentum loss is shorter than the time scale at which matter is being captured, so the accretion time scale dominates the evolution;

3) for binary periods shorter than some critical value, i.e.~$P\lesssim P_{\rm max}$, the NS can reach either the mass-shedding or secular axisymmetric instability (critical mass point), which induces its gravitational collapse to a BH. In systems with $P> P_{\rm max}$ the NS gains both mass and angular momentum but not enough to trigger its collapse to a BH. The value of the critical mass has been calculated including the effects of rotation \citep{2015PhRvD..92b3007C}.

The value of $P_{\rm max}$ was there computed as a function of the initial NS mass, but only for masses larger than $\approx 1.67~M_\odot$, and it was assumed that half of the angular momentum of the disk at the inner disk radius is transferred to the NS.

Although all the above works have already shown that indeed the supernova can induce, by accretion, the gravitational collapse of the NS to a BH, there is still the need of exploring systematically the entire, physically plausible, space of parameters of these systems, as well as to characterize them observationally. The main aims of this work are:
\begin{enumerate}
\item
To improve the estimate of the accretion rate with respect to the one in \citet{2015ApJ...812..100B} by including effects of the finite size/thickness of the density profile and, for different CO core progenitors leading to different ejected masses.
\item
To extend the analysis performed in \citet{2015ApJ...812..100B} and identify the separatrix of systems in which a BH is formed and the ones where there is not BH formation. This is equivalent to improve the determination of $P_{\rm max}$. We extend here the possible range of the initial NS mass and allow for different values of the angular momentum transfer efficiency.
\item
To compute the expected luminosity emitted during the hypercritical accretion process onto the NS for a wide range of binary periods shorter (BdHNe) and longer (XRFs) than $P_{\rm max}$. With this we can establish the energetic budget that characterizes both XRFs and BdHNe.
\end{enumerate}

In parallel,
\begin{itemize}
\item[a)] 
We apply the above considerations to specific GRBs by analyzing in detail the specific case of XRF 060218.
\item[b)]
We estimate in that specific case the asymmetries created by the orbiting and accreting NS on the ejected matter density profile and the structure of the prompt radiation.
\item[c)]
We explore the influence of the prompt X-ray radiation in the late X and optical emission of the supernova and the afterglow.
\end{itemize}

The article is organized as follows. In Sec.~\ref{sec:2} we summarize the framework of the hypercritical accretion of the supernova ejecta onto the NS. Sec.~\ref{sec:3} gives details on the computation of the time evolution of both the (gravitational and baryonic) mass and angular momentum of the accreting NS. In Sec.~\ref{sec:4} we compute the maximum orbital period, $P_{\rm max}$, up to which the induced gravitational collapse of the NS to a BH by accretion can occur. We show in Sec.~\ref{sec:5} the asymmetries that the accreting NS produces on the supernova ejecta. In Sec.~\ref{sec:6} we summarize the hydrodynamics inside the accretion region, including convective instabilities, and the properties of the neutrino emission. We present in Sec.~\ref{sec:7} estimates of the expected luminosities during the hypercritical accretion process. Sec.~\ref{sec:8} shows how the radiation from the accretion process as well as the asymmetries in the ejecta influence the supernova emission both in X-rays and in the optical. Finally in Sec.~\ref{sec:9} we summarize the results of this work. Additional technical details are presented in a series of appendices.

\section{Hypercritical accretion induced by the supernova}\label{sec:2}

In order to model the hypercritical accretion process onto the NS, we use the formalism introduced in \citet{2015ApJ...812..100B}. The accretion rate of the ejected material onto the NS is given by \citep{1939PCPS...35..405H,1944MNRAS.104..273B,1952MNRAS.112..195B}:
\begin{equation}
	\dot{M}_{B}=\pi\rho_{\rm ej}R_{\rm cap}^2\sqrt{v_{\rm rel}^2+c_{\rm s,ej}^2},
	\label{eq:Bondi-HoyleRate}
\end{equation}
where $R_{\rm cap}$ is the NS gravitational capture radius
\begin{equation}
	 R_{\rm cap}=\frac{2 G M_{\rm NS}}{v_{\rm rel}^2+c_{\rm s,ej}^2}.
	\label{eq:CaptureRadius}
\end{equation}
Here $\rho_{\rm ej}$ and $c_{\rm s,ej}$ are the density and sound velocity of the ejecta, $M_{\rm NS}$ the NS mass and $\vec{v}_{\rm rel}=\vec{v}_{\rm ej}-\vec{v}_{\rm orb}$, the  velocity of the ejecta as seen from an observer at the NS, and $G$ is the gravitational constant. The orbital velocity is $v_{\rm orb} = \sqrt{G M/a}$, where $M=M_{\rm NS}+M_{\rm CO}$ is the total binary mass, $M_{\rm CO}=M_{\rm env}+M_{\rm Fe}$ the total mass of the CO core which is given by the envelope mass $M_{\rm env}$ and the central iron core mass $M_{\rm Fe} = 1.5~M_\odot$. The latter is the mass of the new neutron-star formed in the core-collapse supernova process, hereafter indicated as $\nu$NS and its mass $M_{\nu\rm NS}$, i.e. we adopt $M_{\nu\rm NS} = M_{\rm Fe} = 1.5~M_\odot$ in agreement with the range of masses predicted under the convective supernova paradigm \citep{2012ApJ...749...91F}. For the ejecta velocities, we adopt an homologous explosion model for the supernova expansion, i.e. a velocity proportional to the radius:
\begin{equation}
	v_{\rm ej}=n\frac{r}{t}\, ,
	\label{eq:vej_sn}
\end{equation}
where $n$ is the so-called \emph{expansion parameter}. Within this approximation, the density profile evolves as \citep[see, e.g,][]{1968pss..book.....C}:
\begin{equation}
	\rho_{\rm ej}(x,t)=\rho_{\rm ej}^0(x)\left( \frac{R_{\rm star}^0}{R_{\rm star}(t)} \right)^3\frac{M_{\rm env}(t)}{M_{\rm env}^0},
	 \label{eq:rhoej_sn}
\end{equation}
where $x\equiv r/R_{\rm star}(t)$, $M_{\rm env}$ is the mass ejected (i.e. the mass available to be accreted by the NS), $\rho_{\rm ej}^0$ is the pre-supernova density profile and $R_{\rm star}$ is the outermost layer of the supernova ejecta. From the velocity profile law we have that $R_{\rm star}$ evolves as:
\begin{equation}
	R_{\rm star}(t)=R_{\rm star}^0\left(\frac{t}{t_0}\right)^n,
	\label{eq:Rstar_evol}
\end{equation}
where $t_0=n R_{\rm star}^0/v_{\rm star,0}$, being $v_{\rm star,0}$ the velocity of the outermost layer $R_{\rm star}^0$.

The pre-supernova density profile of the CO envelope can be well approximated by a power-law profile, i.e.:
\begin{equation}
	\rho_{\rm ej}^0=\rho_{\rm core}\left( \frac{R_{\rm core}}{r} \right)^m,\qquad \quad R_{\rm core}<r< R_{\rm star}^0.
	\label{eq:prerhosn}
\end{equation}
We show in table~\ref{tb:ProgenitorSN} the properties of the pre-supernova CO cores produced by low-metallicity progenitors with initial zero-age main sequence (ZAMS) masses $M_{\rm ZAMS} = 15$, 20, and 30~$M_\odot$  obtained with the Kepler stellar evolution code \citep{2002RvMP...74.1015W}.
\begin{table*}
\centering
\caption{Properties of the pre-supernova CO cores}\label{tb:ProgenitorSN}
\begin{tabular}{cccccc}
\hline \hline
Progenitor & $\rho_{\rm core}$& $R_{\rm core}$& $M_{\rm env}$&$R^0_{\rm star}$&$m$ \\
$M_{\rm ZAMS}~(M_\odot)$ & ($10^8$~g~cm$^{-3}$) & ($10^7$~cm) & $(M_{\odot})$ & ($10^9$~cm) &\\\hline
$15$  & $3.31$& $5.01$ & $2.079 $  & $4.49$&$2.771$ \\
$20$   &$3.02$ &  $7.59$ & $3.89 $& $4.86$ &$2.946$ \\
$30$   & $3.08$& $8.32$ &  $7.94 $& $7.65$ &$2.801$ \\ \hline 
\end{tabular}
\tablecomments{CO cores obtained for the low-metallicity ZAMS progenitors with $M_{\rm ZAMS} = 15$, $20$, and 30~$M_\odot$ in \citet{2002RvMP...74.1015W}. The central iron core is assumed to have a mass $M_{\rm Fe} = 1.5~M_\odot$, which will be the mass of the $\nu$NS, denoted here as $M_{\nu\rm NS}$, formed out of the supernova process.}
\end{table*}

We now improve the treatment in \citet{2015ApJ...812..100B} taking into account the finite size of the envelope. We thus modify the above density profile by introducing boundaries to the supernova ejecta through density cut-offs at the outermost and innermost layers of the ejecta, namely:
\begin{equation}
	\rho_{\rm ej}^0=\hat{\rho}_{\rm core}\ln\left( \frac{r}{\hat{R}_{\rm core}} \right)\left( \frac{R_{\rm star}}{r}-1 \right)^m,
	\label{eq:prerho02}
\end{equation}
where $\hat{R}_{\rm core}<r<R_{\rm star}$. The condition that the modified profile has the same ejecta mass with respect to the unmodified power-law profile implies $\hat{R}_{\rm core} < R_{\rm core}$.

\begin{figure}
\centering
\includegraphics[width=\hsize,clip]{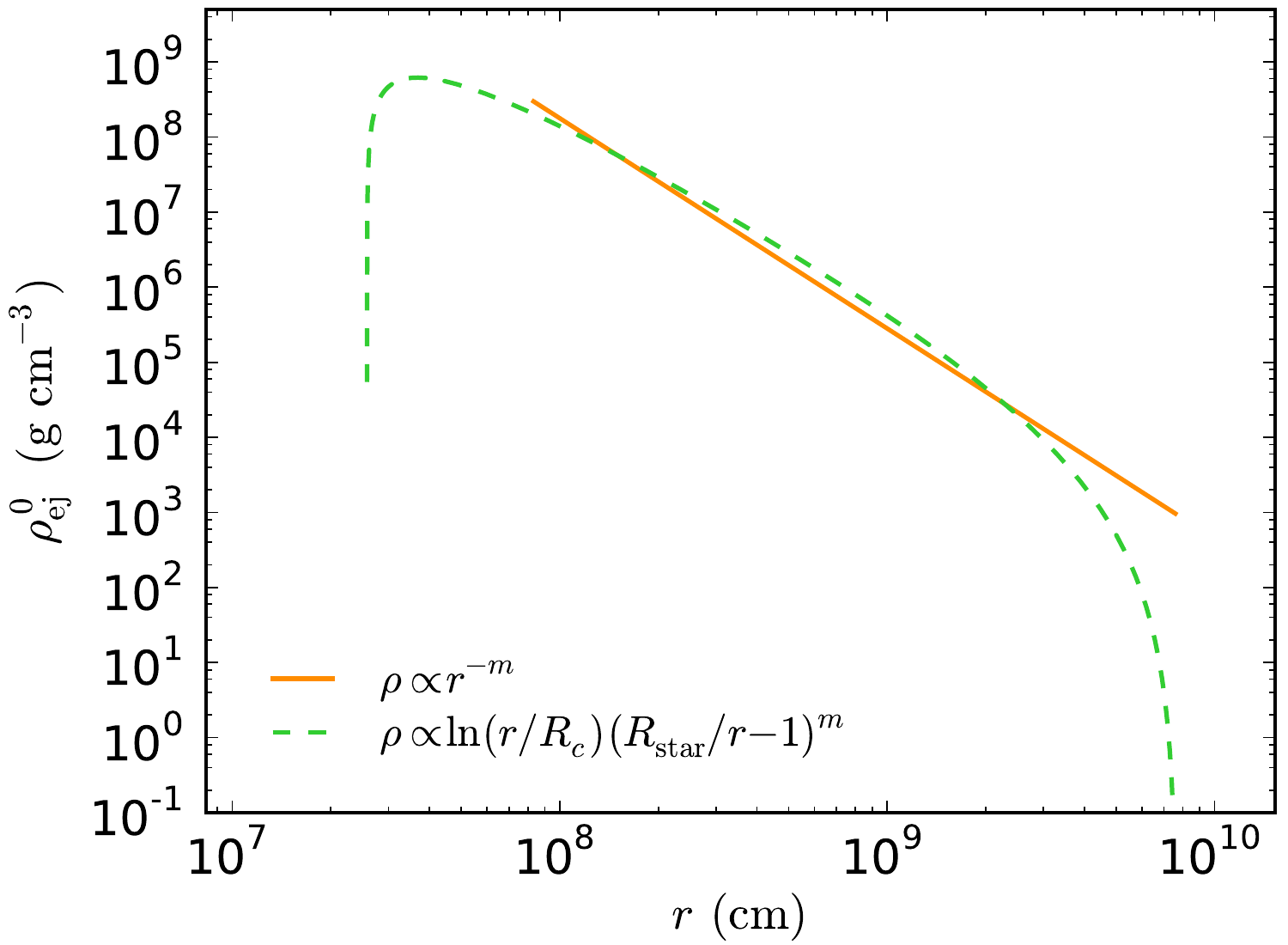}
\caption{Pre-supernova density profile produced by the $M_{\rm ZAMS}=30~M_\odot$ progenitor of table~\ref{tb:ProgenitorSN}. We compare and contrast the power-law density profile (solid curve) with a modified profile (dashed curve) with density cut-offs at the outermost and innermost ejecta layers following Eqs.~(\ref{eq:prerhosn}) and (\ref{eq:prerho02}). The two profiles have the same envelope mass.}
\label{fig:rhoejProfile}
\end{figure}

Fig.~\ref{fig:rhoejProfile} shows the pre-supernova density profile described by equations (\ref{eq:prerhosn}) and (\ref{eq:prerho02}) for the $M_{\rm ZAMS}=30~M_\odot$. For these parameters we have $\hat{R}_{\rm core}=0.31~R_{\rm core}$ and $\hat{\rho}_{\rm core}=567.67$~g~cm$^{-3}$. 

Introducing the homologous expansion for the description of the evolution of the supernova ejecta, equation (\ref{eq:Bondi-HoyleRate}) becomes:
\begin{equation}
	\frac{\dot{\mu}_B(\tau)}{(1-\chi\mu_{B}(\tau))M_{\rm NS}^2}=\frac{\tau^{(m-3)n}}{\hat{r}^m}\ln\left( \frac{\hat{r}}{\hat{r}_c \tau^{n}} \right)\frac{\left(\hat{r}_s-\hat{r}\tau^{-n} \right)^m}{\left[ 1+\eta\hat{r}/\tau\,^2 \right]^{3/2}},
	\label{eq:Mb_evol}
\end{equation}
where
\begin{equation}
\tau\equiv\frac{t}{t_0},\quad  \mu_{B}(\tau)\equiv \frac{M_{B}(\tau)}{\Sigma_B}, \quad \hat{r}\equiv 1-\frac{R_{\rm cap}}{a},
\end{equation}
and the parameters $\chi$, $\Sigma_B$ and $\eta$ depend on the properties of the binary system before the supernova explosion:
\begin{equation}
\Sigma_B=\frac{4\pi \hat{\rho}_c G^2M_\odot^2t_0}{v_{\rm orb}^{3}},\quad 	\chi=\frac{\Sigma_{B}}{M_{\rm env}^0},\quad \eta=\left(\frac{n\,a}{t_0 v_{\rm orb}}\right)^2,
\end{equation}
where $M_{\rm env}^0\equiv M_{\rm env}(t=t_0) = M_{\rm env}(\tau=1)$.

\begin{figure}
	\centering
	\includegraphics[width=\hsize,clip]{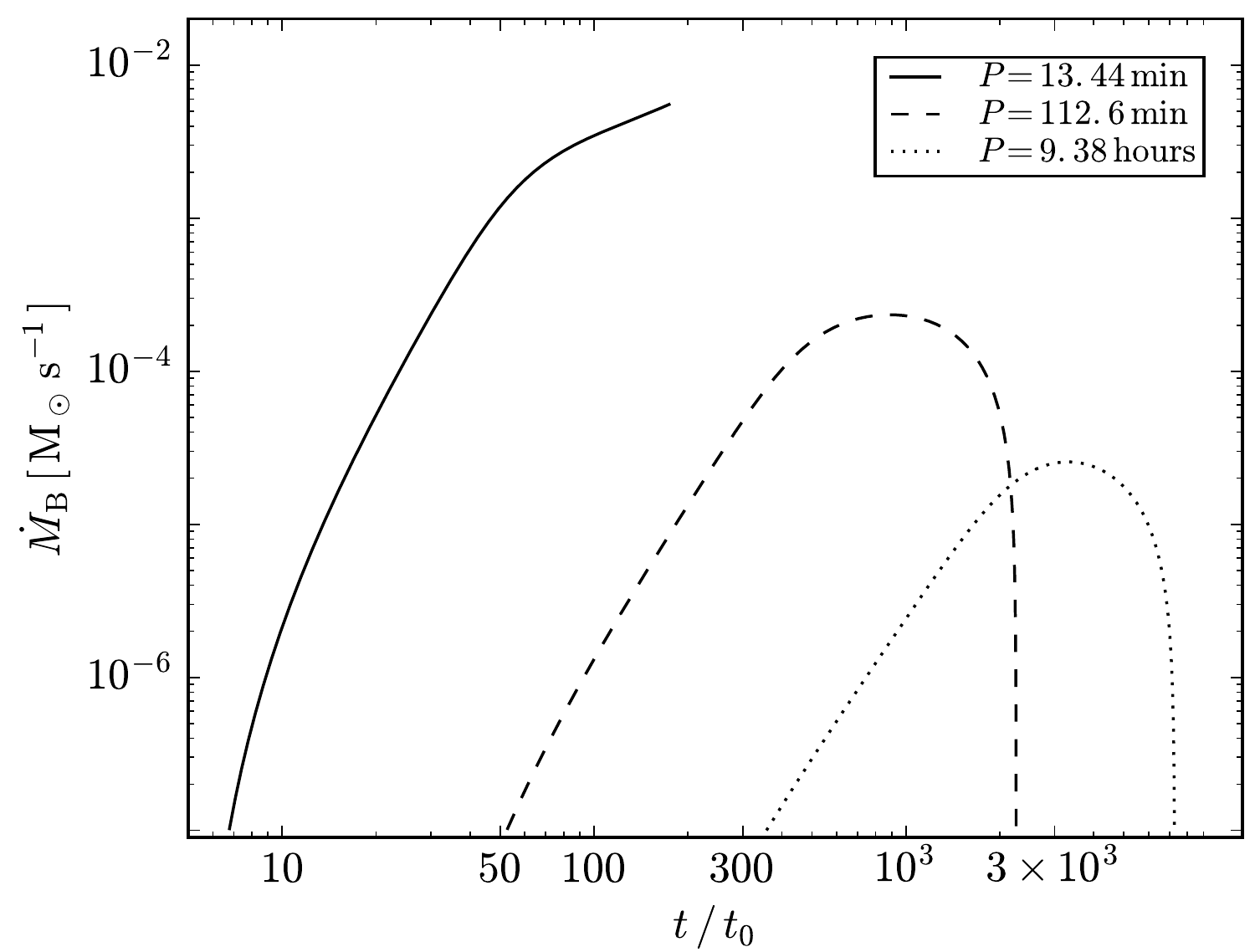}
\caption{Time evolution of the accretion rate onto a NS of initial mass $1.4~M_\odot$ for the following binary parameters: expansion parameter $n=1$, an ejecta outermost layer velocity $v_{\rm star,0}=2\times10^9$~cm~s$^{-1}$ and the supernova ejecta profile is the one obtained for the CO core of the $M_{\rm ZAMS}=20~M_\odot$ progenitor of table~\ref{tb:ProgenitorSN}. For the above progenitor and velocity $t_0=2.4$~s. Three selected orbital periods are shown: $P=13.44$~min, 112.6~min and 9.38~h which correspond to binary separation distances $a=2.46\times 10^{10}$~cm, $1.01\times 10^{11}$~cm, and $2.96\times 10^{11}$~cm, respectively. The solid line corresponds to a case in which the NS reaches the critical mass and collapses to a BH (end point of the curve). In the two other cases, owing to the longer orbital period, there is no induced gravitational collapse of the NS to a BH (see Fig.~\ref{fig:Pmax} in Sec.~\ref{sec:4} for further details).}
	\label{fig:Mdot}
\end{figure}

Fig.~\ref{fig:Mdot} shows the time evolution of the mass accretion rate onto the NS of initial $1.4~M_\odot$ and selected orbital periods. The other binary parameters are: expansion parameter $n=1$, ejecta outermost layer velocity $v_{\rm star,0}=2\times10^9$~cm~s$^{-1}$, and the supernova ejecta profile is the one obtained for the CO core of the $M_{\rm ZAMS}=20~M_\odot$ progenitor of table~\ref{tb:ProgenitorSN}.

We can see from Fig.~\ref{fig:Mdot} that the shorter(smaller) the orbital period(separation) the higher the accretion rate and the shorter the time it peaks. In appendix~\ref{app:A} we derive, following simple arguments, analytic formulas for the peak accretion rate and time which can be useful to get straightforward estimates of these systems.
%

\section{Time evolution of the accreting NS}\label{sec:3}

\citet{2015ApJ...812..100B} showed that the supernova ejecta has enough angular momentum to circularize around the NS in a sort of disk-like structure. Thus, the accreted material will transfer both baryonic mass and angular momentum to the NS. 

The equilibrium NS configurations form a two-parameter family given by the mass (baryonic or gravitational) and angular momentum (or angular velocity). Namely the NS gravitational mass, $M_{\rm NS}$, is in general a function of the NS baryonic mass, $M_b$, and angular momentum, $J_{\rm NS}$. In a similar way the angular momentum contributes to the mass of a BH \citep{1971PhRvD...4.3552C}. It is then clear that the evolution of the NS gravitational mass is given by:
\begin{equation}
	\dot{M}_{\rm NS}(t)=\left(\frac{\partial M_{\rm NS}}{\partial M_b}\right)_J\dot{M}_b+\left(\frac{\partial M_{\rm NS}}{\partial J_{\rm NS}}\right)_{M_b}\dot{J}_{\rm NS},
	\label{eq:Mns_evol}
\end{equation}
We assume that all the (baryonic) mass entering the NS capture region will be accreted by the NS, i.e:
\begin{equation}
	M_{b}(t)=M_{b}(t_0)+M_B(t),
	\label{eq:MbNS}
\end{equation}
then $\dot{M}_b \equiv \dot{M}_B$. 

For the relation between the NS gravitational mass, the baryonic mass, and the angular momentum for the NS equilibrium configurations, namely the NS gravitational binding energy formula, we use the recent result obtained in \citet{2015PhRvD..92b3007C}:
\begin{equation}
	\frac{M_b}{M_\odot}=\frac{M_{\rm NS}}{M_\odot}+\frac{13}{200}\left( \frac{M_{\rm NS}}{M_\odot} \right)^2\left( 1+\frac{1}{137}j_{\rm NS}^{1.7} \right),
	\label{eq:MbMnsjns}
\end{equation}
where $j_{\rm NS}\equiv cJ_{\rm NS}/(GM_\odot^2)$, and which is independent on the nuclear equation of state (EOS).

The torque on the NS by the accreted matter is given by
\begin{equation}
	\dot{J}_{\rm NS}=\xi l(R_{\rm in})\dot{M}_{B},
	\label{eq:Jns_evol}
\end{equation}
where $R_{\rm in}$ is the disk inner boundary radius, $l(R_{\rm in})$ is the angular  momentum per unit mass of the material located at $r=R_{\rm in}$, and $\xi\leq 1$ is a parameter that accounts for the efficiency of the angular momentum transfer. The precise value of $\xi$ depends mainly: 1) on possible angular momentum losses (e.g. by jetted emission during accretion) and 2) on the deceleration of the matter in the disk inner radius zone.

The inner disk radius is given by the maximum between the radius of the last stable circular orbit, $r_{\rm lso}$, and the NS radius, $R_{\rm NS}$. Namely, $R_{\rm in}={\max}(r_{\rm lso}, R_{\rm NS})$. When the disk extends until the NS surface, $l(R_{\rm in})$ is given by the Keplerian orbit with radius equal to the NS equatorial radius. On the other hand, if $R_{\rm NS}<r_{\rm lso}$, $l(R_{\rm in})$ is given by the last stable circular orbit. Summarizing:
\begin{equation}
	l(R_{\rm in})= \begin{cases}
		l_{\rm K}(R_{\rm NS}),& {\rm for}\,R_{\rm NS}>r_{\rm lso}\Rightarrow R_{\rm in}=R_{\rm NS},\\
		 l_{\rm lso},& {\rm for}\,R_{\rm NS}\leq r_{\rm lso}\Rightarrow R_{\rm in}=r_{\rm lso}.
	\end{cases}
	\label{eq:lrin}
\end{equation}
We show hereafter the results for three selected NS nuclear EOS: NL3, TM1 and GM1 \citep{2015PhRvD..92b3007C}. For these EOS and assuming that the NS is initially non-rotating, we have that $r_{\rm lso} = 6 G M_{\rm NS}/c^2> R_{\rm NS}$ for $M_{\rm NS}\gtrsim \left[ 1.78, 1.71, 1.67 \right]~M_\odot$, for the NL3, TM1 and GM1 EOS, respectively. 

For the axially symmetric exterior spacetime around a rotating NS, $l_{\rm lso}$ is well approximated  by \citep{2015ApJ...812..100B}:
\begin{equation}
	l_{\rm lso} \approx2\sqrt{3} \frac{G M_{\rm NS}}{c}\left[1-\frac{1}{10}\left( \frac{j_{\rm NS}}{M_{\rm NS}/M_\odot} \right)^{0.85} \right],
	\label{eq:lmb}
\end{equation}
for co-rotating particles. 

On the contrary, for $M_{\rm NS}\lesssim 1.7~M_{\odot}$ we have $r_{\rm lso} < R_{\rm NS}$ and thus $R_{\rm in}=R_{\rm NS}$. We shall adopt for this case the Hartle's slow-rotation approximation. The angular momentum per unit mass of a Keplerian orbit with a radius equal to the NS radius is, within this approximation, given by \citep{2016IJMPA..3141006B}
\begin{eqnarray}
    l_{\rm K}\left(u\right)&=&\frac{GM_{\rm NS}} {c\sqrt{u\,(1-3u)}} \left[ 1 - j_{\rm NS} \frac{3\,u^{3/2} (1-2u)} {1-3u}\right.\nonumber \\
		&+&\left.j_{\rm NS}^2 \frac{u^4 (3-4u)}{(1-2u)^2(1-3u)} \right],
    \label{eq:lK}
\end{eqnarray}
where $u\equiv G M_{\rm NS}/(c^2 R_{\rm NS})$. This formula can be also obtained by taking the second order slow rotation limit of the angular momentum of the last stable circular orbit around a Kerr BH \citep{1974bhgw.book.....R,2012PhRvD..86f4043B,2016IJMPA..3141006B}.

Therefore, by solving (numerically) simultaneously equations (\ref{eq:Mb_evol}) and (\ref{eq:Jns_evol}), with the aid of Eqs.~(\ref{eq:Mns_evol}--\ref{eq:lK}), it is possible to follow the evolution of the NS mass and angular  momentum during the accretion process.

\section{Induced gravitational collapse of the NS}\label{sec:4}

We proceed now to calculate the binary parameters which discriminate systems in which the NS can reach by accretion its critical mass ($M_{\rm crit}$) and consequently collapse to a BH, from the systems in which the accretion is not sufficient to induce such a collapse. 

The stability of the accreting NS is limited by two main instability conditions: the mass-shedding or Keplerian limit, and the secular axisymmetric instability. Mass-shedding occurs when the centrifugal force balances the gravitational one. Thus, for a given gravitational mass (or central density), it is given by the rotating configuration with angular velocity equal to the the Keplerian velocity of test-particles orbiting at the star's equator. In this limit the matter at the surface is marginally bound to the star and small perturbations will cause mass loss to bring the star stable again or otherwise to bring it to a point of dynamical instability
point \citep{2003LRR.....6....3S}.

At the secular axisymmetric instability point the star the star is unstable against axisymmetric perturbations. It is expected to evolve first quasi-stationarily to then find a dynamical instability point where gravitational collapse takes place \citep{2003LRR.....6....3S}. Using the turning-point method \citep{1988ApJ...325..722F}, \citet{2015PhRvD..92b3007C} computed the critical mass due to this instability point for the NL3, GM1 and TM1 EOS. They showed that the numerical results of the critical NS mass are well fitted, with a maximum error of 0.45\%, by the formula
\begin{equation}\label{eq:Mcrit}
M_{\rm NS}^{\rm crit}=M_{\rm crit}^{J=0}(1 + k j_{\rm NS}^p),
\end{equation}
where the parameters $k$ and $p$ depends of the nuclear EOS and $M_{\rm crit}^{J=0}$ is the critical mass in the non-rotating case (see table \ref{tb:StaticRotatingNS}). It can be checked that the latter is, as expected, below the $3.2~M_\odot$ critical mass upper bound by \citep{1974PhRvL..32..324R}.
\begin{table*}
\centering
\caption{Critical mass (and corresponding radius) for selected parameterizations of nuclear EOS obtained in \citet{2015PhRvD..92b3007C}.}\label{tb:StaticRotatingNS}
\begin{tabular}{cccccccc}
\hline \hline
EOS  &  $M_{\rm crit}^{J=0}$~$(M_{\odot})$ & $R_{\rm crit}^{J=0}$~(km) & $M_{\rm max}^{J\neq 0}$~$(M_{\odot})$ & $R_{\rm max}^{J\neq 0}$~(km) &$p$&$k$ & $f_{K}$~(kHz) \\ 
\hline
NL3 & $2.81$&$13.49$ &$3.38$ & 17.35 & $1.68$&$0.006$ & $1.34$\\
GM1 & $2.39$&$12.56$ &$2.84$& 16.12&$1.69$&$0.011$ & $1.49$\\
TM1 & $2.20$ &$12.07$ &$2.62$ & 15.98 &$1.61$&$0.017$ & $1.40$\\  
\hline
\end{tabular}
\tablecomments{In the last column we have also reported the rotation frequency of the critical mass configuration in the rotating case. This value corresponds to the frequency of the last configuration along the secular axisymmetric instability line, i.e the configuration that intersects the Keplerian mass-shedding sequence.}
\end{table*}

Thus, a NS with initial mass $M_{\rm NS}(t_0)$ can reach $M_{\rm crit}$ if it accretes an amount of mass $\Delta M_{\rm acc}=M_{\rm crit}-M_{\rm NS}(t_0)$ from the supernova ejecta. Given the initial NS mass, the CO core mass, and the supernova ejecta profile and its velocity, the accretion rate increases for shorter binary separation, namely for shorter orbital periods \citep{2015ApJ...812..100B,2014ApJ...793L..36F}. Therefore, there exists a maximum orbital period, denoted here to as $P_{\rm max}$, up to which, given $M_{\rm NS}(t_0)$ (and all the other binary parameters), the NS can accrete this precise amount of mass, $\Delta M_{\rm acc}$.

\begin{figure*}
\includegraphics[width=0.48\hsize,clip]{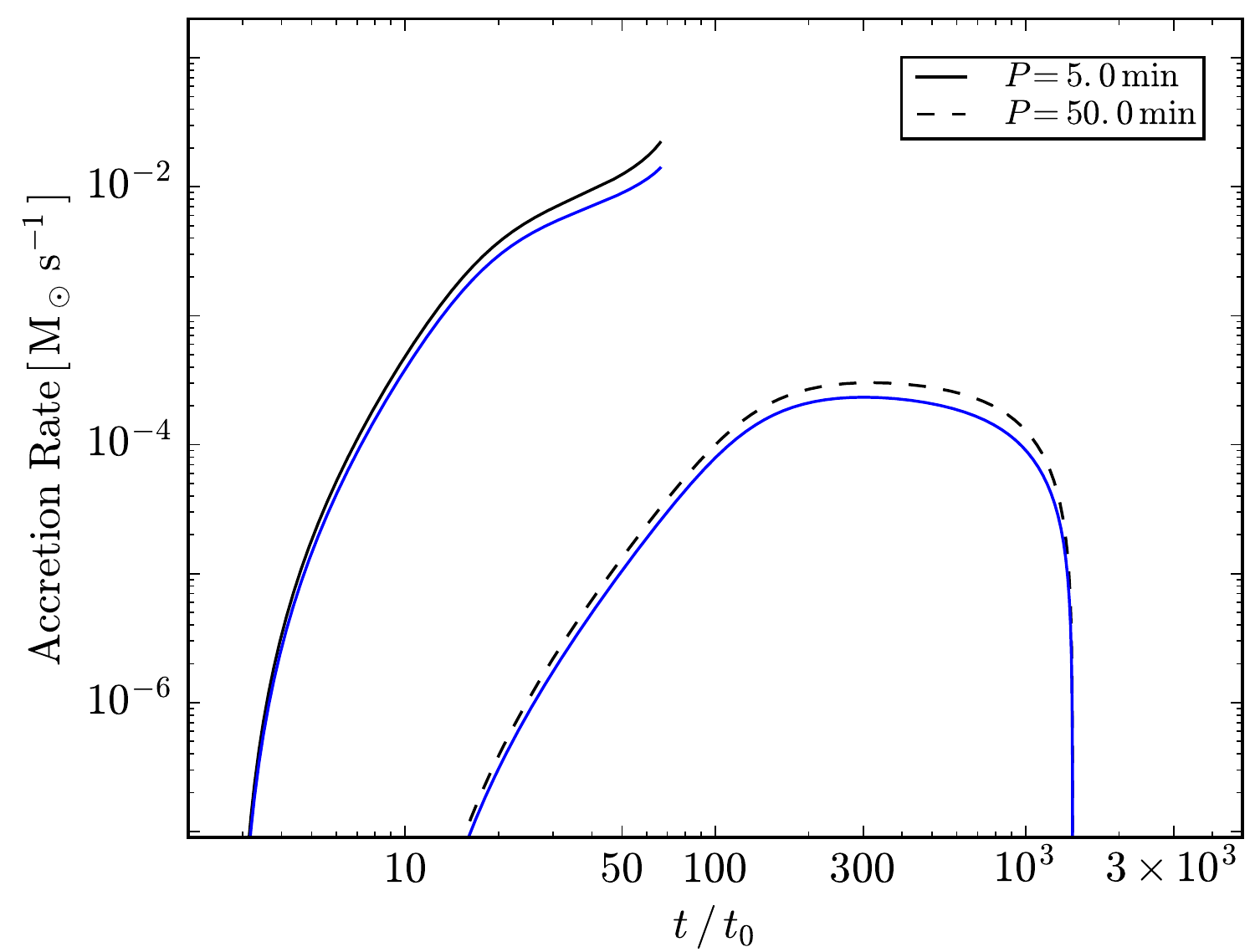}\includegraphics[width=0.48\hsize,clip]{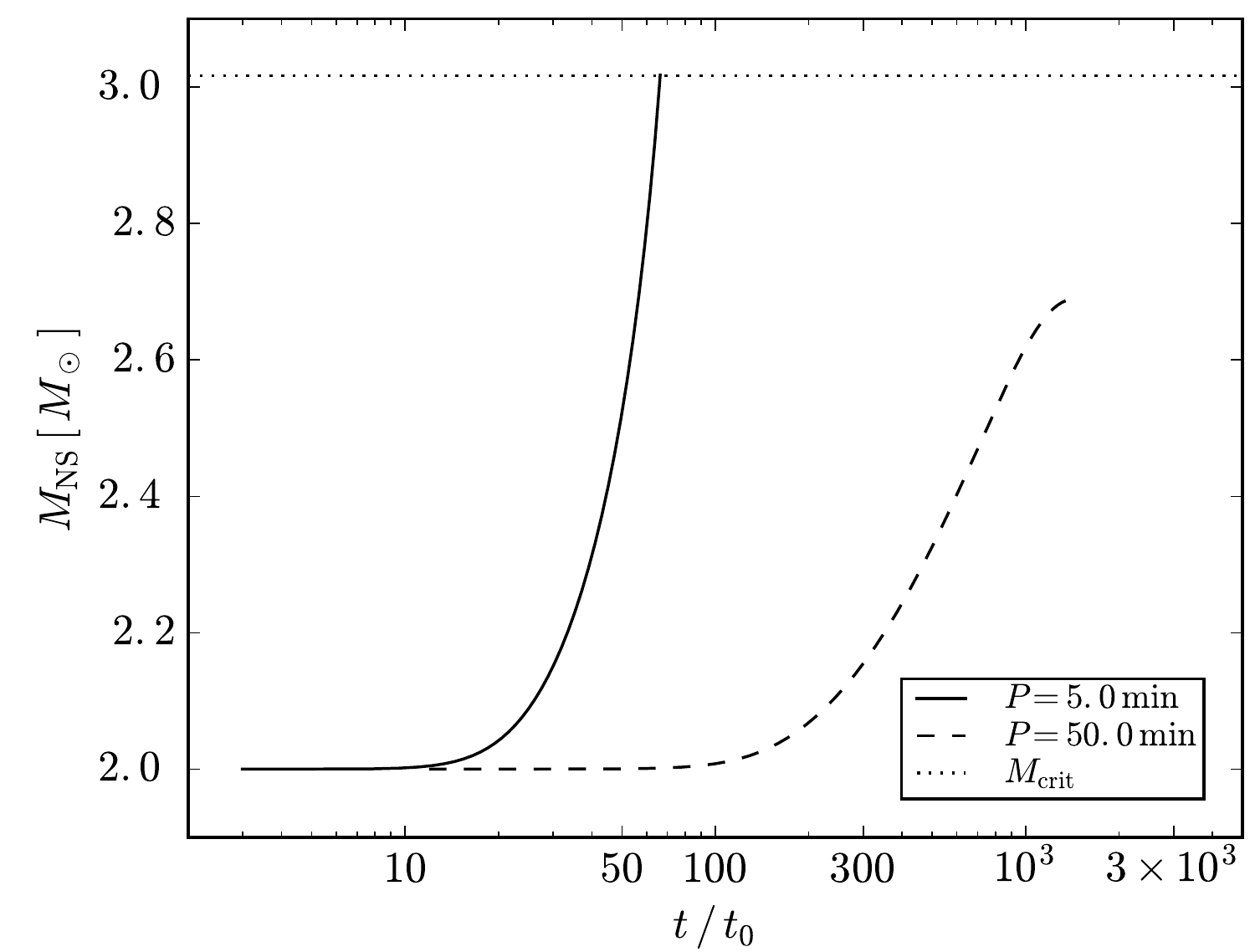}
\caption{
Left panel: time evolution of the baryonic mass accretion rate (black curve) obtained from Eq.~(\ref{eq:Mb_evol}) and rate of increase of the NS gravitational mass obtained from Eq.~(\ref{eq:Mns_evol}). Right panel: Evolution of the NS gravitational mass. We use here the ejecta from the explosion of a CO core by the $30~M_\odot$ ZAMS progenitor, $v_{\rm star,0}=2\times10^9$~cm~s$^{-1}$, expansion parameter $n=1$ and angular momentum transfer efficiency $\xi=0.5$. Two binary periods are here used: $P=5$~min~$<P_{\rm max}$ (solid curves) and $P=50$~min~$>P_{\rm max}$ (dashed curves). For this binary parameters $P_{\rm max}\approx 26$~min. It can be seen only the NS in the system with $P<P_{\rm max}$ accretes enough matter to reach the critical mass (dotted line) for gravitational collapse.
}\label{fig:NSevolution}
\end{figure*}

For example, for a NS with an initial gravitational mass $M_{\rm NS}(t_0)=2~M_\odot$ accreting the ejected material from the supernova explosion of the $30~M_\odot$ ZAMS progenitor (see table~\ref{tb:ProgenitorSN}), $v_{\rm star,0}=2\times10^9$~cm~s$^{-1}$, expansion parameter $n=1$ and angular momentum transfer efficiency $\xi=0.5$, we find $P_{\rm max}\approx 26$~min. Fig.~\ref{fig:NSevolution} shows the evolution of such a NS for two different binary periods, $P=5$~min~$<P_{\rm max}$ and $P=50$~min~$>P_{\rm max}$. We can see that only for the system with $P<P_{\rm max}$ the NS accretes enough matter to reach the critical mass for gravitational collapse, given by Eq.~(\ref{eq:Mcrit}).

\begin{figure*}
	\centering
	\includegraphics[width=0.49\hsize,clip]{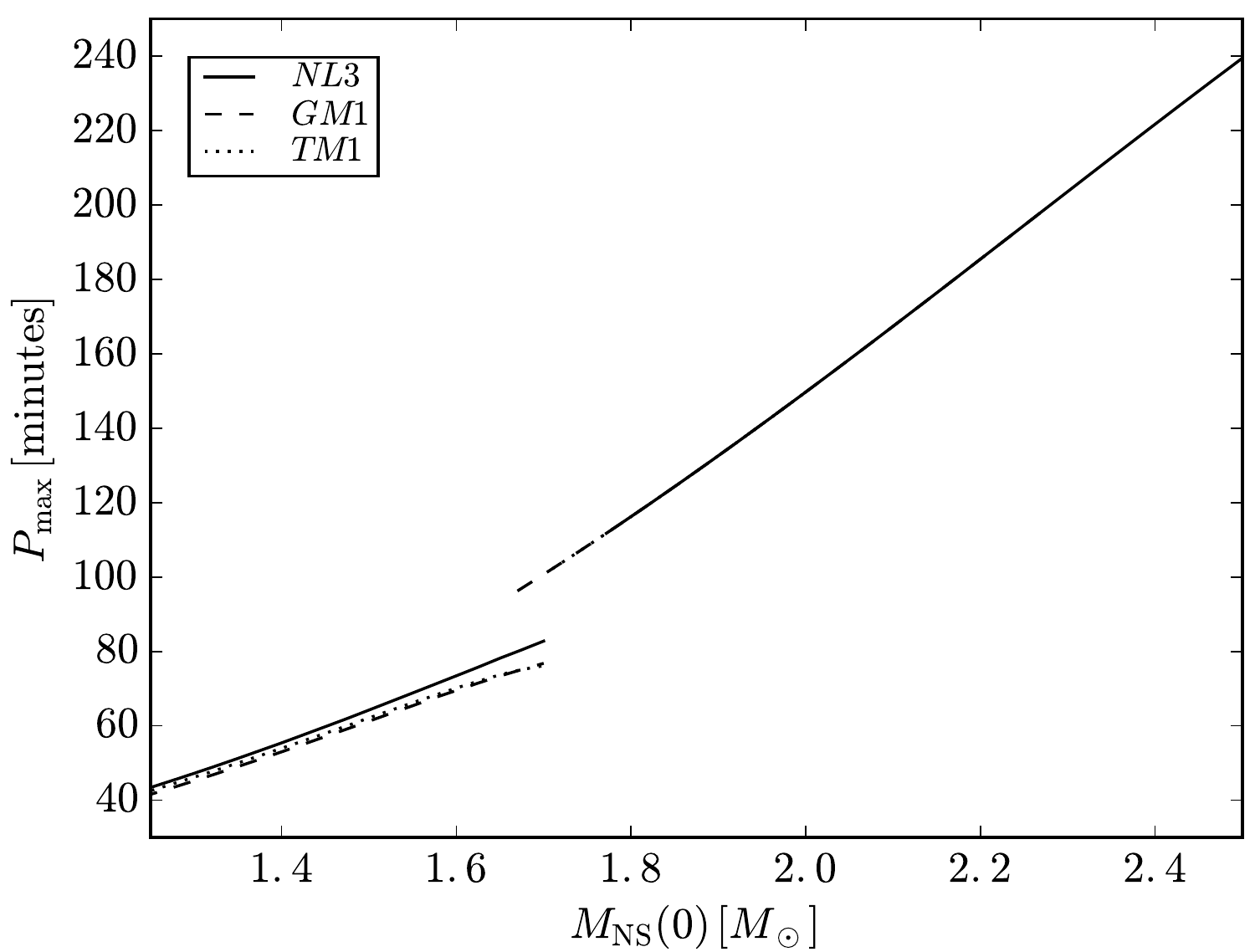}\includegraphics[width=0.49\hsize,clip]{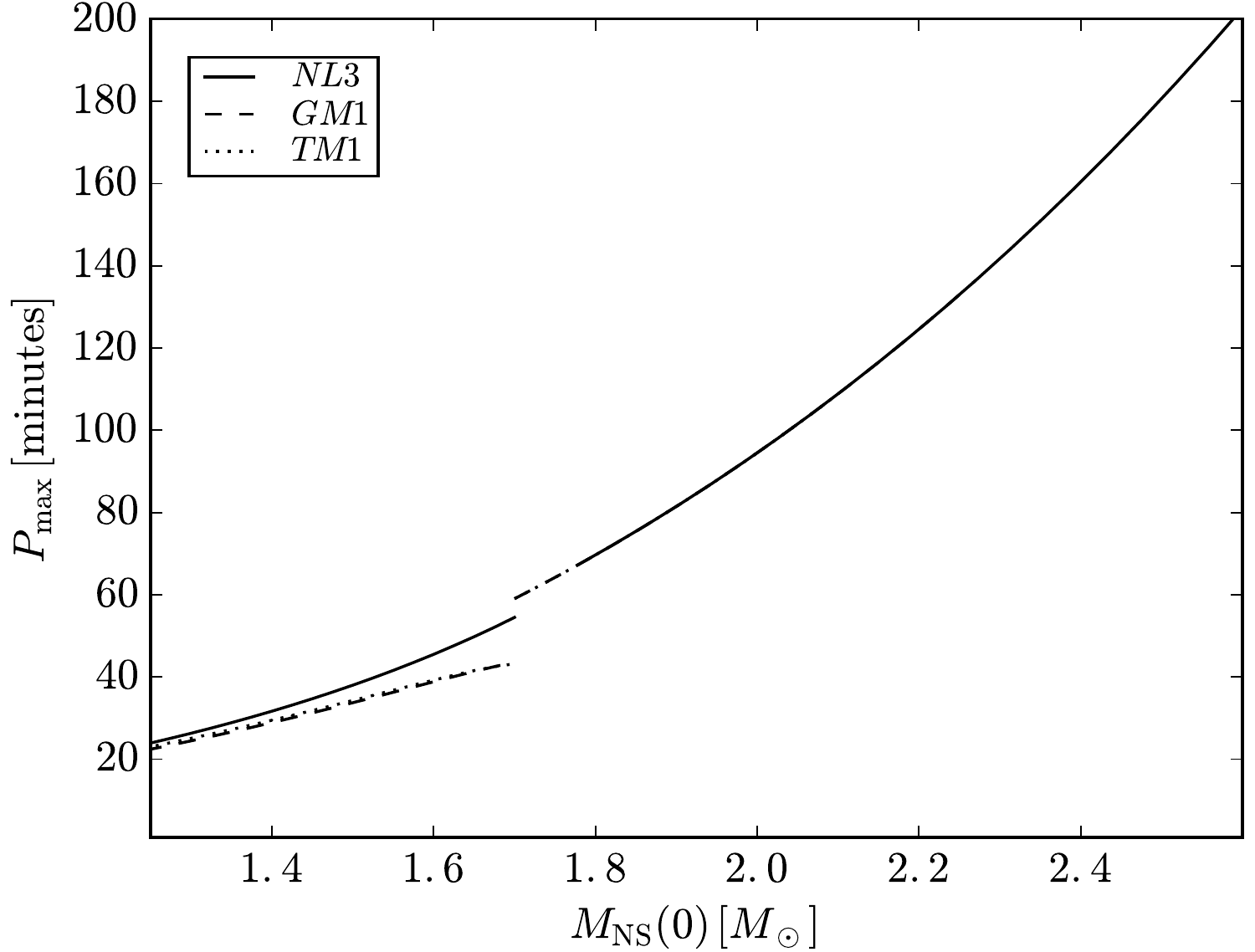}
	\caption{Maximum orbital period for which the NS with initial mass $M_{\rm NS}(0)$ collapses to a BH by accretion of supernova ejecta material. We have adopted a free expansion for the supernova ejecta ($n=1$), an outermost supernova layer velocity $v_{\rm star,0}=2\times 10^9$~cm~s$^{-1}$. The left and right panels show the results for CO cores left by $M_{\rm ZAMS}=20~M_\odot$ and $30~M_\odot$ progenitors, respectively (see table~\ref{tb:ProgenitorSN}). The apparent transition at $M_{\rm NS}(0)\approx 1.7~M_\odot$ is explained as follows: configurations with $M_{\rm NS}(0)\lesssim 1.7~M_\odot$ have the disk extending up to the NS surface, correspondingly we used the angular momentum per unit mass given by Eq.~(\ref{eq:lK}), instead for larger initial masses the accretion occurs from the last stable orbit and we used equation~(\ref{eq:lmb}). Thus, the difference around this transition point are attributable to the use of the slow rotation approximation for masses $M_{\rm NS}(0)< 1.7~M_\odot$. See the text for more details.}
	\label{fig:Pmax}
\end{figure*}

Fig.~\ref{fig:Pmax} shows $P_{\rm max}$, obtained from our numerical simulations, for different values of the NS initial gravitational mass, keeping all the other binary parameters fixed. In this figure we show the results for pre-supernova properties listed in table \ref{tb:ProgenitorSN} for the CO progenitors with $M_{\rm ZAMS}=20~M_\odot$ (left panel) and $30~M_\odot$ (right panel), a free expansion for the supernova explosion ($n=1$), and a velocity of the outermost supernova ejecta layer, $v_{\rm star,0}=2\times10^9$~cm~s$^{-1}$. 

A few comments on Fig.~\ref{fig:Pmax} are in order:
\begin{enumerate}
\item
The increase of $P_{\rm max}$ with the initial NS mass value $M_{\rm NS}(0)$ can be easily understood from the fact that the larger $M_{\rm NS}(0)$ the lower the amount of mass needed to reach the critical NS mass.
\item  
There is a transition in the behavior at $M_{\rm NS}(0)\approx 1.7~M_\odot$. This occurs because configurations with $M_{\rm NS}(0)\lesssim 1.7~M_\odot$ have the disk extending up to the NS surface, correspondingly we used the angular momentum per unit mass given by equation~(\ref{eq:lK}). For larger initial masses, accretion occurs from the last stable orbit and we used equation~(\ref{eq:lmb}). Thus, the difference around this transition point are attributable to the use of the slow rotation approximation for masses $M_{\rm NS}(0)< 1.7~M_\odot$. We recall that \citet{2015ApJ...812..100B} considered only initial NS masses $M_{\rm NS}(0)\gtrsim 1.7~M_\odot$ so this transition is not observed there.
\item
In \citet{2015ApJ...812..100B} an angular momentum transfer efficiency parameter $\xi=0.5$ was adopted. In order to see the effect of such a parameter we adopted here in the simulations the maximum possible value $\xi=1$. Values of $\xi$ lower than unity account for possible angular momentum losses between the inner disk radius and the NS surface. This implies that the values of $P_{\rm max}$ in Fig.~\ref{fig:Pmax} are upper limits to the maximum orbital period for BH formation. Namely, a value $\xi<1$ leads to lower values of $P_{\rm max}$. For instance, in the right panel of Fig.~\ref{fig:Pmax} we see that for $M_{\rm NS}(0)=1.8~M_\odot$ and the NL3 EOS, $P_{\rm max}\approx 70$~min. We checked that, for $\xi=0.5$, the same initial mass and EOS would instead lead to $P_{\rm max}\approx 20$~min. 
\item 
Because of the highly efficient angular momentum transfer ($\xi=1$), the NS in the systems of Fig.~\ref{fig:Pmax} ends at the mass-shedding limit. In the case of lower values of $\xi$, the NS might end directly at the secular axisymmetric instability with a lower values of the critical mass with respect to the maximum mass along the Keplerian mass-shedding sequence. We have checked, for instance in the case of $\xi=0.5$ (the one adopted in \citet{2015ApJ...812..100B}) and the NL3 EOS, that this occurs when the initial NS mass is close to the non-rotating critical mass value, e.g. for $M_{\rm NS}(0)\gtrsim 2.2~M_\odot$.
\item
We recall that in \citet{2015ApJ...812..100B} only the case of the $M_{\rm ZAMS}=30~M_\odot$ progenitor was analyzed. We studied here different progenitors. At first sight, it might appear contradictory that the left panel of Fig.~\ref{fig:Pmax}, which is for a less massive CO core with respect to the one the right panel, shows longer values of the maximum orbital period for BH formation. The reason for this is as follows. First, the binary separation satisfies $a\propto (M_t P^2)^{1/3}$ where $M_t$ is the total binary mass. Thus, for given NS mass and binary period, a less massive CO core implies a less massive binary and a smaller orbit by a factor $a_1/a_2 = (M_{t1}/M_{t2})^{1/3}$. A tighter orbit implies a supernova ejecta density at the NS position higher by a factor $\rho_{\rm ej,1}/\rho_{\rm ej,2} = (a_2/a_1)^3 = M_{t2}/M_{t1}$, hence the accretion rate which is proportional to the density [see equation~(\ref{eq:Bondi-HoyleRate})]. 
\end{enumerate} 

Thus we have shown that in systems with $P\leq P_{\rm max}$ the induced gravitational collapse of the accreting NS to a BH occurs. These systems explain the BdHNe \citep{2015PhRvL.115w1102F,2015ApJ...812..100B,2014ApJ...793L..36F}. 
We do not simulate in this work the complex process of gravitational collapse rather we assume BH formation at the moment when the NS reaches the critical mass value. We also adopt the mass of the newly formed BH as given by the critical NS mass value.

In systems with $P > P_{\rm max}$, the NS does not accrete enough matter from the supernova ejecta and the collapse to a BH does not occur. As we show below, these systems explain XRFs (see also Ruffini et al., in preparation, for more details). 

\section{Supernova ejecta asymmetries induced by the NS}\label{sec:5}

\begin{figure*}
\centering
\includegraphics[width=0.48\hsize,clip]{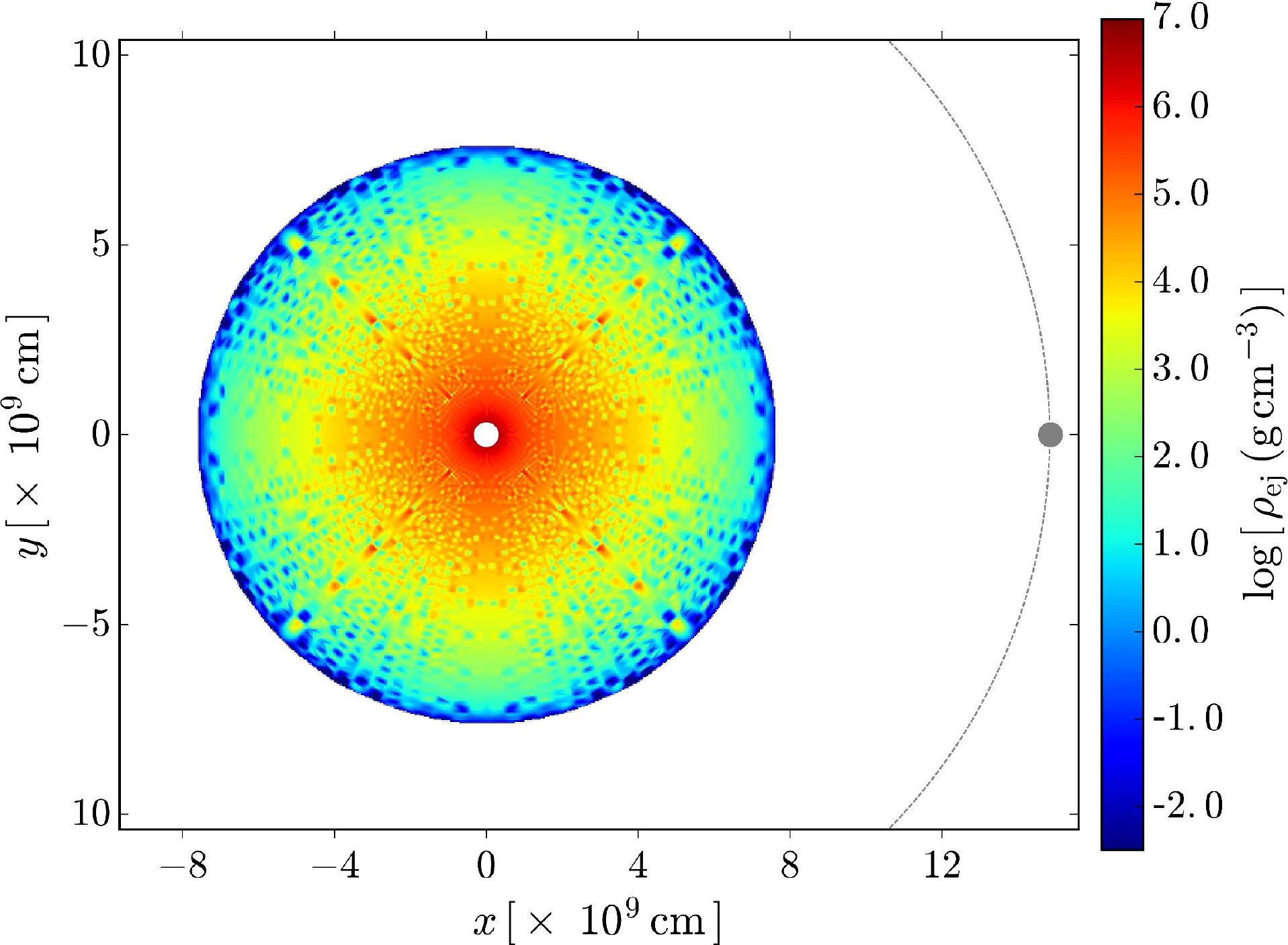}\includegraphics[width=0.48\hsize,clip]{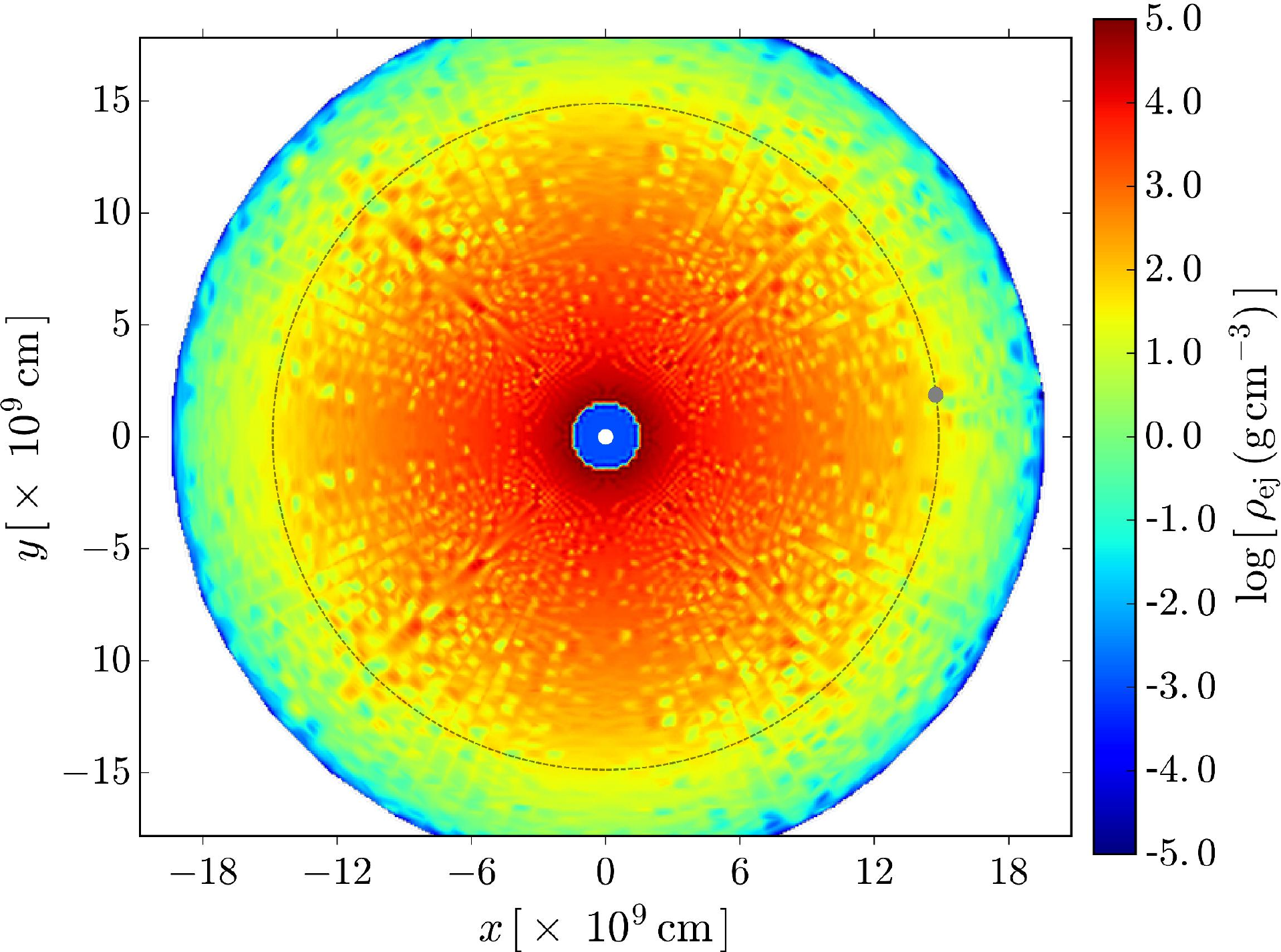}\\
\includegraphics[width=0.48\hsize,clip]{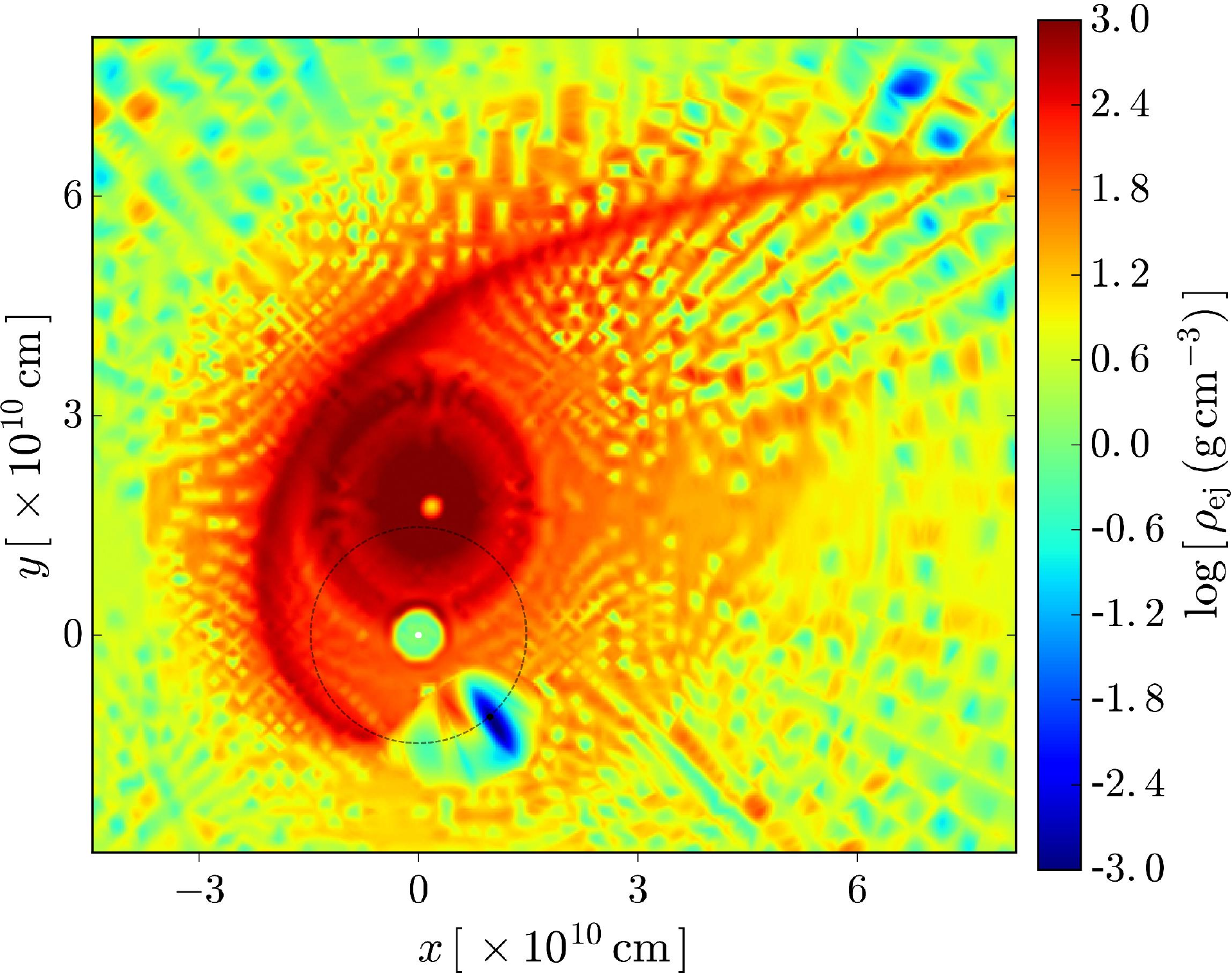}\includegraphics[width=0.48\hsize,clip]{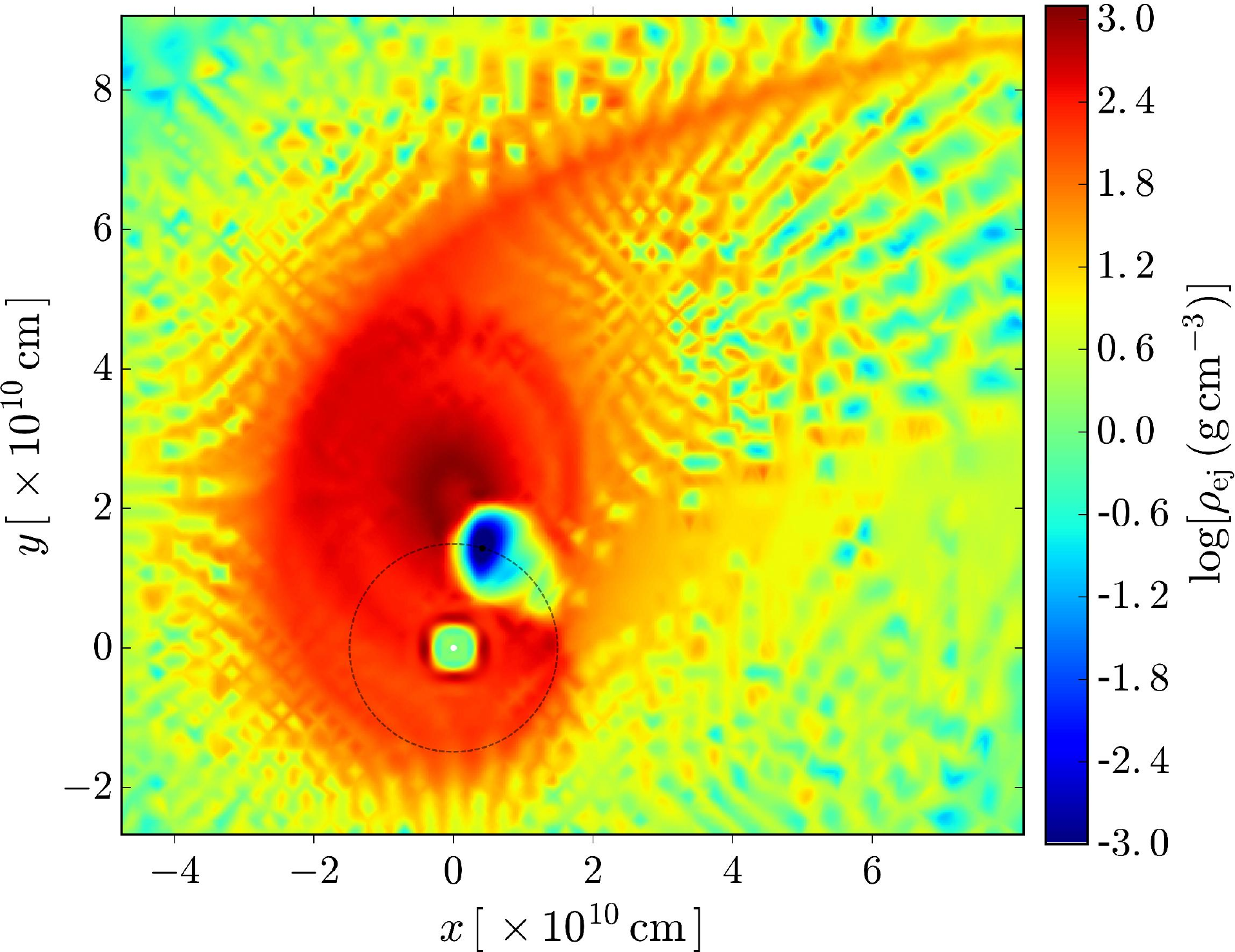}\\
\caption{Snapshots of the IGC binary system and the supernova ejecta density at selected times of the evolution. In this example we have adopted the $M_{\rm ZAMS}=30~M_\odot$ progenitor with an ejecta mass of $7.94~M_\odot$ and a core that left a $\nu$NS of $1.5~M_\odot$. We assume homologous evolution of the supernova ejecta with expansion parameter $n=1$ and ejecta outermost layer velocity $v_{\rm 0,star}=2\times 10^9$~cm~s$^{-1}$. For the NS we adopt an initial mass $2.0~M_{\odot}$. The binary has an orbital period $P\approx 5$~min, i.e.~a binary separation distance $a\approx 1.5\times 10^{10}$~cm. 
The evolution of the accretion rate and the gravitational mass of the NS are shown in Fig.~\ref{fig:NSevolution}.
The coordinate system is centered on the $\nu$NS born in the supernova: it is here represented with a white-filled circle located at $(0,0)$. 
The NS, {\rm represented by the gray-filled circle}, is orbiting counterclockwise in this time-evolution and we have indicated with a thin-dashed circle its trajectory. The colorbar indicates values of ejecta density. We have chosen a thickness $\Delta z$ around the orbital plane to estimate the ejecta density: $\Delta z\approx 0.05 a \approx 7.1\times 10^8$~cm for all these figures. We note that, in order to show better the features of the system at different times, we have chosen for each plot different ranges of the x-y scales and of the colorbar. \emph{Left upper panel:} initial time of the process, $t=t_0=R_{\rm star}^0/v_{\rm star,0}=3.82$~s. The supernova ejecta starts to expand radially outward and the NS (black filled circle) is located at the position $(a,0)$. \emph{Right upper panel:} beginning of the accretion process, i.e.~passage of the first supernova ejecta layers through the NS gravitational capture region. Thus, this time is $t = t_{\rm acc,0}\approx a/v_{\rm 0,star}=7.7$~s. \emph{Left lower panel:} instant when the NS reaches, by accretion, the critical mass and collapses to a BH. This occurs at $t = t_{\rm coll} \approx 254$~s~$\approx 0.85 P$. 
The BH, here represented by the black-filled circle, has a mass set by the critical NS mass, i.e. $M_{\rm BH}=M_{\rm crit}\approx 3~M_\odot$ (see Fig.~\ref{fig:NSevolution}).
It can be seen here the asymmetry of the supernova ejecta density induced that have been generated by the nearby presence of the NS and the accretion process onto it. We can also see at this stage a stream of matter (of negligible mass) being expelled (i.e. reaching scape velocity) from the system. \emph{Right lower panel:} system $100$~s after the BH formation, namely $t = t_{\rm coll} + 100$~s = 354~s~$\approx 1.2 P$. This figure shows the new binary system composed by the $\nu$NS [white-filled circle at the $(0,0)$ position] out of the supernova, and a BH [black-filled circle at the $(0.5,1.7)$ position] out of the gravitational collapse of the NS due to the hypercritical accretion process. The asymmetry of the supernova ejecta is now even larger than the one showed by the left lower panel figure. The asymmetry of the supernova ejecta is such that its `center' has been displaced, from the explosion site originally at the position $(0,0)$, to the approximate position $(0,2)$, due to the action of the orbiting NS.}
\label{fig:denassym1a}
\end{figure*}

For isolated supernova explosions, or for very wide binaries with negligible accretion rates, the density of the supernova ejecta would approximately follow the homologous evolution given by equation~(\ref{eq:rhoej_sn}) with constant ejecta mass, i.e. $M_{\rm env}(t)=M_{\rm env}^0$, i.e. the density will decrease with time following a simple power-law $\rho_{\rm ej} (t)\propto t^{-3n}$, with $n$ the expansion parameter, keeping its spherical symmetry about the explosion site (see, e.g., Fig.~\ref{fig:denassym1a}). However, for explosions occurring in close binaries with compact companions such as NSs or BHs, the supernova ejecta is subjected to a strong gravitational field which produces at least two non-negligible effects: 1) an accretion process on the NS that subtracts part of the ejecta mass, and 2) a deformation of the supernova fronts closer to the accreting NS companion. As we show below, the conjunction of these effects can generate large changes in the density profile of the ejecta in a region around the orbital plane.

\begin{figure}
\centering
\includegraphics[width=\hsize,clip]{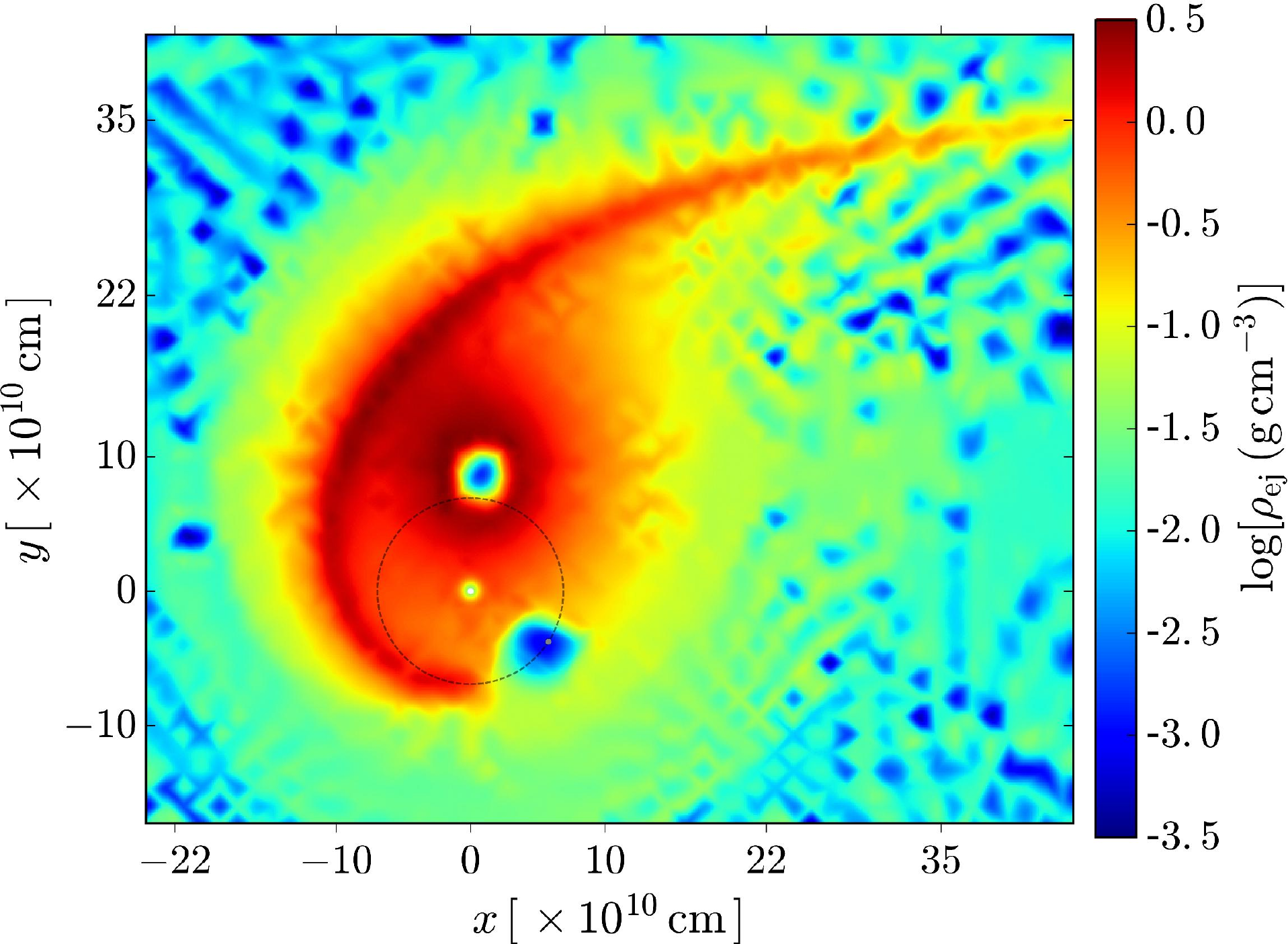}
\caption{Snapshot of an IGC binary system analogous to the one of Fig.~\ref{fig:denassym1a} but with an orbital period of $P\approx 50$~min (i.e.~binary separation $a\approx 7\times 10^{10}$~cm). In this case the accreting NS (gray-filled circle), which is orbiting counterclockwise (thin-dashed circle), does not collapse to the BH. The $\nu$NS left by the supernova is represented by the white filled circle at the position $(0,0)$. It is clear from this figure the asymmetry of the supernova ejecta: indeed the supernova `center' has been displaced, from the explosion site originally at the position $(0,0)$, to the approximate position $(0,9)$, due to the action of the orbiting NS. The snapshot corresponds to a time $t = 2667$~s~$\approx 44$~min, which corresponds to roughly 1/4 the total accretion process. To estimate the ejecta density we have adopted $\Delta z = 0.08 a \approx 5.3\times 10^9$~cm.}\label{fig:denassym2}
\end{figure}

In order to visualize the above effects we have simulated the evolution of the supernova layers in presence of the NS during the accretion process (see Figs.~\ref{fig:denassym1a} and \ref{fig:denassym2}). Thus, we followed the three-dimensional motion of $N$ particles in the gravitational field of the orbiting NS. 
Following \citet{2015ApJ...812..100B}, we consider the gravitational field of the NS on the supernova ejecta including the effects of the orbital motion as well as the changes in the NS gravitational mass as described above in Sec.~\ref{sec:2} via the Bondi formalism. The supernova matter is described as formed by point-like particles whose trajectory was computed by solving the Newtonian equation of
motion. We plan to do an SPH simulation of this process and expect to present the results in a forthcoming publication.

The initial conditions of the supernova ejecta are computed assuming the supernova layers move via homologous velocity distribution in free expansion (i.e.~evolving with $n=1$).
The initial power-law density profile of the CO envelope is simulated by populating the inner layers with more particles, as follows. The total number of particles is $N=N_r\times N_\theta \times N_\phi$ and for symmetry reasons, we simulate only the north hemisphere of the supernova; thus the polar and azimuthal angles are divided as $\Delta \theta = (\pi/2)/N_\theta$ and $\Delta \phi = 2\pi/N_\phi$, respectively. For the radial coordinate we first introduce the logarithmic coordinate $x=\log(r)$ and $\Delta x=(x_s-x_c)/N_r$, where $x_s = \log(R_{\rm star})$ and $x_c = \log(R_{\rm core})$. Thus,the thickness of each layer is $\Delta r=r_i (10^{\Delta x}-1)$, where $r_i$ is the location of the layer. The mass of each particle of the $i$-layer is: $m_i=4\pi r_i^3 \ln(10)\Delta x \rho(r_i)/(2 N_\theta N_\phi)$. 

Let us assume, for the sake of example, the $M_{\rm ZAMS}=30~M_\odot$ progenitor of table~\ref{tb:ProgenitorSN} which gives a CO core with envelope profile $\rho_{\rm ej}^0 \approx 3.1\times 10^8 (8.3\times 10^7/r)^{2.8}$~g~cm$^{-3}$ and $R^0_{\rm star}=7.65\times 10^9$~cm. This implies that, for a total number of $N=10^6$ particles in the simulation, the particles of the innermost radius $r_i=R_{\rm core} = 8.3\times 10^7$~cm with density $\rho^0_{\rm ej}(r_i) = 3.1\times 10^8$~g~cm$^{-3}$ have mass $m_i \approx 2\times 10^{-5}~M_\odot$ while, the particles of the outermost radius $r_i = R^0_{\rm star}$, would have $m_i \approx 6\times 10^{-6}~M_\odot$. In addition, we assume that particles crossing the Bondi-Hoyle radius are captured and accreted by the NS so we removed them from the system as they reach that region. We removed these particles according to the results obtained from the numerical integration of Eq.~(\ref{eq:Mb_evol}).

Fig.~\ref{fig:denassym1a} shows in detail the orbital plane of an IGC binary at selected times of its evolution. The NS has an initial mass of $2.0~M_\odot$; the CO core is the one obtained by the $M_{\rm ZAMS}=30~M_\odot$ progenitor (see table~\ref{tb:ProgenitorSN}), which leads to a total ejecta with mass $7.94~M_\odot$ and an iron core that left a $\nu$NS of $1.5~M_\odot$. The orbital period of the binary is $P\approx 5$~min, i.e. a binary separation $a\approx 1.5\times 10^{10}$~cm, and we have adopted an angular momentum transfer efficiency parameter $\xi=0.5$. 
The evolution of the accretion rate and the gravitational mass of the NS in this system are the ones shown in Fig.~\ref{fig:NSevolution}.
As it can be seen, for the above parameters the NS reaches the critical mass and collapses to form a BH (see also Fig.~\ref{fig:Pmax} and conclusion 3 in Sec.~\ref{sec:6}).

In the simulation shown in Fig.~\ref{fig:denassym1a} we adopted two millions of particles per solar mass of ejecta so in this simulation we have followed the three-dimensional motion of $N=2\times 10^6 (M_{\rm env}^0/M_\odot) \approx 1.6\times 10^7$ particles in the gravitational field of the orbiting NS. To estimate the ejecta density we have chosen a thickness $\Delta z$ around the orbital plane. For the plots in Fig.~\ref{fig:denassym1a} we have adopted $\Delta z\approx 0.05 a \approx 7.1\times 10^8$~cm. 

The left upper panel shows the binary at the initial time of the process, i.e. $t=t_0=R_{\rm star}^0/v_{\rm star,0}=3.82$~s, the first instant of the ejecta radial expansion.

The right upper panel shows the instant at which the accretion process begins, namely at $t = t_{\rm acc,0}\approx a/v_{\rm 0,star}=7.7$~s.
Owing to their fast velocity, the accretion rate of the first layers is low and they escape almost undisturbed, so the supernova ejecta at these times keeps its original spherical symmetry.

The left lower panel shows the binary at the instant in which the accreting NS reaches the critical mass, hence the instant of formation of the BH, at $t = t_{\rm coll}= 254$~s~$\approx 0.85 P$. 
The BH mass is thus set by the critical NS mass, i.e. $M_{\rm BH}=M_{\rm crit}\approx 3~M_\odot$ (see Fig.~\ref{fig:NSevolution}).
This figure also evidences the asymmetry on the supernova density as induced by the presence of the companion and its increasing gravitational field due to the ongoing accretion process onto it. Indeed, it can be seen how the supernova `center' has been shifted from the explosion site originally at the $(0,0)$ position (see left upper panel), to the approximate position $(0,2)$. 
Thus, the layers of the ejecta are displaced as a result of the gravitational attraction of the orbiting NS. This can be understood as follows. When the NS passes over the northern hemisphere, it attracts the northern region of the ejecta towards it. Consequently, that part of the ejecta gain velocity in the northern direction. The same effect occurs in the other regions of the orbit, however the effect is asymmetric because, by the time the NS passes, say over the southern hemisphere, it attracts layers moving at slower velocity with respect to the ones it had attracted in the northern hemisphere before. The reason for this is that the southern fastest layers have moved further already while the NS were passing the northern side. This effect is indeed incorporated in our simulation which follows the trajectory of each of the 14 million particles.

It can be also seen in this figure a stream of matter {\bf (one-armed flow)} of negligible mass with respect to the total mass escaping from the system. 
As we have mentioned above, the NS attracts some layers increasing their velocity. As a result some material can reach escape velocity to leave the binary system forming this unbound debris. The appearance of a one-armed flow (instead of two-armed flows) is because the center-of- mass is located roughly at the CO core position, thus the NS is in practice orbiting the CO core. If the two masses were of comparable masses, (e.g.~as in the case of binary NS mergers), they would move around a common center-of-mass lying in between them. In such systems the momentum transfer is more symmetric leading to a symmetric two-armed flow structure. The one-armed flow in our system is, in this sense, more similar to the one that appears in the tidal disruption of a small body by a supermassive BH.

The right lower panel shows the system $100$~s after the BH formation, namely at $t = t_{\rm coll} + 100$~s =354~s~$\approx 1.2 P$. Thus, this figure shows the new binary system formed by the $\nu$NS, out of the supernova, and the BH from the gravitational collapse of the NS. The $\nu$NS is at the $(0,0)$ position and it is represented by the white filled circle. The BH is in this instant of time located at the $(0.5,1.7)$ position and it is represented by the black filled circle. It can be seen the increasing asymmetry of the supernova ejecta around the orbital plane. We note the presence of ejecta in the vicinity of the newly formed BH, the latter sited at the approximate position $(0.5,1.5)$. It is interesting that part of these ejecta can indeed cause a subsequent accretion process onto the newly formed BH. The possible outcomes of this process deserve further attention which will be analyzed elsewhere (Ruffini et al., in preparation).

We have shown above the evolution of an IGC binary with very short orbital period of $P\approx 5$~min, for which it occurs the gravitational collapse of the NS of the CO core. Besides the formation of a BH, we have evidenced the asymmetry caused by the presence and accretion onto the NS on the supernova ejecta density. It is natural to ask if these asymmetries also appear for less compact binaries. For comparison, we show in Fig.~\ref{fig:denassym2} the results of a numerical simulation for a binary with orbital period $P\approx 50$~min, in which the NS does not reach the critical mass during the entire accretion process (see Sec.~\ref{sec:6}). 
The evolution of the accretion rate and the gravitational mass of the NS in this system are shown in Fig.~\ref{fig:NSevolution}.

In these kind of systems, all the ejecta layers passed the NS position. Thus, the total duration of the accretion process, denoted here $t_{\rm acc}$, is approximately given by the time it takes to the innermost layer of the ejecta to overcome the NS position, i.e.~$t_{\rm acc}\approx a/v_{\rm inner}$, where $v_{\rm inner} = (\hat{R}_{\rm core}/R_{\rm star}^0)v_{\rm star,0}$ using the homologous expansion assumption. The snapshot corresponds to a time $t = 2667$~s~$\approx 44$~min~$\approx t_{\rm acc}/4$. To estimate the ejecta density we have chosen in this example $\Delta z = 0.08 a \approx 5.3\times 10^9$~cm. It is interesting that although the NS is in this case farther away from the CO core, it still induces a high asymmetry on the supernova ejecta. We shall investigate elsewhere if this mechanism could explain the asymmetries observed in some type Ibc supernovae \citep[see, e.g.,]{2009ApJ...699.1119T,2009MNRAS.397..677T}. We here constrain ourselves in section \ref{sec:8}, instead, to the consequences that the ejecta asymmetries have on the supernova emission (both in X-rays and in the optical). 

\section{Hydrodynamics inside the accretion region}\label{sec:6}

We turn now to analyze in detail the properties of the system inside the Bondi-Hoyle accretion region. Fig.~\ref{fig:Mdot} shows the mass accretion rate onto the NS of initial mass $1.4~M_\odot$. We can see that the accretion rate can be as high as $\sim 10^{-1}~M_{\odot}$~s$^{-1}$. For these high accretion rates we can draw some general properties:
\begin{enumerate}
\item 
We can neglect the effect of the NS magnetic field since for $\dot{M}_B>2.6\times 10^{-8}~M_{\odot}$~s$^{-1}$ the magnetic pressure remains much smaller than the random pressure of the infalling material \citep{1996ApJ...460..801F,2012ApJ...758L...7R}.
\item 
The photons are trapped in the accretion flow. The trapping radius, defined at which the photons emitted diffuse outward at a slower velocity than the one of the infalling material, is \citep{1989ApJ...346..847C}:
\begin{equation}
	r_{\rm trapping}={\rm min}\{\dot{M}_B\kappa/(4\pi c),R_{\rm cap}\},
	\label{eq:rtrapping}
\end{equation}
where $\kappa$ is the opacity. For the CO core, \citet{2014ApJ...793L..36F} estimated a Rosseland mean opacity roughly $5\times 10^3$~cm$^2$~g$^{-1}$. For the range of accretion rates, we obtain that $\dot{M}_B\kappa/(4\pi c)\sim 10^{13}$--$10^{19}$~cm, a radius much bigger than the NS capture radius which is in our simulations at most 1/3 of the binary separation. Thus, in our systems the trapping radius extends all the way to the Bondi-Hoyle region, hence the Eddington limit does not apply and hypercritical accretion onto the NS occurs. See Figs.~\ref{fig:taus} and \ref{fig:Trho} in appendix~\ref{app:B4} for details.
\item 
Under these conditions of photons being trapped within the accretion flow, the gain of gravitational energy of the accreted material is mainly radiated via neutrino emission \citep{1972SvA....16..209Z,1973PhRvL..31.1362R,1996ApJ...460..801F,2012ApJ...758L...7R,2014ApJ...793L..36F}. See Figs.~\ref{fig:taus} and \ref{fig:Trho} in appendix~\ref{app:B4} for details.
\item 
Fig.~\ref{fig:Mdot} shows that the evolution of the accretion rate has a shape composed of a rising part, followed by an almost flat maximum and finally it decreases with time. The rising part corresponds to the passage and accretion of the first layers of the ejecta. The sharpness of the density cut-off of the outermost ejecta layer defines the sharpness of this rising accretion rate. The maximum rate is given by the accretion of the ejecta layers with velocities of the same order as the orbital velocity of the NS. These layers are located very close to the innermost part of the supernova ejecta. Then, the rate start to decrease with the accretion of the innermost layers whose density cut-off determines the sharpness of this decreasing part of the mass accretion rate. See also appendix~\ref{app:A} for further details.
\item 
The longer the orbital period/larger binary separation, the lower the accretion rate (see Fig.~\ref{fig:Mpeak} in appendix~\ref{app:A} for details.); hence the lower the accretion luminosity and the longer the time at which peak luminosity occurs. These features confirm what advanced in \citet{2016arXiv160202732R,2015ApJ...812..100B,2015ApJ...798...10R}, namely that less energetic long GRBs correspond to the binaries with wider orbits. Specifically, XRFs correspond to the binaries in which the NS does not reach the point of gravitational collapse to a BH (see below). Since there is a limiting orbital period, $P_{\rm max}$, up to which the NS can reach the critical mass and collapse to a BH (these systems are the BdHNe \citealp{2015PhRvL.115w1102F,2014ApJ...793L..36F}), the XRFs are the binaries with $P>P_{\rm max}$ (see Sec.~\ref{sec:5} for details).
\end{enumerate}

\subsection{Convective instabilities}\label{sec:6a}

As the material piles onto the NS and the atmosphere radius, the accretion shock moves outward. The post-shock entropy is a decreasing function of the shock radius position which creates an atmosphere unstable to Rayleigh-Taylor convection during the initial phase of the accretion process (see appendix~\ref{app:B} for additional details). These instabilities can accelerate above the escape velocity driving outflows from the accreting NS with final velocities approaching the speed of light \citep{2006ApJ...646L.131F,2009ApJ...699..409F}. Assuming that radiation dominates, the entropy of the material at the base of the atmosphere is \citep{1996ApJ...460..801F}:
\begin{equation}
	S_{\rm bubble} \approx 16\left( \frac{M_{\rm NS}}{1.4\,M_\odot} \right)^{7/8}\left( \frac{\dot{M}_{\rm B}}{{M_\odot\,{\rm s}^{-1}}} \right)^{-1/4}\left( \frac{r}{10^6\, {\rm cm}} \right)^{-3/8},
	\label{eq:Sbuble}
\end{equation}
in units of $k_B$ per nucleon. 

This material will rise and expand, cooling adiabatically, i.e.~$T^3/\rho$ = constant, for radiation dominated gas. If we assume a spherically symmetric expansion, then $\rho \propto 1/r^3$ and we obtain
\begin{equation}
	k_B T_{\rm bubble}=195\, S_{\rm bubble}^{-1}\left( \frac{r}{10^6\, {\rm cm}} \right)^{-1}\,{\rm MeV}.
	\label{eq:Tbubble}
\end{equation}
However, it is more likely that the bubbles expand in the lateral but not in the radial direction \citep{2009ApJ...699..409F}, thus we have $\rho \propto 1/r^2$, i.e.
\begin{equation}\label{eq:Tbubble2}
T_{\rm bubble} = T_0 (S_{\rm bubble}) \left(\frac{r_0}{r}\right)^{2/3},
\end{equation}
where $T_0(S_{\rm bubble})$ is given by equation~(\ref{eq:Tbubble}) evaluated at $r=r_0\approx R_{\rm NS}$. 

This temperature implies a bolometric blackbody flux at the source from the bubbles
\begin{eqnarray}\label{eq:Lbubble}
F_{\rm bubble} &=& \sigma T_{\rm bubble}^4 \approx 2\times 10^{40} \left(\frac{M_{\rm NS}}{1.4\,M_\odot} \right)^{-7/2}\left( \frac{\dot{M}_{\rm B}}{M_\odot\,{\rm s}^{-1}} \right)\nonumber \\
&\times&\left( \frac{R_{\rm NS}}{10^6\,{\rm cm}} \right)^{3/2}\left(\frac{r_0}{r}\right)^{8/3}\,{\rm erg\,s}^{-1}{\rm cm}^{-2},
\end{eqnarray}
where $\sigma$ is the Stefan-Boltzmann constant.

In \citet{2014ApJ...793L..36F} it was shown that the above thermal emission from the rising bubbles produced during the hypercritical accretion process can explain the early ($t\lesssim 50$~s) thermal X-ray emission observed in GRB 090618 \citep{2012A&A...543A..10I,2012A&A...548L...5I}. In that case $T_{\rm bubble}$ drops from 50~keV to 15~keV expanding from $r\approx 10^9$~cm to $6\times 10^9$~cm, for an accretion rate $10^{-2}~M_\odot$~s$^{-1}$. 

From the above formulas we can explain the blackbody emission observed in XRF 060218~\citep{2006Natur.442.1008C}. The observed temperature ($k_B T\approx 0.2$~keV) and radius of the emitter (a few $10^{11}$~cm) are consistent with the temperature and surface radius of the above bubbles formed in a system with a NS of initial mass $1.4~M_{\odot}$, supernova-progenitor of $20~M_{\rm ZAMS}$, and orbital period 2.5~h: it can be easily checked via equation~(\ref{eq:Tbubble2}) that for $r\sim 10^{11}$~cm and an accretion rate of the order of $10^{-6}~M_\odot$~s$^{-1}$, the bubbles would have a temperature consistent with the one observed in XRF 060218. Further details on this specific case and additional examples will be presented elsewhere (Ruffini et al., in preparation). 

It is worth mentioning the possibility that, as discussed in \citet{2006ApJ...646L.131F}, r-process nucleosynthesis occurs in these outflows. This implies that long GRBs can be also r-process sites with specific signatures from the decay of the produced heavy elements, possibly similar as in the case of the \emph{kilonova} emission in short GRBs \citep[see, e.g.,][and references therein]{2013Natur.500..547T}. The signatures of this phenomenon in XRFs and BdHNe, and its comparison with kilonovae, deserves to be explored. However, this is out of the scope of the present work and it will be presented elsewhere.

\subsection{Neutrino emission during hypercritical accretion}\label{sec:6b}

Most of the energy from the accretion is lost through neutrino emission and the neutrino luminosities are proportional to the
accretion rate (see appendix \ref{app:B3} for details). For the accretion rate conditions characteristic of our models $\sim 10^{-4}$--$10^{-2}~M_\odot$~s$^{-1}$ (see Sec.~\ref{sec:2} and appendix~\ref{app:A}), pair annihilation dominates the neutrino emission and electron neutrinos remove the bulk of the energy (see \citealp{2009ApJ...699..409F}, and Figs.~\ref{fig:taus} and \ref{fig:Trho} in appendix~\ref{app:B4} for details).  The temperature of these neutrinos can be roughly approximated by assuming that the inflowing material
generally flows near to the NS surface before shocking and emitting neutrinos (see appendix \ref{app:B1} and \ref{app:B3}).  The pressure from this shock is given by \citep{2006ApJ...646L.131F}: 
\begin{equation} 
P_{\rm shock} = \frac{1}{2} (\gamma+1) \rho_{\rm acc} v^2_{\rm acc}
\end{equation} 
where, if we assume the accretion occurs nearly at free-fall, 
\begin{equation}
v_{\rm acc} = \left(\frac{2 G M_{\rm NS}}{R_{\rm NS}}\right)^{1/2}
\end{equation} 
and 
\begin{equation}
\rho_{\rm acc} = \frac{\dot{M}_{\rm acc}}{4 \pi R^2_{\rm NS} v_{\rm
acc}}
\end{equation}
where $\dot{M}_{\rm acc}=\dot{M}_B$ is the accretion rate onto the NS. The equation of state $\gamma =4/3$ for the radiation-dominated conditions in this material, leads to the temperature of this material:

\begin{equation} \label{eq:Tacc}
T_{\rm acc}=\left(\frac{3 P_{\rm shock}}{4 \sigma/c}\right)^{1/4}=\left(\frac{7}{8} \frac{\dot{M}_{\rm acc}
v_{\rm acc} c}{4 \pi R^2_{\rm NS} \sigma}\right)^{1/4}.
\end{equation}
The electron-positron pairs producing the neutrinos are thermalized at this temperature and the resulting neutrino temperature
can be estimated by this formula. For accretion rates lying between $\sim 10^{-4}$--$10^{-2}~M_\odot$~s$^{-1}$, we estimate neutrino
temperatures lying between 1.7--5.2~MeV (i.e.~neutrino energies $E_\nu \approx 3 k_B T\approx 5$--15~MeV; see appendix~\ref{app:B3}), predicting energies only slightly below those produced by detailed calculations \citep{2009ApJ...699..409F}. A detailed study of the neutrino emission will be the presented elsewhere (Ruffini et al., in preparation; see also appendix \ref{app:B1} and \ref{app:B3}). 

As we show in appendix~\ref{app:B3}, for the developed temperatures (say $k_B T\sim 1$--10~MeV) near the NS surface (see Figs.~\ref{fig:profilesAtm} and \ref{fig:Atm2}), the dominant neutrino emission process is the electron-positron annihilation leading to neutrino-antineutrino. This process produces a neutrino emissivity proportional to the nineth power of the temperature [see Eq.~(\ref{eq:L_neutrinos})]. The accretion atmosphere near the NS surface is characterized by a temperature gradient (see Fig.~\ref{fig:profilesAtm} in appendix~\ref{app:B1}) with a typical scale height $\Delta r_{\rm ER}\approx 0.7~R_{\rm NS}$, obtained from Eq.~(\ref{eq:H_T}). Owing to the aforementioned strong dependence of the neutrino emission on temperature, most of the neutrino emission occurs in the region $\Delta r_{\rm ER}$ above the NS surface.

These conditions lead to the neutrinos to be efficient in balancing the gravitational potential energy gain, as indicated in Eq.~(\ref{eq:bal_neu_grav(rel)}), allowing the hypercritical accretion rates. We show in appendix~\ref{app:B4} (see Figs.~\ref{fig:taus} and \ref{fig:Trho}) photons are trapped within the flow the Eddington limit does not apply in this system. As discussed in \citet{1996ApJ...460..801F}, the neutrinos can balance efficiently the gravitational energy gain. The effective accretion onto the NS can be estimated as: 
\begin{equation}\label{eq:Mdoteff}
\dot{M}_{\rm eff} \approx \Delta M_{\rm ER} \frac{L_{\rm ER}}{E_{\rm er}},
\end{equation}
where $\Delta M_{\rm ER}$, $L_{\rm ER}$ are the mass and neutrino luminosity in the emission region (i.e.~$\Delta r_{\rm ER}$), and $E_{\rm ER}$ is half the gravitational potential energy gained by the material falling from infinity to the $R_{\rm NS}+\Delta r_{\rm ER}$. Since $L_{\rm ER}\approx 2\pi R_{\rm NS}\Delta r_{\rm ER} \epsilon_{\rm e^{-}e^{+}}$ with $\epsilon_{\rm e^{-}e^{+}}$ the electron-positron pair annihilation process emissivity given by Eq.~(\ref{eq:L_neutrinos}), and $E_{\rm ER}=(1/2) G M_{\rm NS} \Delta M_{\rm ER}/(R_{\rm NS}+\Delta r_{\rm ER})$, it can be checked that for $M_{\rm NS}=1.4~M_\odot$ this accretion rate leads to values $\dot{M}_{\rm eff} \approx 10^{-9}$--$10^{-1}~M_\odot$~s$^{-1}$ for temperatures $k_B T = 1$--10~MeV.

The neutrino signal from this accretion can be similar to the one from accretion in supernovae with fallback. Fallback begins immediately after the launch of the supernova explosion and, after peaking, decays with time ($t^{-5/3}$: \citealp{1989ApJ...346..847C}).  Depending upon the total fallback mass, the fallback accretion rate can remain above $10^{-4}~M_\odot$~s$^{-1}$ for $10^3$--$10^4$~s \citep{2014arXiv1401.3032W}.

\section{Accretion luminosity}\label{sec:7}

In order to make a comparison with observed light-curves we need to estimate the luminosity produced during the accretion process. The gain of gravitational potential energy in the accretion process is the total one available to be released e.g. by neutrinos and photons. The total energy released in the star in a time-interval $dt$ during the accretion of an amount of mass $dM_b$ with  angular  momentum $l \dot{M}_b$, is given by \citep[see, e.g.,][]{2000AstL...26..772S,2015ApJ...812..100B}:
\begin{eqnarray}
&&L_{\rm acc}= (\dot{M}_b - \dot{M}_{\rm NS})c^2 \nonumber \\
&&=\dot{M}_b c^2 \left[1-\left(\frac{\partial M_{\rm NS}}{\partial J_{\rm NS}}\right)_{M_b}\,l -\left(\frac{\partial M_{\rm NS}}{\partial M_b}\right)_{J_{\rm NS}}\right],
\label{eq:DiskLuminosity}
\end{eqnarray}
where we have used Eq.~(\ref{eq:Mns_evol}). This upper limit to the energy released is just the amount of gravitational energy gained by the accreted matter by falling to the NS surface and which is not spent in changing the gravitational binding energy of the NS. 
The total energy releasable during the accretion process, $\Delta E_{\rm acc} \equiv \int L_{\rm acc}dt$, is thus given by the difference in binding energies of the initial and final NS configurations. The typical luminosity will be $L_{\rm acc}\approx \Delta E_{\rm acc}/\Delta t_{\rm acc}$ where $\Delta t_{\rm acc}$ is the duration of the accretion process.

The duration of the accretion process is given approximately by the flow time of the slowest layers of the supernova ejecta to the NS. If the velocity of these layers is $v_{\rm inner}$, then $\Delta t_{\rm acc}\sim a/v_{\rm inner}$, where $a$ is the binary separation. For $a\sim 10^{11}$~cm and $v_{\rm inner}\sim 10^8$~cm~s$^{-1}$ we obtain $\Delta t_{\rm acc}\sim 10^3$~s, while for shorter binary separation, e.g.~$a\sim 10^{10}$~cm ($P\sim 5$~min), $\Delta t_{\rm acc}\sim 10^2$~s, as validated by the results of our numerical integrations shown e.g.~in Figs.~\ref{fig:Mdot} and \ref{fig:NSevolution}. See also appendix~\ref{app:A}. 

We have shown in Fig.~\ref{fig:NSevolution} the evolution of both the baryonic mass $\dot{M}_b$ and the gravitational mass $\dot{M}_{\rm NS}$ for a specific example. We have seen that these two quantities show a similar behavior, therefore we should expect the difference between them, which gives the available energy to be released (\ref{eq:DiskLuminosity}), evolves with time analogously. Besides, we can see that the NS in the system with $P=5$~min accretes $\approx 1~M_\odot$ in $\Delta t_{\rm acc}\approx 100$~s. With the aid of Eq.~(\ref{eq:MbMnsjns}) we can estimate the difference in binding energies between a $2~M_\odot$ and a $3~M_\odot$ NS, i.e. $\Delta E_{\rm acc}\approx 13/200 (3^2-2^2)~M_\odot c^2\approx 0.32~M_\odot c^2$ leading to a maximum luminosity $L_{\rm acc}\approx 3\times 10^-3~M_\odot c^2 \approx 0.1 \dot{M_b} c^2$.

\begin{figure}
	\centering
	\includegraphics[width=\hsize,clip]{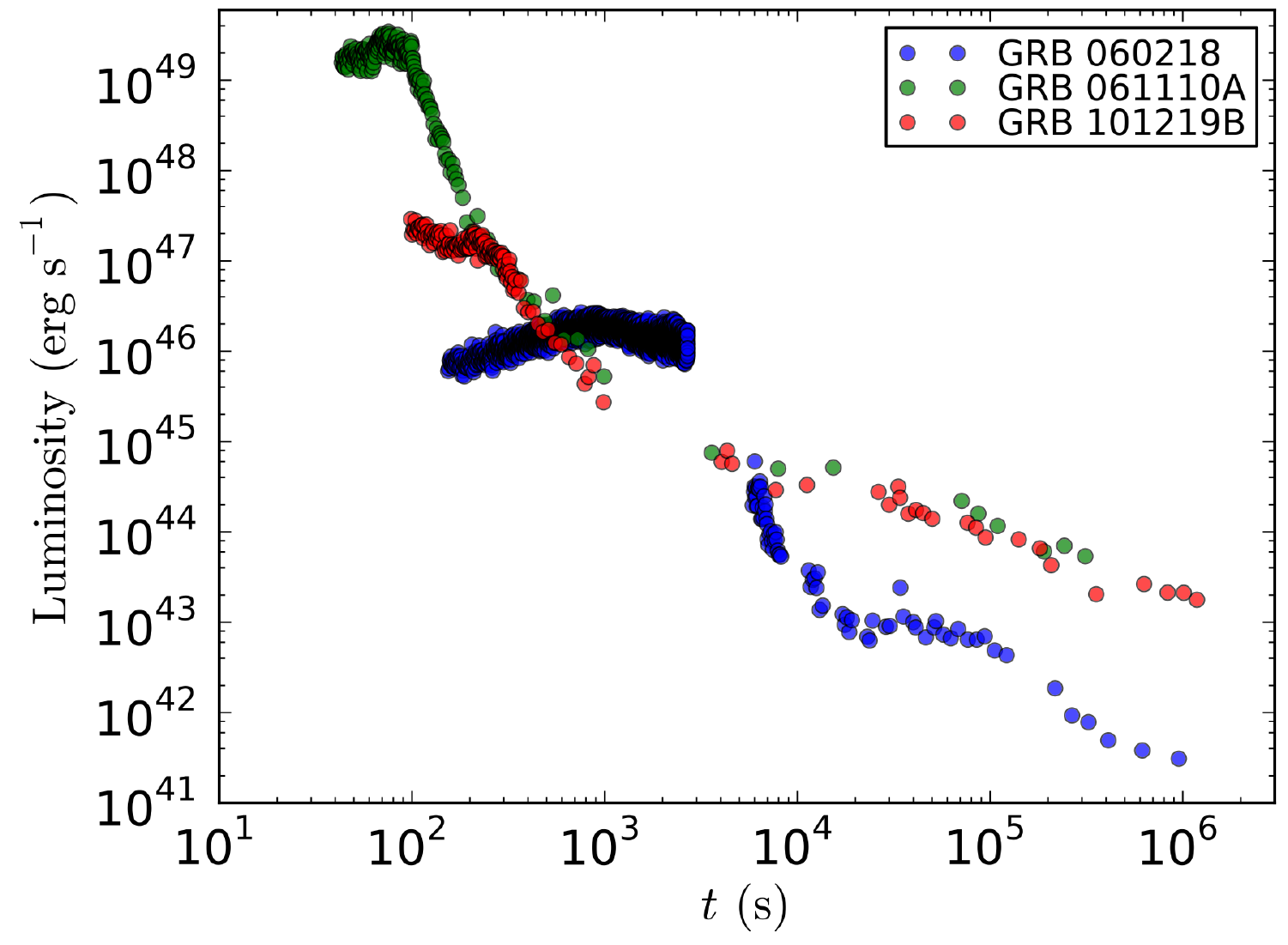}
\caption{Observed 0.3--10~keV XRT light-curves of some XRFs.}
	\label{fig:lum}
\end{figure}

Such an accretion power could lead to signatures observable in long GRBs \citep[see, e.g.,][]{2012A&A...548L...5I,2014ApJ...793L..36F} since it could be as high as $L_{\rm acc}\sim 0.1 \dot{M_b} c^2\sim 10^{47}$--$10^{51}$~erg~s$^{-1}$ for accretion rates in the range $\dot{M_b}\sim 10^{-6}$--$10^{-2}~M_\odot$~s$^{-1}$. Fig.~\ref{fig:lum} shows a few examples of light-curves of XRFs. It can be shown that the accretion luminosity can explain the observed early emission ($t\lesssim 10^3$~s) in these examples (see Sec.~\ref{sec:8} and Becerra et al., in preparation). 

\section{Influence of the hypercritical accretion on the supernova emission}\label{sec:8}

The duration of the accretion is shorter than the one of the long-lasting X-ray emission (at times $t\sim 10^3$--$10^6$~s). We shall show below (see Sec.~\ref{sec:8} and Becerra et al., in preparation) such a long-lasting emission can be explained from the supernova powered by the prompt radiation, i.e.~the X-ray radiation occurring during the accretion process onto the NS. In the case of the tightest binaries leading to BdHNe, the supernova is in addition powered by the prompt radiation following the BH formation. 

We now analyze the emission of the supernova at early stages. Traditionally, the supernova shock breaks out of the star producing a burst of X-ray emission which, in a spherically symmetric model, behaves as a sharp rise and equally fast decay as the forward shock cools. However, in our models, the supernova shock has distinct asymmetries caused by the accretion onto the NS (see Figs.~\ref{fig:denassym1a}--\ref{fig:denassym2} in Sec.~\ref{sec:5}). In addition, the X-rays emitted from this hypercritical accretion add energy to the explosion. To calculate the shock breakout luminosity, we use the simplified light-curve code described in \citet{2015ApJ...805...98B} and \citet{delarosa2016}. This code assumes homologous outflow for the ejecta velocities, modeling the radiative transport using a single group diffusion scheme with prescriptions for recombination opacities and energies. Energy released in the accretion onto the neutron star is injected as an energy source at the base of the explosion.  Because these calculations are 1-dimensional, we mimic the asymmetry in the explosion by modeling a series of spherical explosions with different densities.  Each of these densities produces a different light-curve with the more massive models producing later shock breakout times.  

\begin{figure}
	\centering
	\includegraphics[width=\hsize,clip]{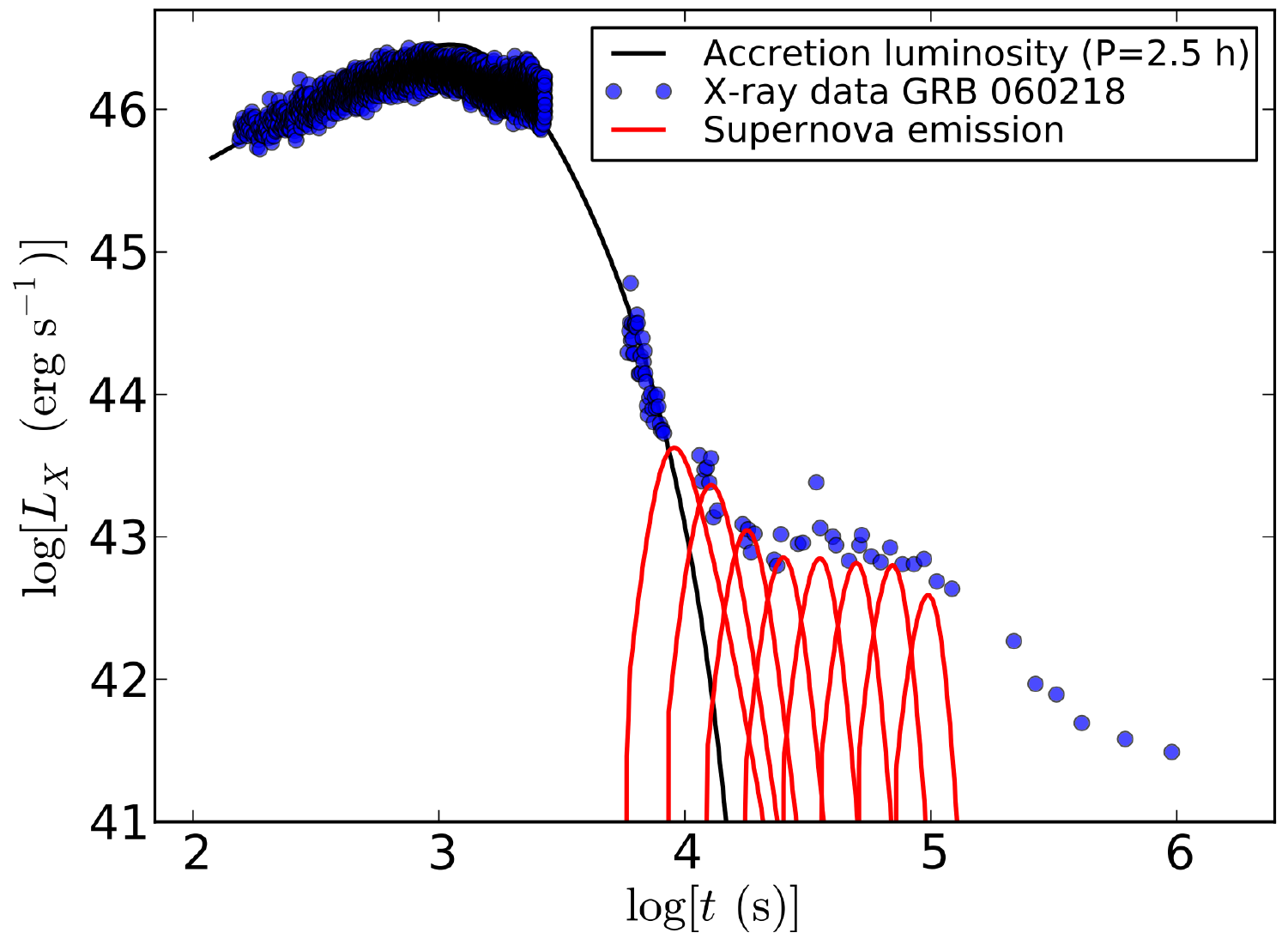}
\caption{Comparison of the accretion luminosity given by equation~(\ref{eq:DiskLuminosity}) and the supernova luminosity with the observed X-ray luminosity of XRF 060218. The binary system has the following parameters: supernova velocity $v_{\rm star,0}=2\times10^9$~cm~$s^{-1}$, a pre-supernova core obtained from the $M_{\rm ZAMS}=20~M_\odot$ evolution (see table~\ref{tb:ProgenitorSN}), initial NS mass $M_{\rm NS}(t_0)=1.4~M_\odot$, and orbital period of 2.5~h. In this example the initial explosion energy is $2\times10^{51}$~erg, ranging the spherical equivalent-mass from 0.05--4~$M_\odot$. It can be seen that at early times $t\lesssim 10^4$~s the luminosity is dominated by the accretion process. The supernova X-ray light-curves rise quickly at $t\approx 10^4$~s for the lowest mass, to $t\sim 10^5$~s for the 4~$M_\odot$ explosion, which corresponds to total ejecta mass from our supernova.}
	\label{fig:Lx060218}
\end{figure}

In Fig.~\ref{fig:Lx060218} we compare and contrast the luminosity expected from the accretion process given by equation~(\ref{eq:DiskLuminosity}) and from the accretion-powered supernova, with the observed X-ray luminosity of XRF 060218. The parameters characterizing the binary are: orbital period of 2.5~h, supernova velocity $v_{\rm star,0}=2\times10^9$~cm~s$^{-1}$, a pre-supernova core obtained from the $M_{\rm ZAMS}=20~M_\odot$ evolution which leads to a CO core envelope mass $\sim 4~M_\odot$ (see table~\ref{tb:ProgenitorSN}), and initial NS mass $M_{\rm NS}(t_0)=1.4~M_\odot$. For these binary parameters, the NS does not collapse to a BH, in agreement with the fact that XRFs, as XRF 060218, should be explained by these kind of binaries.

For this burst, our model assumes an initial explosion energy of $2\times10^{51}$~erg, ranging the spherical equivalent-mass from 0.05--4~$M_\odot$.  Fig.~\ref{fig:Lx060218} shows light-curves rising quickly at $t\lesssim 10^4$~s for the lowest mass to $\sim 10^5$~s for the 4~$M_\odot$ explosion. This maximum mass corresponds to the ejecta mass from our supernova. The corresponding CO core mass of our progenitor is this ejecta mass plus the mass of the $\nu$NS, roughly 5.4~$M_\odot$. It is possible that the mass is slightly larger for our progenitors, and the emission from the breakout could be longer, but peak X-ray emission from shock breakout beyond a few times $10^5$~s will be difficult to achieve. The observed emission would come from the sum of this full range of explosions.  The close match of our models (fitted to our expected progenitor mass) to this X-ray plateau demonstrates that this sequence of shock breakouts is certainly a viable and natural explanation for this emission (see Fig.~\ref{fig:Lx060218}). 

We have shown that the X-ray plateau in the afterglow is powered almost entirely by a sequence of shock breakouts and the expanding photosphere. We turn now to the the optical emission which is more complex. The optical emission can be powered by the expanding photosphere, $^{56}$Ni decay and the energy deposited by the accreting NS. For XRF 060218, the light-curve in the optical and UV exhibits a double-peaked structure suggestive of multiple power sources and, using our light-curve code, we can test out different scenarios. Just like the X-ray, geometry effects will modify the optical light-curve. Here we merely probe the different emission mechanisms to determine the viability of each to explain the XRF 060218 optical light-curve.

\begin{figure}
	\centering
	\includegraphics[width=\hsize,clip]{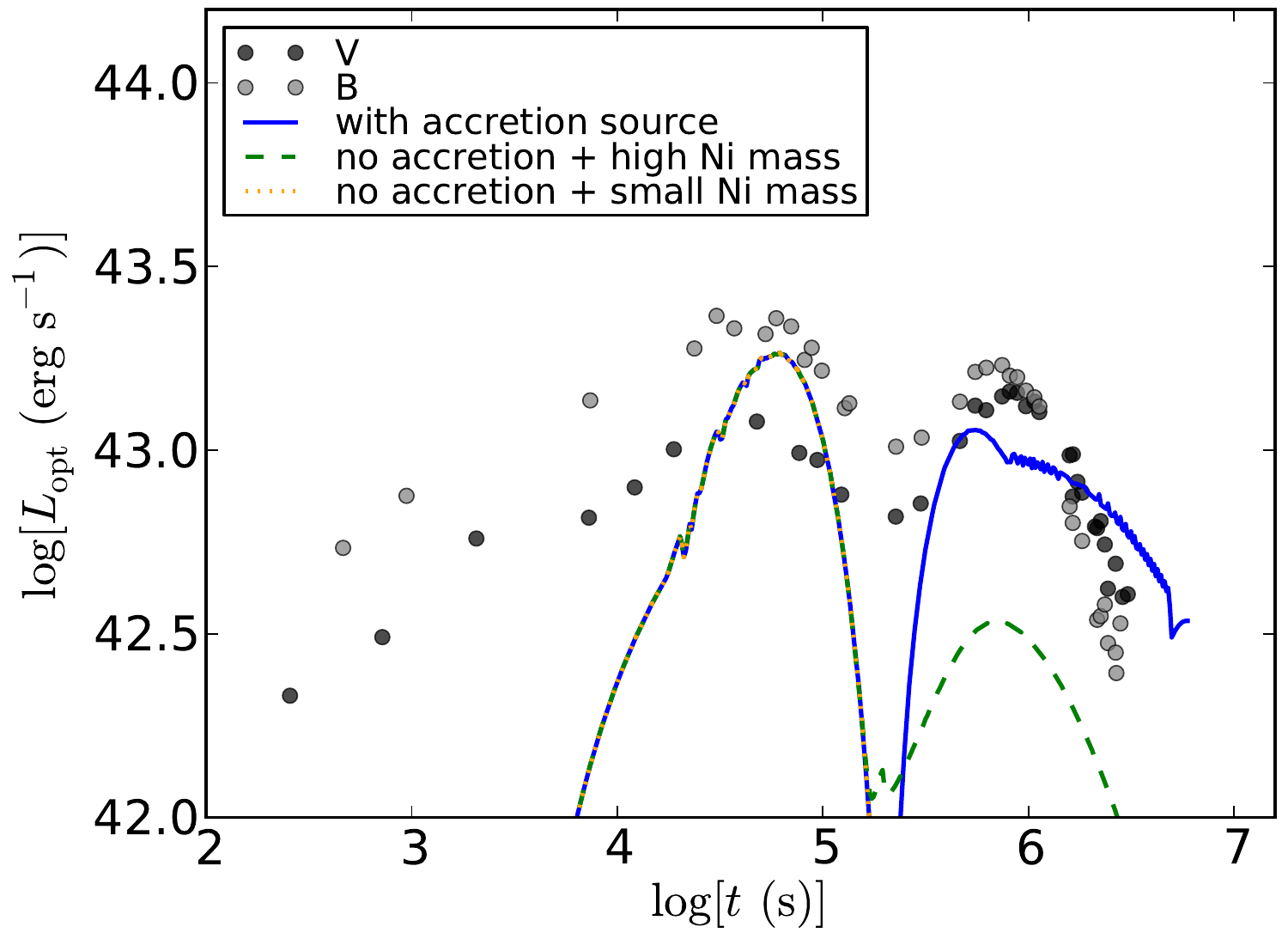}
\caption{Optical and UV luminosity of XRF 060218 \citep{2006Natur.442.1011P}. The light-curve shows a double-peaked structure. The red dotted curve shows the supernova optical emission without either $^{56}$Ni decay or accretion energy; it can be seen that it explains only the first peak. The blue solid curve includes the energy deposition from the accretion onto the NS (which is a power source of $4\times10^{46}$~erg~s$^{-1}$ over a 2500~s duration). This simulation reproduces both the first peak at $\sim 50,000$~s as well as the second peak at $\sim 500,000$~s. The dashed green curve shows that a second peak can also be produced without accretion power by increasing the total $^{56}$Ni yield. However, even if we assume half of the total ejecta is $^{56}$Ni, the produced second peak cannot explain the observational data.}
	\label{fig:optical}
\end{figure}

Fig.~\ref{fig:optical} shows the V and B band light-curves for XRF 060218 \citep{2006Natur.442.1011P}. The light-curve in both bands peaks first near 50,000~s and then again at 500,000~s. Using our 1~$M_\odot$ 1D model from our X-ray emission, we simulate the V and B band light-curves.  Without either $^{56}$Ni decay or accretion energy, the supernova explosion only explains the first peak.  However, if we include the energy deposition from the accretion onto the NS (for our energy deposition, we use $4\times10^{46}$~erg~s$^{-1}$ over a 2500~s
duration), our simulations produce a second peak at roughly 500,000~s. A second peak can also be produced by increasing the 
total $^{56}$Ni yield.  However, even if we assume half of the total ejecta is $^{56}$Ni, the second peak remains too dim to 
explain the observations.   

The accretion energy in our model provides a natural explanation for the double-peaked features observed in the optical emission of XRF 060218. However, our simple model makes a series of approximations: e.g., we use gray transport, estimating the V and B emission assuming a blackbody, we assume the opacities are dominated by electron scattering, etc. Our simplified picture cannot reproduce accurately the first slowly rising part of the optical data which can be due to a combination of 1) the low-energy tail of the X-ray bubbles and 2) the geometry asymmetries which, just like for the X-rays, cause 1D effective mass ejecta to be lower along some lines of sight leading to some optical emission. The simulation of these details are out of the scope of the present article and will be the subject of future simulations. We have shown that, although approximate, the accretion mechanism can power the observed XRF 060218 light-curve.

\section{Conclusions}\label{sec:9}

We have analyzed in detail the IGC paradigm of GRBs associated with supernovae. The progenitor is a binary system composed of CO core and a NS in which the explosion of the CO core as supernova triggers a hypercritical accretion process onto the NS. For the given supernova parameters (total CO core mass, density profile, ejecta mass and velocity) and an initial mass of the NS, the fate of the NS depends only on the binary separation/orbital period. The picture that arises from the simulation of the accretion process as a function of the orbital period is as follows. 
\begin{enumerate}

\item
Since the accretion rate decreases for increasing values of the orbital period, there exists a specific value of it over which BH formation is not possible because the NS does not accrete sufficient matter to reach the critical mass. We denoted this maximum period for gravitational collapse as $P_{\rm max}$ and computed it as a function of the initial NS mass for selected pre-supernova CO cores (see Fig.~\ref{fig:Pmax}). Therefore, in systems with $P\leq P_{\rm max}$ BH formation occurs and these systems, within the IGC paradigm, can explain BdHNe \citep{2015PhRvL.115w1102F,2015ApJ...812..100B,2014ApJ...793L..36F}. In systems with $P>P_{\rm max}$, the NS does not accrete enough matter from the supernova ejecta and the collapse to a BH does not occur: these systems, within the IGC paradigm, are used to explain the nature of XRFs.

\item 
We have shown that the early emission ($t\lesssim 10^3$~s) of an IGC binary is powered by the accretion luminosity. This luminosity explains the prompt emission of XRFs as presented here in the case of XRF 060218 (see Fig.~\ref{fig:Lx060218}). We are planning to extend this conclusion to additional XRFs (Ruffini et al., in preparation).

\item 
We have shown that convection instabilities arising from the NS accretion atmosphere can drive hot outflows emitting X-rays observable in the early emission of GRBs. It has been shown in \cite{2014ApJ...793L..36F} that the emission from such an outflows is consistent with the early ($t\lesssim 50$~s) thermal X-ray emission observed in the BdHN GRB 090618 \citep{2012A&A...548L...5I,2012A&A...543A..10I}. We have shown here the consistency with the thermal emission of XRF 060218. The observational verification in the case of additional XRFs will be presented elsewhere (Ruffini et al., in preparation). Details of the structure of the accretion region are presented in appendix~\ref{app:B}.

\item
Neutrino emission is the main energy sink of the system, allowing the hypercritical accretion to occur. We have given estimates of the neutrino flux and energy. 
Typical neutrino energies are in the range 1--15~MeV. A detailed study of the neutrino emission will be presented elsewhere (Ruffini et al., in preparation). Details are also presented in appendix~\ref{app:B3}.

\item 
We have shown that the presence of the NS in very compact orbit produces large asymmetries in the supernova ejecta around the orbital plane (see Figs.~\ref{fig:denassym1a} and \ref{fig:denassym2}). These asymmetries are the combined effect of the accretion and of the action of the gravitational field of the NS on the supernova layers.

\item
The above supernova asymmetries lead to observable effects in the supernova emission. The shocked material becomes transparent at different times with different luminosities along different directions owing to the asymmetry created in the supernova ejecta by the orbiting and accreting NS (see Figs.~\ref{fig:denassym1a} and \ref{fig:denassym2}). The sequence of shock breakout luminosities are thus influenced by the asymmetries in the explosion: the light-curve produced along the more massive directions produce later shock breakout times. We have shown that the observed long-lasting, $t>t_{\rm acc}$, afterglow X-ray emission observed in XRFs can be powered by this mechanism and presented as an example XRF 060218 (see Fig.~\ref{fig:Lx060218}). The specific example is here presented for XRF 060218 and evidence that this mechanism is also observed in additional XRFs will be presented elsewhere (Ruffini et al., in preparation).

\item We have exemplified the above mechanism for late time X-ray emission observed in XRF 060218. The supernova ejecta asymmetries are even more pronounced in more compact binaries in which the NS, by accretion, reaches the critical mass and collapses to a BH (see Fig.~\ref{fig:denassym1a}). This implies that this mechanism is also at work in the X-ray afterglow of BdHNe with specific additional features in the spike, in the plateau and in the late power-law emission (Ruffini et al., in preparation).

\item We have shown that not only asymmetries caused by the close accreting NS modify the classic picture of supernova emission. The X-rays emitted from the accretion add energy to the supernova explosion. We have simulated the optical emission of the supernova and compared and contrasted our theoretical expectation with the optical luminosity of XRF 060218 which shows a peculiar double-peaked shape. We have shown that without either $^{56}$Ni decay or accretion energy, the supernova explosion can explain only the first peak. We then showed that the inclusion of $^{56}$Ni decay produces indeed a double-peaked light-curve but with a second peak which is too dim to explain the observed optical emission. This conclusion holds even adopting unphysical high amounts of $^{56}$Ni mass of up to half of the ejecta mass. Instead, we demonstrated that the source of energy given by the hypercritical accretion onto the NS provides a double-peaked light-curve consistent with the observational data. See Fig.~\ref{fig:optical} and Ruffini et al. (in preparation), for details.

\item We have shown how the radiation during the continuous accretion process affects the supernova emission both in X-rays and in the optical. We have simulated this effect for binaries in which the NS does not collapse to a BH, namely for XRFs (e.g.~XRF 060218). For systems with shorter orbital periods in which a BH is formed, namely for BdHNe, besides the initial interaction of the supernova with the radiation from the accretion process, the supernova interacts with the radiation from the prompt radiation following the BH formation. The interaction of the electron-positron pairs (moving with Lorentz factor $\Gamma\sim 10^2$) with the supernova material at a distance of $r\sim 10^{12}$~cm and moving at $\Gamma \sim 1$ can originate the flare observed around $t\sim 100$~s after the GRB trigger time in the X-ray data of BdHNe. The theoretical and observational details of this process will be presented elsewhere (Ruffini et al., in preparation).
\end{enumerate}

It is interesting that in parallel to the above conclusions we can also draw some inferences on the astrophysics of NS-NS binaries. Our results suggest that the systems in which the accreting NS does not reach the critical mass (i.e the XRFs) are natural candidates to produce such binaries \citep{2016arXiv160202732R}. We have shown that this will occur for CO-NS binaries with long orbital periods; thus it is possible that many of these systems become unbound by the supernova explosion produced by the CO core. 
The XRF to BdHN occurrence rate ratio can shed light on the ratio of bound/unbound IGC binaries \cite{2015PhRvL.115w1102F}. The short orbital periods $P<P_{\rm max}$ needed for BdHNe obtained from our theoretical model imply that XRF must be much more common than BdHNe, as it is indeed observed \citep[see, e.g.,][and references therein]{2007ApJ...657L..73G,2016arXiv160202732R}. 
The few systems which will keep bound become NS-NS binaries where at least one of the components can be massive and with a rotation period in the millisecond region. If the NS accretes from the LSO, then at the end of the process it will have an angular momentum $J_{\rm NS}\sim 2\sqrt{3} G M_{\rm acc}M_{\rm NS}/c \approx 4.3\times 10^{48} [M_{\rm acc}/(0.1 M_\odot)] [M_{\rm NS}/(1.4 M_\odot)]$~g~cm$^2$~s$^{-2}$, where $M_{\rm acc}$ is the total accreted mass. Thus, the NS will have a rotation period $P=2\pi I_{\rm NS}/J_{\rm NS} \approx 1.6\,(0.1 M_\odot/M_{\rm acc})(R_{\rm NS}/10^6\,{\rm cm})^2$~ms, where $I_{\rm NS}\sim 2/5 M_{\rm NS} R_{\rm NS}^2$ is the NS moment of inertia. That known binary millisecond pulsars could be formed in XRFs is a very exciting result that deserves further scrutiny (Ruffini et al., in preparation).

\acknowledgements{
We thank the referee for the comments and suggestions. L.B. acknowledges the support given by the International Relativistic Astrophysics Ph. D Program (IRAP-PhD). J.A.R. acknowledges the support by the International Cooperation Program CAPES-ICRANet financed by CAPES, Brazilian Federal Agency for Support and Evaluation of Graduate Education within the Ministry of Education of Brazil. J.A.R. acknowledges partial support of the project No.~3101/GF4 IPC-11/2015 and the target program of the Ministry of Education and Science of the Republic of Kazakhstan.
}

\appendix

\section{Analytic approximation for the peak accretion rate}\label{app:A}

We can see from Fig.~\ref{fig:Mdot} that the shorter(smaller) the orbital period(separation) the higher the peak accretion rate $\dot{M}_{\rm peak}$ and the shorter the peak time, $t_{\rm peak}$. Indeed, we can derive such a feature from simple arguments. The accretion rate (\ref{eq:Bondi-HoyleRate}) increases for higher densities and lower velocities, so we should expect as indeed shown in Fig.~\ref{fig:Mdot}, it increases with time as the inner ejecta layers, which are denser and slower [see Eqs.~(\ref{eq:rhoej_sn}) and (\ref{eq:vej_sn})], reach and passed the accretion region. The accretion rate starts to peak at the passage of the innermost densest layer, $r_{\rm inner}$, through the capture region. Such a layer moves with velocity $v_{\rm inner} = (r_{\rm inner}/R_{\rm star}^0)v_{\rm star,0}$ as given by the homologous expansion assumption. 

Thus, the accretion rate peaks around the peak time:
\begin{equation}\label{eq:tpeak1}
t_{\rm peak} = \frac{a-R_{\rm cap}}{v_{\rm inner}} = \frac{(a-R_{\rm cap})R_{\rm star}^0}{r_{\rm inner} v_{\rm star,0}},
\end{equation}
namely the time when $r_{\rm inner}$ reaches the capture region which is located at a distance $r = a-R_{\rm cap}$ from the CO core center.

The radius $r_{\rm inner}$ is the maximum of the density profile (\ref{eq:prerho02}), namely the root of the equation:

\begin{equation}\label{eq:eqrinner}
r_{\rm inner}-R_{\rm star}^0 + R_{\rm star}^0 m \ln\left(\frac{r_{\rm inner}}{\hat{R}_{\rm core}}\right) = 0,
\end{equation}
where we recall $\hat{R}_{\rm core}\approx 0.31 R_{\rm core}$. Since $r_{\rm inner}\approx \hat{R}_{\rm core}$, we can obtain the approximate solution:
\begin{equation}\label{eq:rinner}
r_{\rm inner}\approx \eta R_{\rm core},\qquad \eta \equiv \frac{R_{\rm star}^0}{R_{\rm core}}\frac{1+m}{1+m (R_{\rm star}^0/\hat{R}_{\rm core})}.
\end{equation}

Since $v_{\rm inner} < v_{\rm orb}$, we can approximate the relative velocity as given only by the orbital one, i.e. $v_{\rm rel}\approx v_{\rm orb}$, and within this approximation, the capture radius reduces to $R_{\rm cap}\approx (2 M_{\rm NS}/M) a$. Then, equation~(\ref{eq:tpeak1}) becomes
\begin{equation}\label{eq:tpeak2}
t_{\rm peak} \approx \left(1-\frac{2 M_{\rm NS}}{M}\right)\left(\frac{G M}{4\pi^2}\right)^{1/3} \left(\frac{R_{\rm star}^0}{\eta R_{\rm core}}\right) \frac{P^{2/3}}{v_{\rm star,0}}.
\end{equation}

We can now evaluate equation~(\ref{eq:Bondi-HoyleRate}) at the above $t=t_{\rm peak}$ and applying the same approximations, we obtain for the peak accretion rate
\begin{equation}\label{eq:Mpeak}
    \dot{M}_{\rm peak}\approx  2\pi^2\frac{(2 M_{\rm NS}/M)^{5/2}}{(1-2 M_{\rm NS}/M)^3}\eta^{3-m}\frac{\rho_{\rm core}\,R_{\rm core}^3}{P},
\end{equation}
where we recall $M=M_{\rm CO}+M_{\rm NS}$ is the total binary mass, being $M_{\rm CO} = M_{\rm env} + M_{\nu\rm NS}$ the total mass of the CO core given by the envelope mass and the central iron core mass leading to the formation of the $\nu$NS.

\begin{figure}
	\centering
	\includegraphics[width=0.5\hsize,clip]{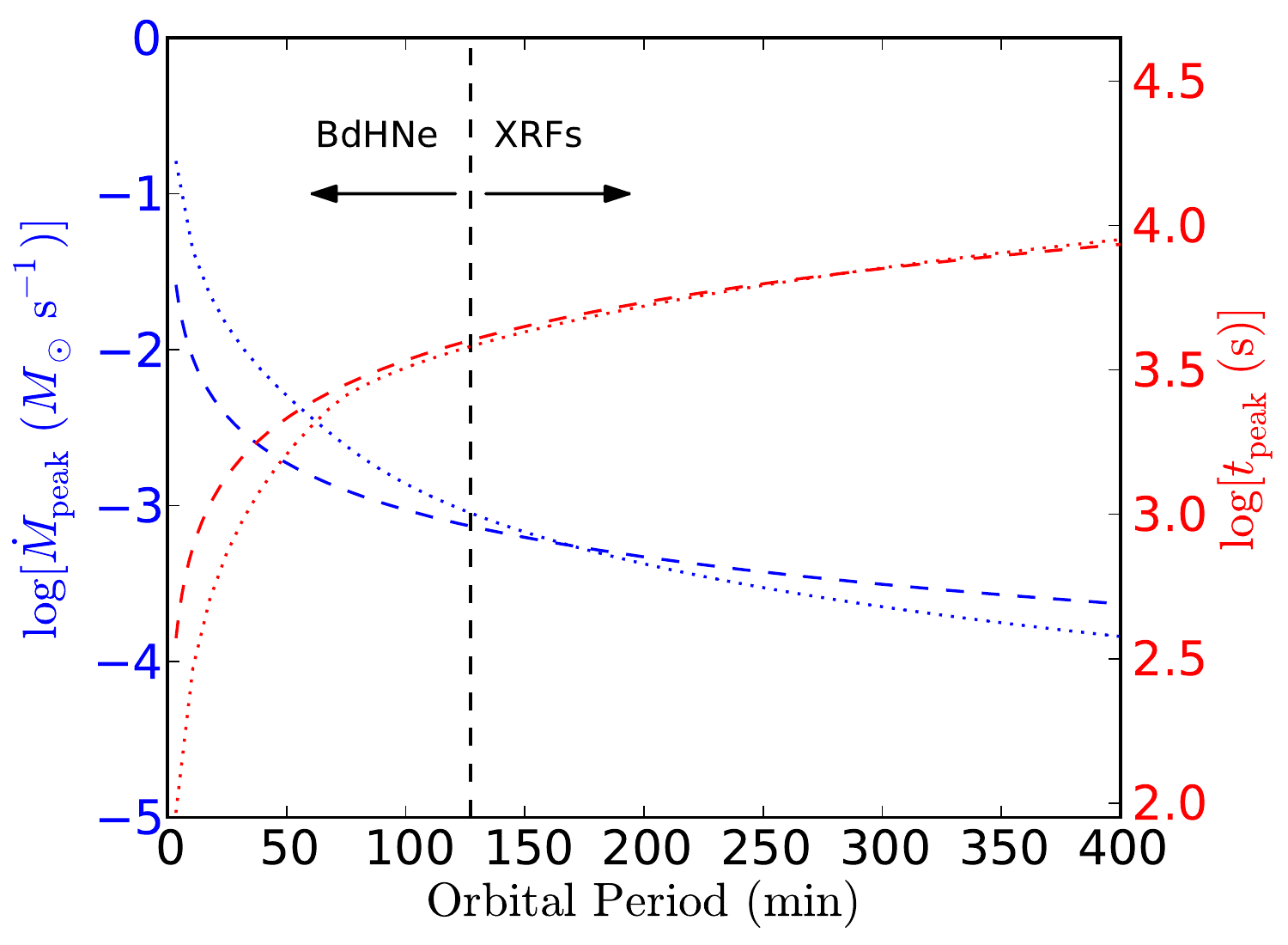}
\caption{(Color online) Peak accretion rate ($\dot{M}_{\rm peak}$, blue curves and left y-scale) and peak time ($t_{\rm peak}$, red curves and right y-scale) as a function of the orbital period. The dashed curves give the analytic peak accretion rate and time (\ref{eq:Mpeak}) and (\ref{eq:tpeak2}), respectively, while the dotted curves correspond to the values obtained from the numerical integration of the equations in Sec.~\ref{sec:2}. This example corresponds to the following binary parameters: a CO core from the $M_{\rm ZAMS}=20~M_\odot$ progenitor of table~\ref{tb:ProgenitorSN}, an initial NS mass $2.0~M_\odot$, and a velocity of the outermost ejecta layer $v_{\rm star,0}=2\times10^9$~cm~s$^{-1}$. For these parameters we have $\eta\approx 0.41$ from equation~(\ref{eq:rinner}). The black dashed vertical line marks the maximum orbital period (for these system parameters, $P_{\rm max}\approx 127$~min) for which the NS reaches, by accretion, the critical mass and collapses to a BH (see Fig.~\ref{fig:Pmax} in Sec.~\ref{sec:4}). We recall that within the IGC interpretation systems with $P<P_{\rm max}$ lead to BdHNe while systems with $P>P_{\rm max}$ lead to XRFs.}
	\label{fig:Mpeak}
\end{figure}

Fig.~\ref{fig:Mpeak} shows the behavior of Eqs.~(\ref{eq:Mpeak}) and (\ref{eq:tpeak2}) as a function of the orbital period and compare them with the corresponding values obtained from the numerical integration of the accretion equations presented in Sec.~\ref{sec:2}. This example is for the binary parameters: a CO core from the $M_{\rm ZAMS}=20~M_\odot$ progenitor of table~\ref{tb:ProgenitorSN}, an initial NS mass $2.0~M_\odot$, and a velocity of the outermost ejecta layer $v_{\rm star,0}=2\times10^9$~cm~s$^{-1}$. For these parameters, $\eta\approx 0.41$ from equation~(\ref{eq:rinner}). It can be seen that the accuracy of the above simple analytic formulas increases for the systems with $P>P_{\rm max}$. This is expected since, as we have mentioned, only for these systems the innermost ejecta layers passed the NS position. In systems with $P<P_{\rm max}$, the NS collapses to a BH before the passage of the innermost layers. In those cases, the maximum accretion rate is not reached at the passage of $r_{\rm inner}$ but at the passage of a layer located at $r_{\rm max} > r_{\rm inner}$, hence with velocity $v_{\rm max} = v (r=r_{\rm max})>v_{\rm inner}$, and thus $v_{\rm max}\gtrsim v_{\rm orb}$. In any case, it is clear the above formulas for $\dot{M}_{\rm peak}$ and $t_{\rm peak}$ remain valid to obtain typical (order-of-magnitude) estimates of the accretion process in these binaries. The consistency of the numerical and analytic results (within their range of validity) shown here serves as well as an indicator of the reliability of the numerical results (see also appendix~\ref{app:C}).

\section{Physics inside the accretion region}\label{app:B}

\begin{figure}
\centering
\includegraphics[width=0.35\hsize,clip]{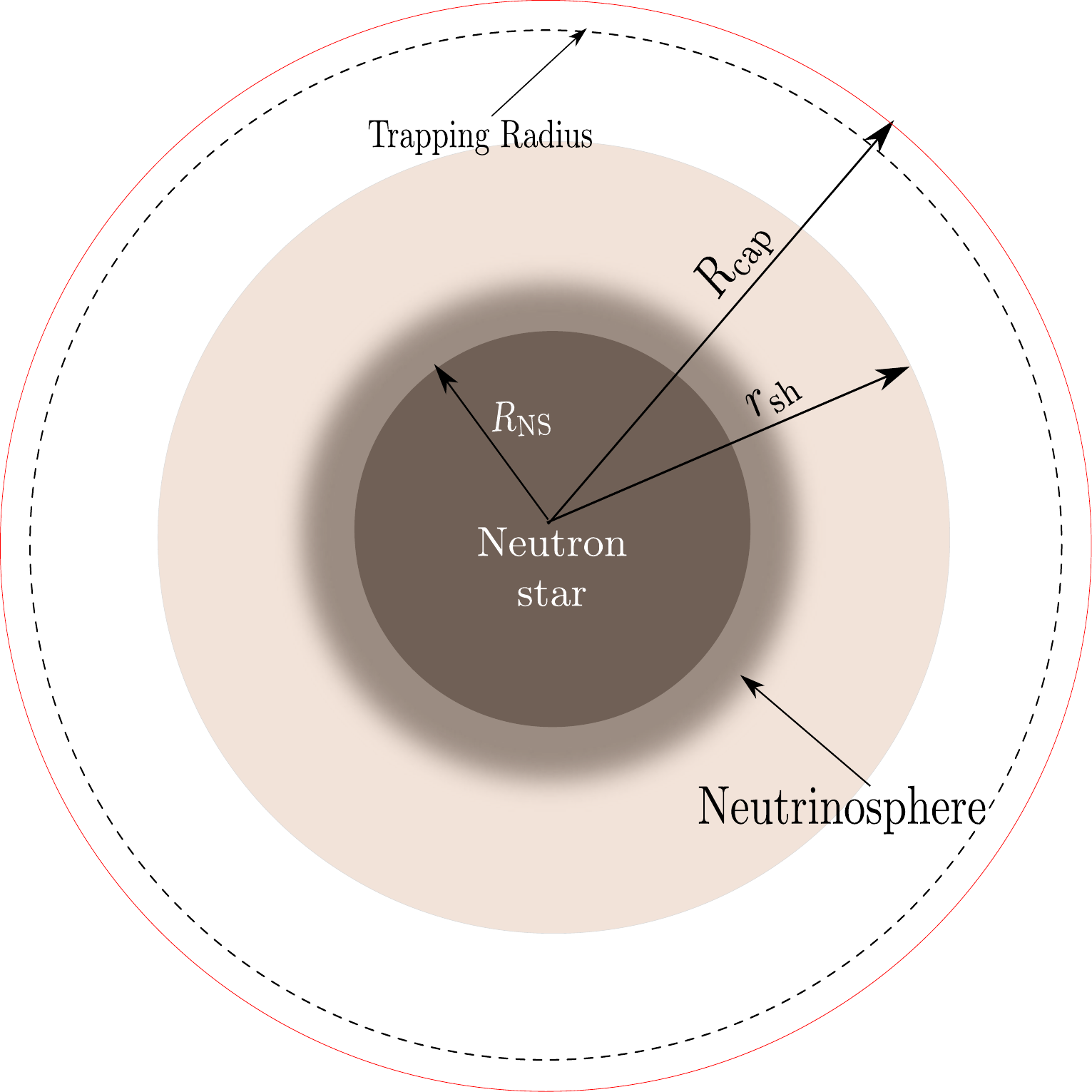}
\caption{Structure of the NS acrretion atmosphere. The ejecta from the supernova enter the NS capture region (red circle) at a distance $r=R_{\rm cap}$ [see equation~(\ref{eq:CaptureRadius})] from the NS center and start to fall to the NS surface. The material shocks as it piles on top the NS surface. The shock decelerates the material while it moves towards the NS and near the surface, at the \emph{neutrinosphere}, it looses energy by the emission of neutrinos. The neutrino emission allows the material to reduce its entropy to be finally accreted by the NS.}
	\label{fig:NSatmosphere}
\end{figure}

In this appendix we analyze the NS accretion zone following the theoretical framework established for supernova fallback accretion \citep{1989ApJ...346..847C,1991ApJ...376..234H,1996ApJ...460..801F}. Fig.~\ref{fig:NSatmosphere} shows schematically the structure of the NS atmosphere: the supernova material entering the NS capture region shocks as it piles up onto the NS surface.  As the atmosphere compresses, it becomes sufficiently hot to emit neutrinos allowing the matter to reduce its entropy and be incorporated into the NS. 
%

\subsection{Accretion zone structure and equation of state}\label{app:B1}

In order to model the evolution of the NS accretion zone, we assume that it passes through a sequence of quasi-steady state envelopes, each characterized by the mass accretion rate $\dot{M}$, the NS mass, $M_{\rm NS}$ and its radius $R_{\rm NS}$. The spacetime outside the NS is described by the Schwarzschild metric:
\begin{equation}
	ds^2=-\left(1-\frac{r_{{\rm sch}}}{r}\right)dt^2+\left(1-\frac{r_{{\rm sch}}}{r}\right)^{-1}dr^2+r^2\left(d\theta^2+\sin^2\theta d\phi^2\right),
\label{eq:sch_metric}
\end{equation}
where $r_{\rm sch}=2 G M_{\rm NS}/c^2$ is the Schwarzschild radius. The steady-state relativistic fluid equations for mass, momentum and energy conservation in this geometry are:
\begin{eqnarray}
	\frac{1}{r^2}\frac{d}{dr}\left( r^2\rho u \right)&=&0,\\
	\frac{1}{2}\frac{d}{dr}\left( \frac{u}{c} \right)^2+\frac{r_{\rm sch}}{2r}+\frac{1}{w}\frac{dP}{dr}\left[ \left( \frac{u}{c} \right)^2+1-\frac{r_{\rm sch}}{r} \right]&=&0,\\
	\frac{d}{dr}\left(\rho c^2+U\right)-\frac{w}{\rho}\frac{d\rho}{dr}+\frac{Q_\nu}{u}&=&0,
	\label{eq:steadyequations}
\end{eqnarray}
where $u$ is the radial component of the four-velocity,  $Q_\nu$ is the total energy loss rate per unit volume by neutrino cooling, $w=\rho c^2+U+P$ is the relativistic enthalpy, $\rho$ is the mass density, $P$ is the pressure and $U$ is the internal energy density.

The boundary conditions are determined by the the conservation of mass, momentum and energy flows through the shock front at $r=R_s$. These one are expressed by the Rankine-Hugoniot conditions \citep{1959flme.book.....L}:
\begin{eqnarray}
	\rho_pu_p-\rho_{\rm sh}u_{\rm sh}&=&0,\\
	w_pu^t_pu_p-w_{\rm sh}u^t_{\rm sh}u_{\rm sh}&=&0,\\
	w_pu_p^2+P_p-w_{\rm sh}u_{\rm sh}^2-P_{\rm sh}&=&0,
	\label{eq:shock_cond}
\end{eqnarray}
where $u^t$ is the time component of the four-velocity, determined by the condition $g_{\mu\nu}u^\mu u^\nu=-c^2$. The indexes `$p$' and `sh' denote the quantities in the pre-shock and  post-shock zone, respectively. Outside the shock front, the material is in approximate free fall, thus:
\begin{equation}
	u_p=\sqrt{\frac{2GM_{\rm ns}}{r}},\qquad \rho_p=\frac{\dot{M}}{4\pi r^2 v_p},\qquad P_p=\frac{1}{2}\rho_p v_p^2.
	\label{eq:pre_shock}
\end{equation}

We consider a gas of electrons, positrons, ions and photons. Then, the total pressure and density energy are: 
\begin{eqnarray}
P_{{\rm tot}}(\rho,T)&=&P_{\gamma}+P_{{\rm ion}}+P_{{\rm e^-}}+P_{{\rm e^+}},
\label{eq:Ptotal}\\
U_{{\rm tot}}(\rho,T)&=&U_{\gamma}+U_{\rm ion}+U_{{\rm e^-}}+U_{{\rm e^+}}.
\label{eq:Utotal}
\end{eqnarray}
For the pressure and the internal energy of the radiation field, we adopt a blackbody in thermodynamical equilibrium:
\begin{equation}
P_{\gamma}=\frac{1}{3}a T^4,\qquad U_{\gamma}=3 P_{\gamma},
\end{equation}
with $a=4\sigma/c=7.56\times 10^{-15}$~erg~cm$^{-3}$~K$^{-4}$, where $\sigma$ is the Stefan-Boltzmann constant. 

For the ion gas, we assume a perfect gas:
\begin{equation}
n_{\rm ion}=\frac{\rho}{A m_u},\qquad P_{{\rm ion}}=n_{{\rm ion}}\kappa_B T,\qquad U_{{\rm ion}}=\frac{3}{2}P_{{\rm ion}},
\end{equation}
where $n_{\rm ion}$ is the ion number density, $m_u=1.6604\times 10^{-24}$~g is the atomic mass unit and $\kappa_B$ is the Boltzmann constant.

Finally, the electrons and positrons are described by the Fermi-Dirac distributions:
\begin{eqnarray}
	n_{{\rm e^\pm}}=\frac{m_e^3c^3}{\pi^2\hbar^3}\,\sqrt{2}\,\beta^{3/2}\left[\mathcal{F}_{1/2}(\eta_{\rm e^{\pm}},\beta)+\beta\mathcal{F}_{3/2}(\eta_{\rm e^{\pm}},\beta)\right],\\
	P_{{\rm e^\pm}}=\frac{8\,m_e^4c^5}{3\sqrt{2}\, \pi^2\hbar^3}\beta^{5/2}\left[\mathcal{F}_{3/2}(\eta_{\rm e^{\pm}},\beta)+\frac{1}{2}\beta\mathcal{F}_{5/2}(\eta_{\rm e^{\pm}},\beta)\right],\\
	U_{{\rm e^\pm}}=\frac{m_ec^2}{\pi^2}\frac{2\sqrt{2}\,m_e^3c^3}{\hbar^3} \beta^{5/2}\left[\mathcal{F}_{3/2}(\eta_{\rm e^{\pm}},\beta)+\beta\mathcal{F}_{5/2}(\eta_{\rm e^{\pm}},\beta)\right],
\end{eqnarray}
where $\mathcal{F}_k(\eta,\beta)\equiv \int_0^\infty\frac{x^k(1+0.5x\beta)^{1/2}dx}{{\rm e}^{x-\eta}+1}$ is the relativistic Fermi-Dirac integral, $\beta\equiv\kappa_BT/(m_e c^2)$ is the relativity parameter and $\eta\equiv(\mu-m_ec^2)/\kappa_B T$ is the degeneracy parameter, with $\mu$ the chemical potential. Since the electrons and positrons are in equilibrium with radiation ($e^{+}+e^{-}\rightarrow \gamma+\gamma$), their chemical potential are related by $\mu_{\rm e^{-}}+\mu_{\rm e^{+}}=0$ and then $\eta_{\rm e^{+}}=-\eta_{\rm e^{-}}-2/\beta$. Finally, for each value of density and temperature, $\eta_{\rm e^{-}}$ is determined from the charge neutrality condition:
\begin{equation}\label{eq:EoS_NeutralityCondition}
n_{\rm e^{-}}-n_{\rm e^{+}}=\frac{Z}{A}\frac{\rho}{m_u}=Z n_{{\rm ion}}.
\end{equation}

\begin{figure}
\centering
\includegraphics[width=0.32\hsize,clip]{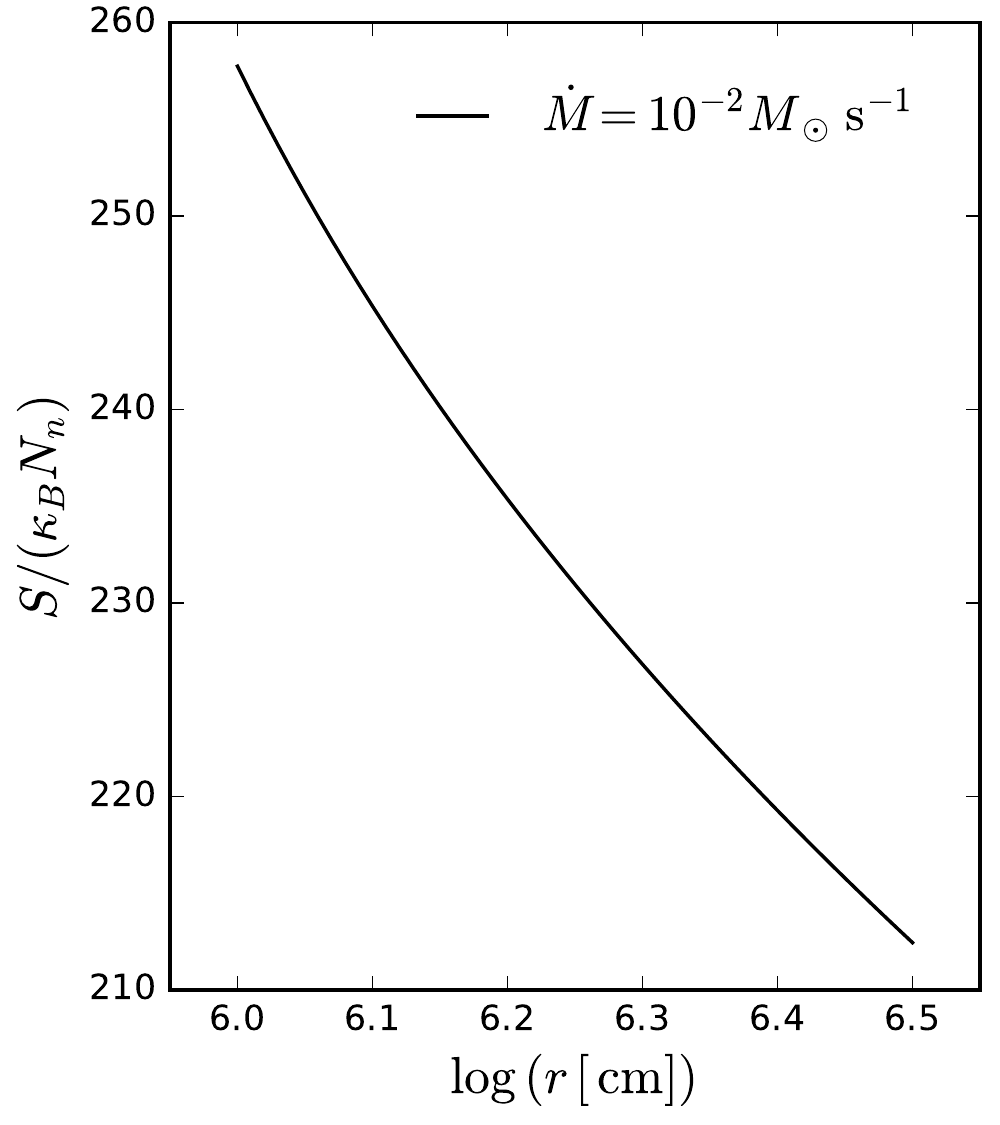}\includegraphics[width=0.32\hsize,clip]{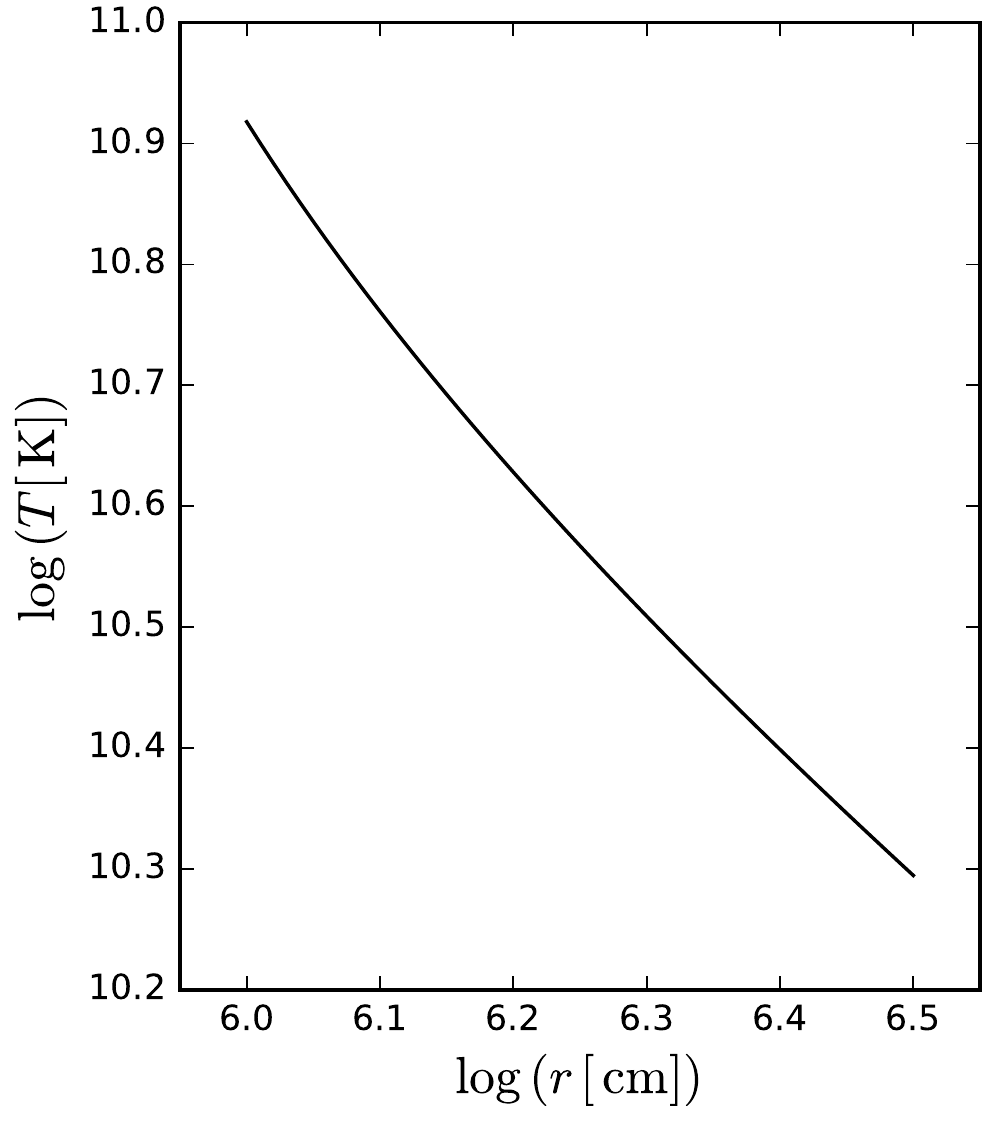}\includegraphics[width=0.32\hsize,clip]{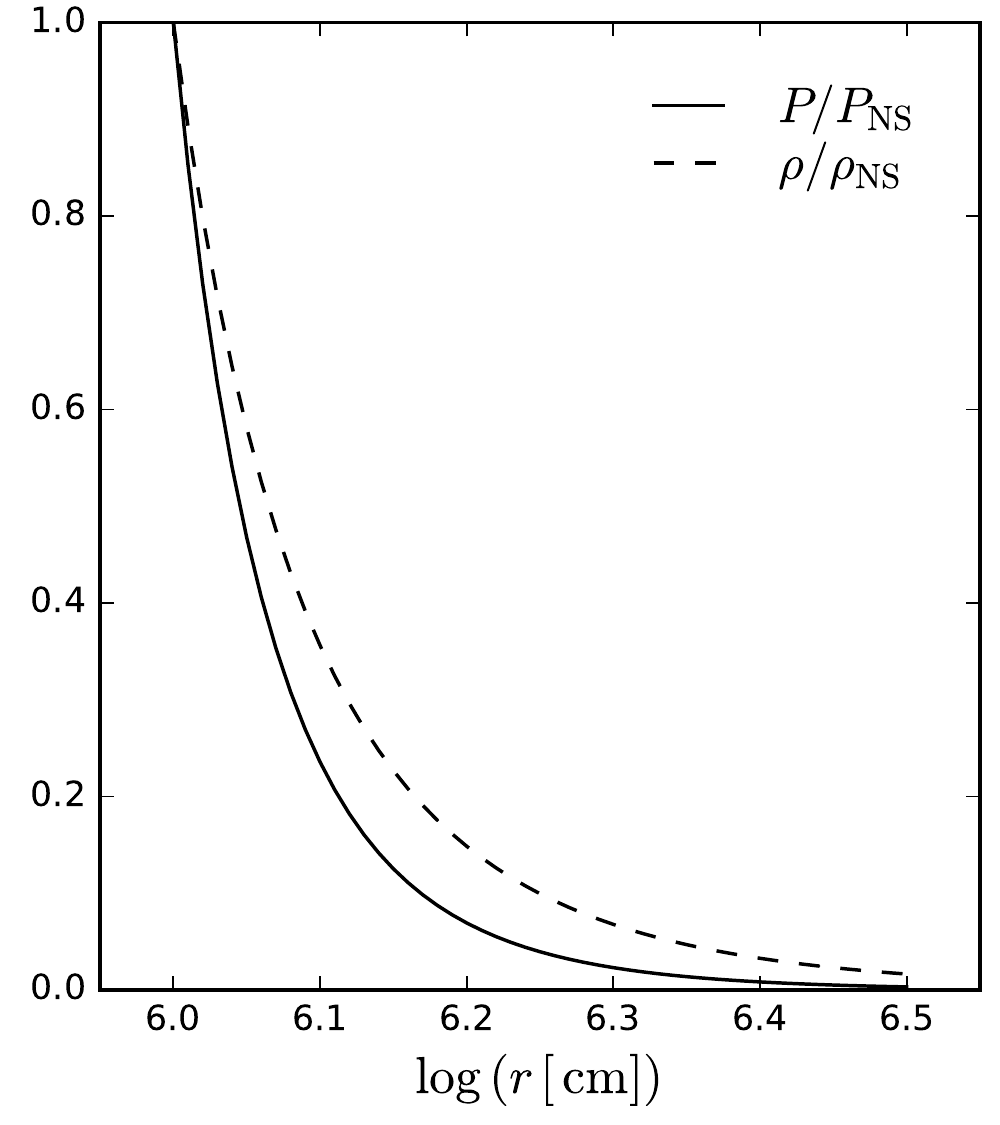}
\caption{Entropy, temperature, density and pressure profile for a NS accreting atmosphere for $\dot{M}=10^{-2}\,M_\odot$~s$^{-1}$. The pressure and density are normalized to $P_{\rm NS}\approx 3.28\times 10^{29}$~dyn~cm$^{-2}$ and $\rho_{\rm NS}\approx 7.5\times 10^8$~g~cm$^{-3}$, respectively.}\label{fig:profilesAtm}
\end{figure}

As an example, we show in Fig.~\ref{fig:profilesAtm} the entropy, temperature, density and pressure profile from the NS surface (we have assumed a NS of $M_{\rm NS}=2.0\, M_\odot$ with $R_{\rm NS}=10^6$~cm) to the shock radius for a specific value of the mass accretion rate $10^{-2}\,M_\odot$~s$^{-1}$. For the ions we adopt here $Z=6$ and $A=12$. It can be seen here that the entropy gradient of the NS atmosphere is negative, and it is thus subjected to convective instabilities (see Sec.~\ref{sec:6a}). 

\subsection{Neutrino emission and shock position}\label{app:B3}

We turn now to discuss the neutrino emission processes taken into account in our calculations. We follow the results reported in \citet{1996ApJS..102..411I} for the neutrino energy loss rates computed within the Weinberg-Salam theory \citep{1967PhRvL..19.1264W,salam1968}. We use here the formulas which fit the numerical results in the following regime of density and temperature: $10^0$~g~cm$^{-3}<\rho<10^{14}$~g~cm$^{-3}$ and $10^7$~K$<T<10^{11}$~K \citep{1996ApJS..102..411I}.
.

We consider the following channels of neutrino emission. i) Pair annihilation: $e^{+}+e^{-}\rightarrow \nu+\bar{\nu}$ \citep{1985ApJ...296..197M,1989ApJ...339..354I}; this neutrino energy loss rate is here denoted by $\epsilon_{e^-e^+}$. ii) Photo-neutrino process: $\gamma+e^{\pm}\rightarrow e^{\pm}+\nu+\bar{\nu}$ \citep{1985ApJ...296..197M,1989ApJ...339..354I}, denoted by $\epsilon_\gamma$. iii) Plasmon decay: $\bar{\gamma}\rightarrow \nu+\bar{\nu}$ \citep{1986ApJ...310..815K,1994ApJ...431..761K}, denoted by $\epsilon_{\rm pl}$. iv) Bremsstrahlung processes \citep{1983ApJ...275..858I,1984ApJ...285..304I,1984ApJ...280..787I,1984ApJ...279..413I}, denoted by $\epsilon_{\rm BR}$, which can be due to electron-nucleon interaction $e^{\pm}+N\rightarrow N+\nu+\bar{\nu}$ or to nucleon-nucleon interaction $N+N\rightarrow N+N+\nu+\bar{\nu}$.  It is important to mention that two different expressions for the total Bremsstrahlung emission are shown in \citet{1996ApJS..102..411I} depending if the Coulomb parameter, $\Gamma \equiv (Z e)^2/(r_i k_B T)$ where $r_i=[3/(4\pi n_{\rm ion})]^{1/3}$, is higher or lower than the critical value $\Gamma\approx 180$, over which the system crystallizes. So the total energy loss rate per unit volume due to neutrino emission is $Q_\nu=\epsilon_{e^-e^+}+\epsilon_\gamma+\epsilon_{\rm pl}+\epsilon_{\rm BR}$.

Since the infalling material is strongly decelerated by the accretion shock, the post-shock kinetic energy is much less that the internal and gravitational energy. Then, assuming a polytropic gas [$P=(\gamma-1)U\propto \rho^\gamma$] and subsonic velocities inside the shock radius, $(v/c)^2\ll 1$, Eqs.~(\ref{eq:steadyequations}) can be solved for the radial dependence of the fluid variables $\rho$, $P$ and $u$ as \citep{1991ApJ...376..234H}:
\begin{equation}\label{eq:prho(rel)}
\rho=\rho_{{\rm sh}}f(r)^{\frac{1}{\gamma-1}},\quad P=P_{{\rm sh}}f(r)^{\frac{\gamma}{\gamma-1}},\quad u=\frac{u_{{\rm sh}}}{r^2}f(r)^{\frac{1}{1-\gamma}}\quad f(r)\equiv \frac{\left(1-\frac{r_{{\rm sch}}}{r}\right)^{-1/2}-1}{\left(1-\frac{r_{{\rm sch}}}{R_{{\rm ns}}}\right)^{-1/2}-1}\, .
\end{equation}
The approximation of a polytropic equation of state was validated by numerical simulations in \citet{1996ApJ...460..801F}, who showed the infall NS atmosphere is well approximated by a polytropic gas of index $\gamma=1.4$.

Since neutrinos are the main energy sink of the system (see below), the position of the shock can be estimated from the balance between the neutrino emission and the release of the potential gravitational energy due to the accretion process, i.e.:
\begin{equation}\label{eq:bal_neu_grav(rel)}
\left(\frac{4\pi R_{{\rm NS}}^2\Delta r_{{\rm ER}}}{\sqrt{1-\frac{2 G M_{\rm NS}}{c^2 R_{\rm NS}}}}\right) Q_\nu\approx c^2\dot{M}\left[\left(1-\frac{2GM_{\rm NS}}{c^2R_{\rm NS}}\right)^{-1/2}-1\right],
\end{equation}
where we have assumed the rate at which gravitational energy is released as the kinetic energy gained in the free fall from infinity, and we have considered the proper volume of the cooling region and the proper cooling rate. We have also introduced the thickness of the neutrino emission region at the base of the atmosphere, $\Delta r_{\rm ER}$, which in view of the strong dependence of the neutrino emission processes on the temperature, can be estimated as one temperature scale height, i.e.:
\begin{equation}\label{eq:H_T}
\Delta r_{\rm ER} \approx H_T=\frac{T}{\left|(dT/dr)\right|},\qquad \frac{dT}{dr}=\left( \frac{\partial\, {\rm ln}\, T}{\partial\, {\rm ln}\, \rho} \right)_P\frac{d{\rm ln}\, \rho}{dr}+\left( \frac{\partial\, {\rm ln}\, T}{\partial\, {\rm ln}\, P} \right)_\rho\frac{d{\rm ln}\, P}{dr}.
\end{equation}

\begin{figure}
\centering
\includegraphics[width=0.4\hsize,clip]{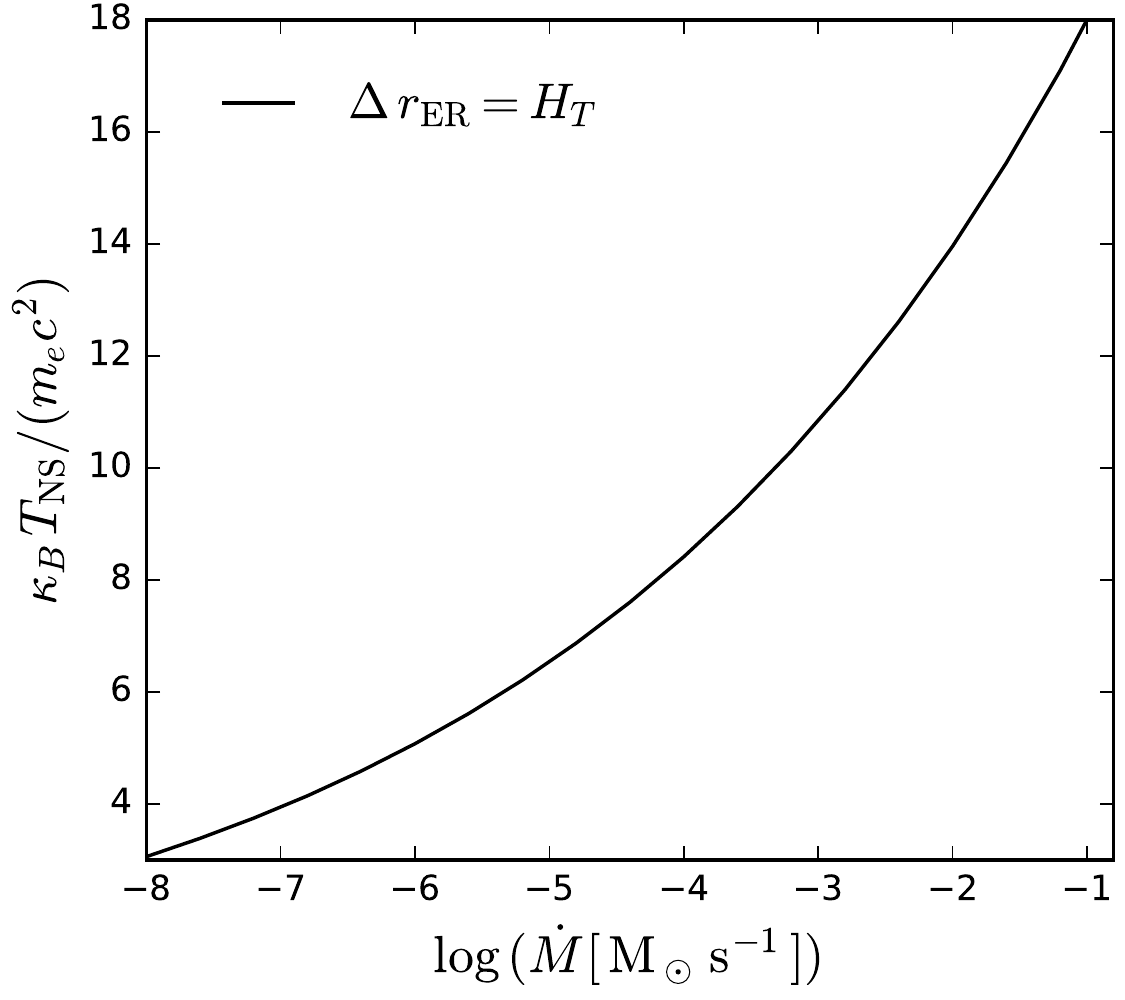}\includegraphics[width=0.4\hsize,clip]{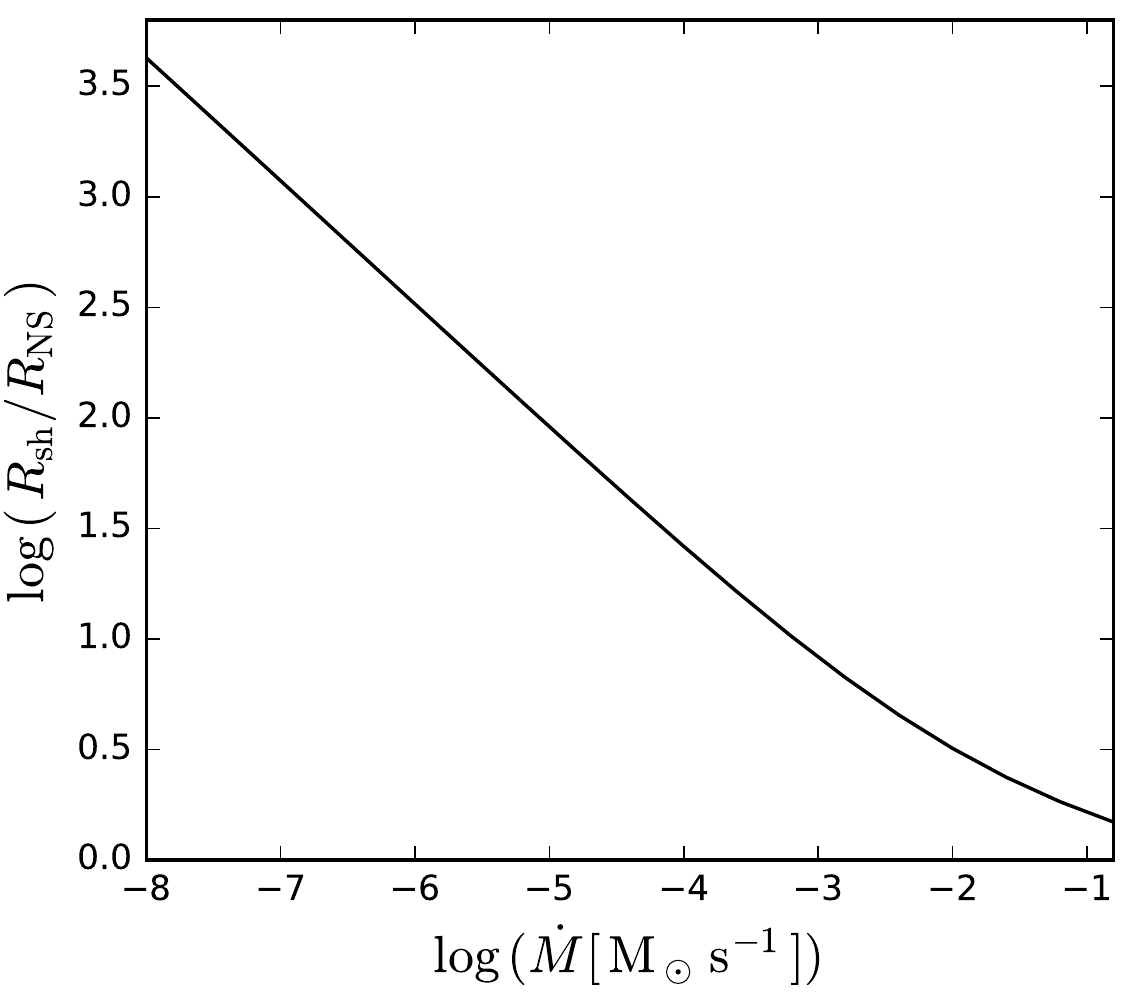}
\caption{Temperature of the NS surface (left panel) and ratio between the shock radius and the NS radius (right plot) as a function of the mass accretion rate in the range $\dot{M}=10^{-8}$--$10^{-1}$~$M_\odot$~s$^{-1}$.}\label{fig:Atm2}
\end{figure}

Fig.~\ref{fig:Atm2} shows the NS surface temperature and the shock position as a function of the mass accretion rate. The thickness of the neutrino emission region is very poorly dependent on the accretion rate; indeed equation~(\ref{eq:H_T}) gives $\Delta r_{\rm ER} \approx 0.76$--$0.77\,R_{\rm NS}$ for $\dot{M}=10^{-8}$--$10^{-1}$~$M_\odot$~s$^{-1}$.

Under the conditions (non-degenerate, relativistic, hot plasma) of our hypercritically accreting NS, the most efficient neutrino emission is given by the $e^{+}e^{-}$ pair annihilation (see Fig.~\ref{fig:Trho}). In these $T$-$\rho$ conditions, $\epsilon_{\rm e^{-}e^{+}}$ reduces to the simple expression \citep{2001PhR...354....1Y}:
\begin{equation}\label{eq:L_neutrinos}
\epsilon_{\rm e^{-}e^{+}}=1.39\times 10^{25}\left(\frac{k_B T}{1\,{\rm MeV}}\right)^9\quad {\rm erg}\,{\rm cm}^{-3}\,{\rm s}^{-1}.
\end{equation}  

\subsection{Neutrino and photon optical depth}\label{app:B4}

We have assumed that the neutrinos produced at the base of the NS surface are the main sink of the gravitational potential energy gained by the infalling material. We proceed now to assess the validity of this statement through the calculation of the neutrino opacity. 

The total neutrino opacity is:
\begin{equation}
	\kappa_\nu=\kappa_{\nu,{\rm abs}}+\kappa_{\nu,{\rm scat}},
	\label{eq:kappatot}
\end{equation}
where $\kappa_{\nu,{\rm abs}}$ and $\kappa_{\nu,{\rm scat}}$ correspond to the opacity produced by absorption and scattering processes. In general, the opacity can be written as
\begin{equation}
	\kappa=\frac{\sigma_in_i}{\rho},
	\label{eq:kappa}
\end{equation}
where $n_i$ is the particle density and $\sigma_i$ is the process cross section. We adopt the following scattering and absorption process:

{\bf \emph{Scattering processes}}: neutrinos transfer momentum to the matter by the scattering off nuclei and electrons and positrons:
\begin{itemize}
\item 
Coherent neutrino nucleus scattering: $\nu+(A,Z)\rightarrow \nu+(A,Z)$ \citep{1975ApJ...201..467T} 
\begin{equation}
\sigma_A=\frac{1}{16}\sigma_0\left( \frac{E_\nu}{m_ec^2} \right)^2A^2\left[ 1-\frac{Z}{A}+(4{\rm sin}^2\theta_w-1)\frac{Z}{A} \right]^2\quad {\rm with}\quad \sigma_0=\frac{4G_F^2(m_ec^2)^2}{\pi(\hbar c)^4}\approx 1.71\times 10^{-44}\, {\rm cm^2}
\label{eq:sigmaA}
\end{equation}
where $G_F$ is the Fermi weak neutrino coupling constant and $\theta_w=$ is the Weinberg angle, $\sin^2\theta_w=0.23$. The scattering is coherent in the sense that nucleus acts as a single particle and the initial and final neutrino energy are nearly equal.
\item Neutrino-electron scattering \citep{1982ApJS...50..115B,2002astro.ph.11404B}:
\begin{equation}
	\sigma_{e}(E)=\frac{3}{8}\sigma_0\beta \frac{E}{m_ec^2}\left( 1+\frac{\eta_e}{4} \right)\left[ (C_v+C_a)^2+\frac{1}{3}(C_v+Ca)^2 \right]
	\label{eq:sca_e}
\end{equation}
where $C_v=1/2+2\sin^2\theta_w $ for electron neutrino and antineutrino types, $C_a=1/2$ for neutrino and $C_a=-1/2$ for antineutrinos.
\end{itemize}
	
{\bf \emph{Absorption processes}}: Since we have shown that the most efficient neutrino cooling process near the NS surface is the electron-positron annihilation, the inverse process namely the annihilation of neutrinos, $\nu+\bar{\nu}\rightarrow e^{-}+e^{+}$, represents the main source of opacity. The total average cross sections are given by \citep{1987ApJ...314L...7G}:
\begin{equation}
	\sigma_\nu(E_\nu)=\frac{4}{3}K_{\nu\bar{\nu}}\sigma_0 \langle E_\nu\rangle \langle E_{\bar{\nu}}\rangle,\qquad 
	\sigma_{\bar{\nu}}(E_{\bar{\nu}})=\frac{4}{3}K_{\nu\bar{\nu}}\sigma_0 \langle E_{\bar{\nu}}\rangle \langle E_\nu\rangle,
	\label{eq:sigmaNu}
\end{equation}
where $K_{\nu\bar{\nu}}=(1+4\sin^2\theta_w+8\sin^4\theta_w)/12=0.195$. The energy of the neutrino and antineutrinos are calculated assuming they are described by the Fermi-Dirac distribution with zero chemical potential:
\begin{equation}
	\langle E_\nu\rangle=\langle E_{\bar{\nu}}\rangle = \frac{U_\nu}{n_\nu}=\frac{\mathcal{F}_{3}(0,0)}{\mathcal{F}_{2}(0,0)}k_B T=3.15 k_BT,
	\qquad \langle E^2_{\nu}\rangle=\frac{\mathcal{F}_{4}(0,0)}{\mathcal{F}_{2}(0,0)}(k_BT)^2=12.93(k_B T)^2.
	\label{eq:Enucuad}
\end{equation}
Then, the total neutrino opacity is:
\begin{equation}
	\kappa_\nu=\left[\sigma_A\left( \frac{\rho}{A m_u} \right)+\sigma_e(E_\nu)\,n_{e^{-}}+\sigma_\nu(E_\nu)n_\nu\right]/\rho,
	\label{eq:tot_kappa}
\end{equation}

The neutrino optical depth can then be obtained as:
\begin{equation}
	d\,\tau_\nu= \kappa_\nu\,\rho\, dr= \frac{dr}{\lambda_\nu},
	\label{eq:opdepth}
\end{equation}
where $\lambda_\nu$ is the neutrino mean free path:
\begin{equation}
	\lambda_\nu=\frac{1}{\kappa_\nu\,\rho}.
	\label{eq:meanfreepath}
\end{equation}
Thus, the optical depth at the base of the neutrino emission region can be estimated as: $\tau_{\nu,{\rm ER}}\approx \kappa_\nu\, \rho_{\rm NS}\Delta r_{\rm ER}=\Delta r_{\rm ER}/\lambda_{\nu,{\rm ER}}$. Large values for the optical depth means  ($\tau_\nu\gg1$) implies that the neutrinos are reabsorbed by the matter and cannot freely scape from the system.

In order to verify that photons are trapped in the infalling material, we evaluate the photon mean free path and photon emissivity:
\begin{equation}
\tau_\gamma=\kappa_R\,\rho\Delta\, r_{\rm ER},\qquad	\dot{q}_\gamma\approx \frac{1}{\Delta\, r_{\rm ER}}\frac{\sigma T^4}{\tau_\gamma},
	\label{eq:qphoton}
\end{equation}
where $\sigma$ is the Stefan-Boltzmann constant, $\tau_\gamma$ is the photon optical depth, and $\kappa_R$ is the Rosseland mean opacity:
\begin{equation}
	\kappa_R=0.4+0.64\times 10^{23}\left( \frac{\rho}{\rm g\, cm^{-3}} \right)\left( \frac{T}{K} \right)^{-3}\, {\rm g^{-1}\, cm^2},
	\label{eq:op_photon}
\end{equation}
being the first term due to the electron scattering and the second one to the free-free absorption. 

\begin{figure}
\centering
\includegraphics[width=0.5\hsize,clip]{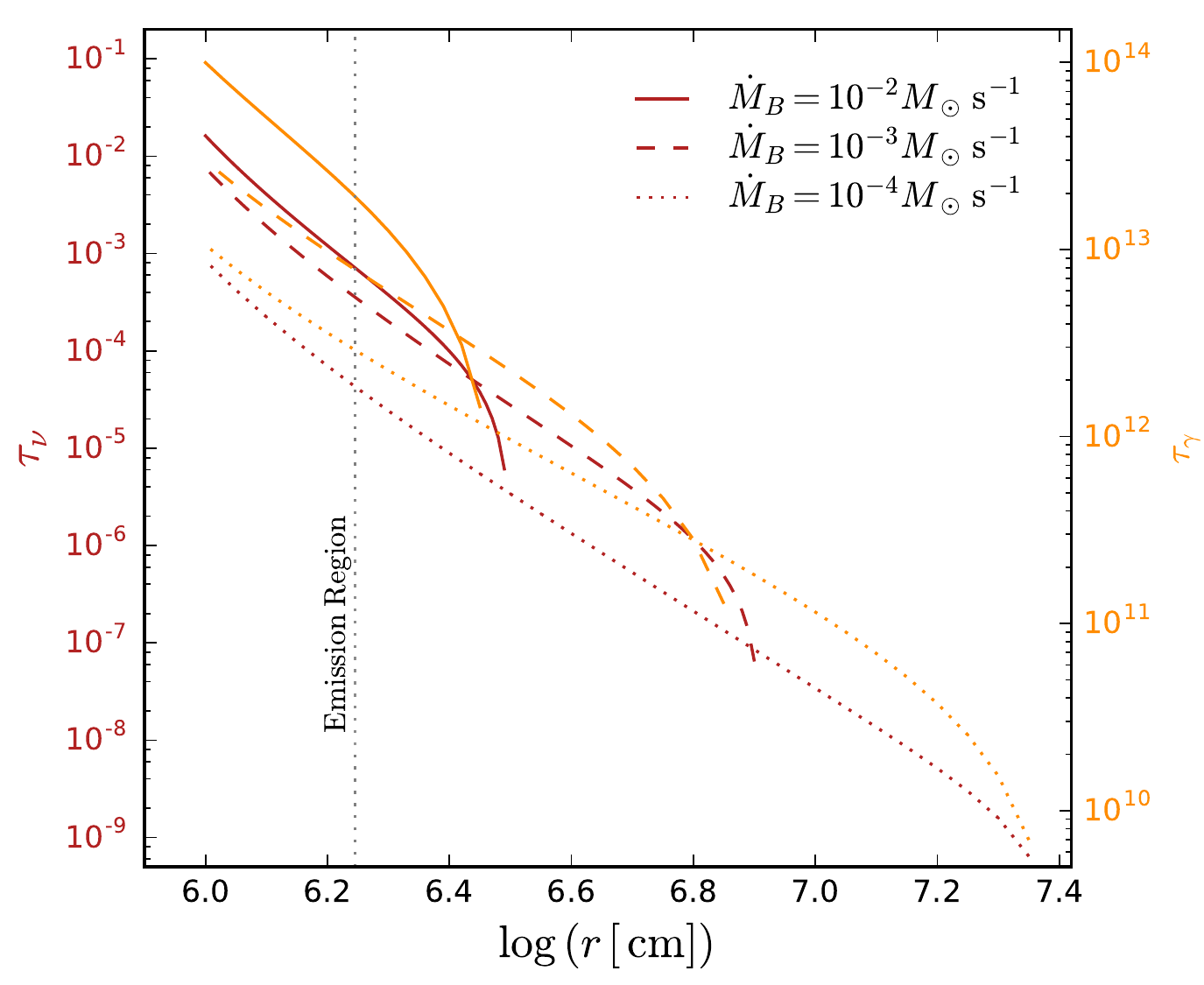}
\caption{Neutrino (left y-scale, $\tau_\nu$) and photon (right y-scale, $\tau_\gamma$) optical depths in the NS star accretion region (from the shock radius to the NS surface) for selected accretion rates.}\label{fig:taus}
\end{figure}

We show in Fig.~\ref{fig:taus} the neutrino and photon optical depth profile in the NS accretion region for three different values of the mass accretion rate. We can see the photon optical depth is much higher than unity for photons, implying they are indeed trapped at any radius. On the contrary, the neutrino optical depth is much lower than unity, implying they efficiently cool the atmosphere which allows the system to proceed the accretion at hypercritical rates.

\begin{figure}
\centering
\includegraphics[width=0.5\hsize,clip]{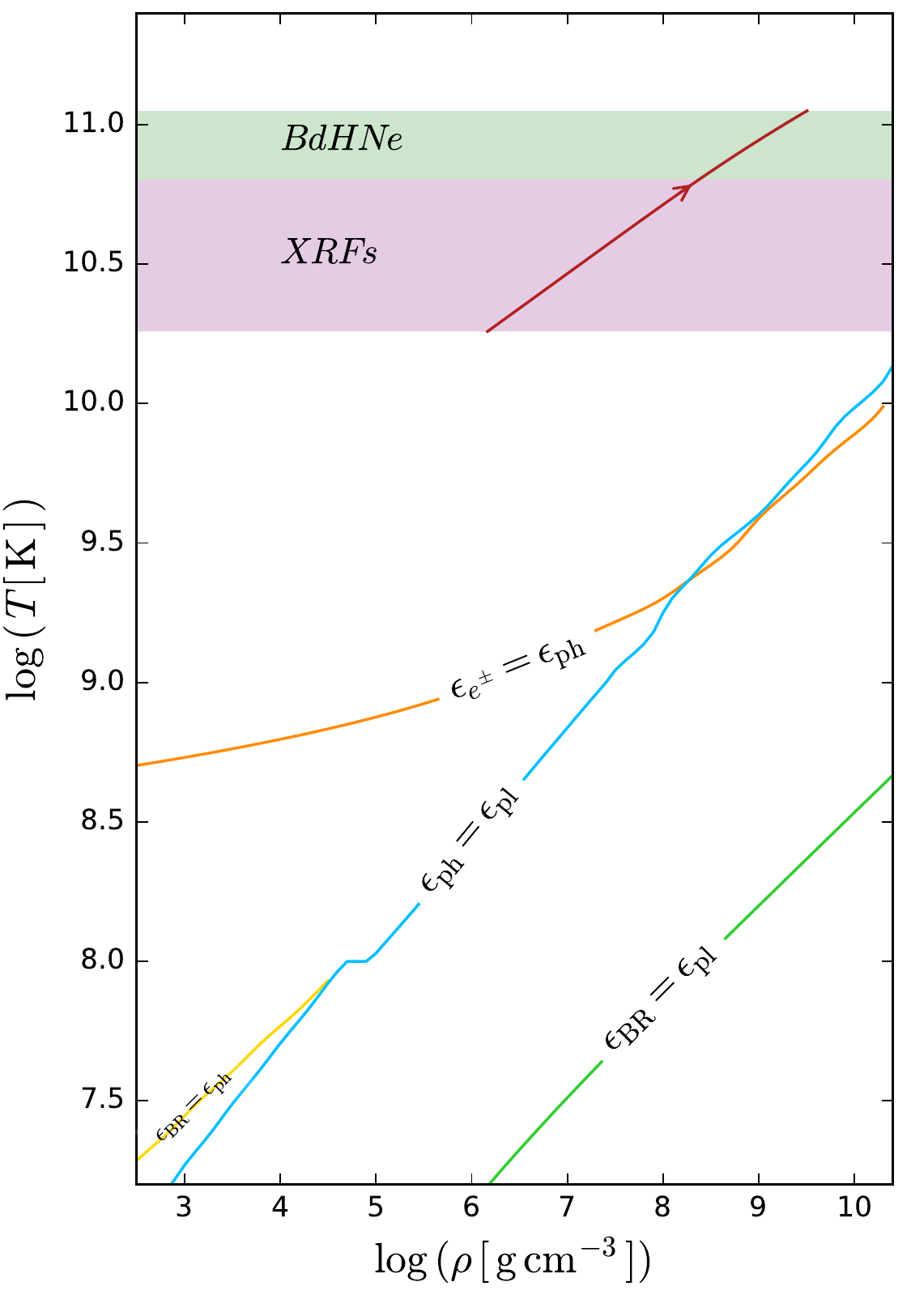}
	\caption{Temperature-density diagram of the accreting atmosphere equation of state. In order to see the dominant neutrino processes, we show contours at which the  emissivities of the different neutrinos process become equal: $\epsilon_{e^\pm}$ correspond to the pair annihilation process, $\epsilon_\gamma$ to the photo-neutrino emission, $\epsilon_{\rm pl}$ to the plasmon decay and $\epsilon_{\rm BR}$ to the Bremsstralung emission. 
The solid red curve shows the corresponding $T$-$\rho$ values of the NS surface in the range of accretion $\dot{M}=10^{-8}$--$10^{-1}$~$M_\odot$~s$^{-1}$ which covers typical rates achieved in XRFs and BdHNe (see Fig.~\ref{fig:Mpeak}). The arrow indicates the direction of increasing accretion rate. Thus, while accreting, the NS moves from the left lower part of the red curve to the right upper part of it. It is clear that in this regime of XRFs and BdHNe, the electron-positron pair annihilation dominates the neutrino emission. 
}
\label{fig:Trho}
\end{figure}

We show in Fig.~\ref{fig:Trho} the $T$-$\rho$ diagram of the NS surface for accretion rates $\dot{M}=10^{-8}$--$10^{-1}$~$M_\odot$~s$^{-1}$ which covers both XRFs and BdHNe (see, e.g., Fig.~\ref{fig:Mpeak}). Higher temperatures and densities correspond to higher accretion rates. We show contours indicating where the neutrino emissivities of the different neutrino emission processes are equal
. It can be seen from these two figures that: 1) pair annihilation neutrino process are highly dominant over the other neutrino emission mechanisms; 2) neutrinos can efficiently escape taking away most of the energy (high emissivity); 3) photons are trapped hence they have negligible emissivity; 4) even for the largest accretion rates the neutrino optical depth in the accretion zone is below unity and so the system is not opaque to neutrinos.

\section{Convergence tests}\label{app:C}

We proceed now to perform a convergence test of the results of our numerical integration. We will perform the test for four important quantities, as a function of the dimensionless time $\tau=t/t_0$: the NS accreted mass $M_{\rm acc}(\tau)$, gravitational mass $M_{\rm NS}(\tau)$, angular momentum $J_{\rm NS}(\tau)$ and gravitational capture radius measured from the supernova center normalized to the binary separation, i.e. $\hat{r} = 1-R_{\rm cap}(\tau)/a$. 

\begin{figure}
\centering
\includegraphics[width=0.45\hsize,clip]{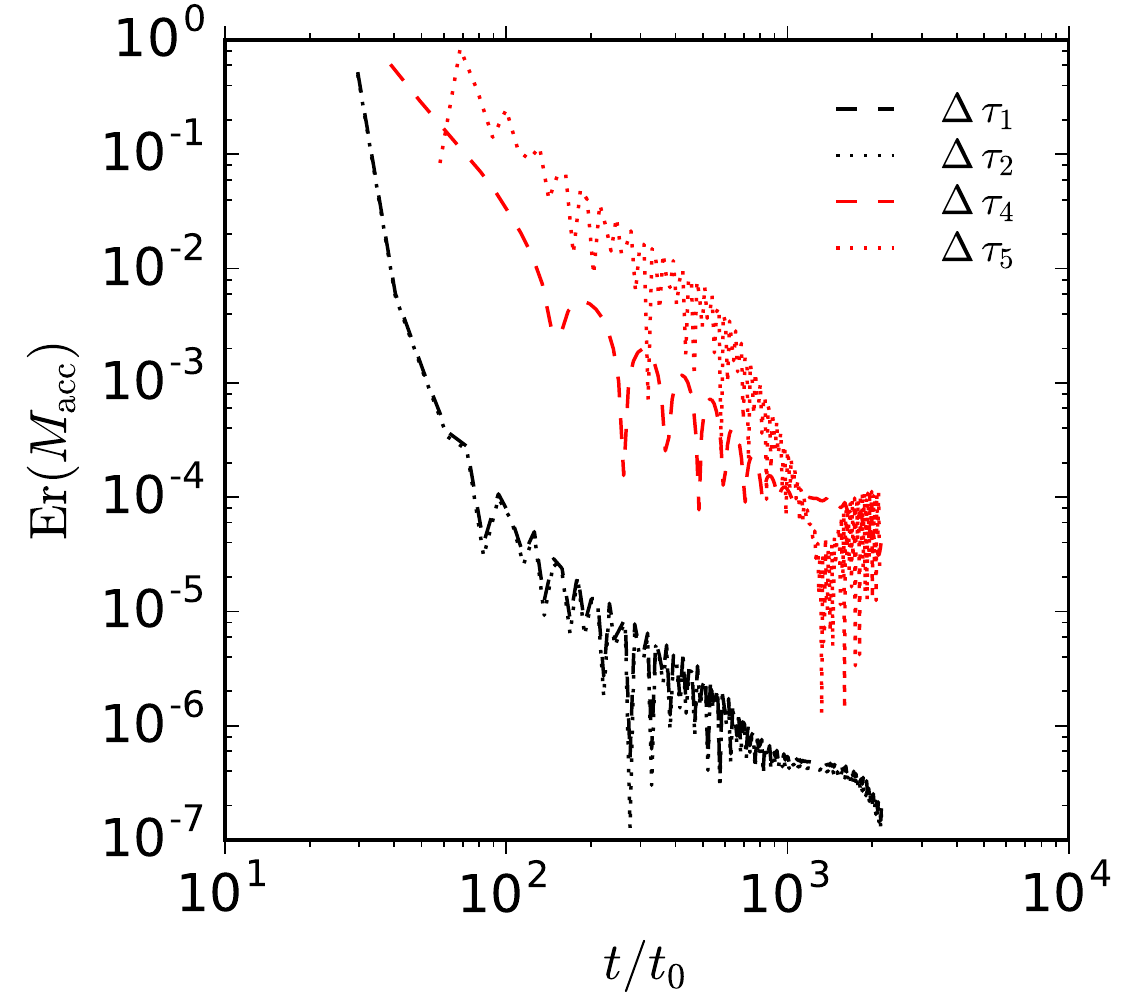}\includegraphics[width=0.45\hsize,clip]{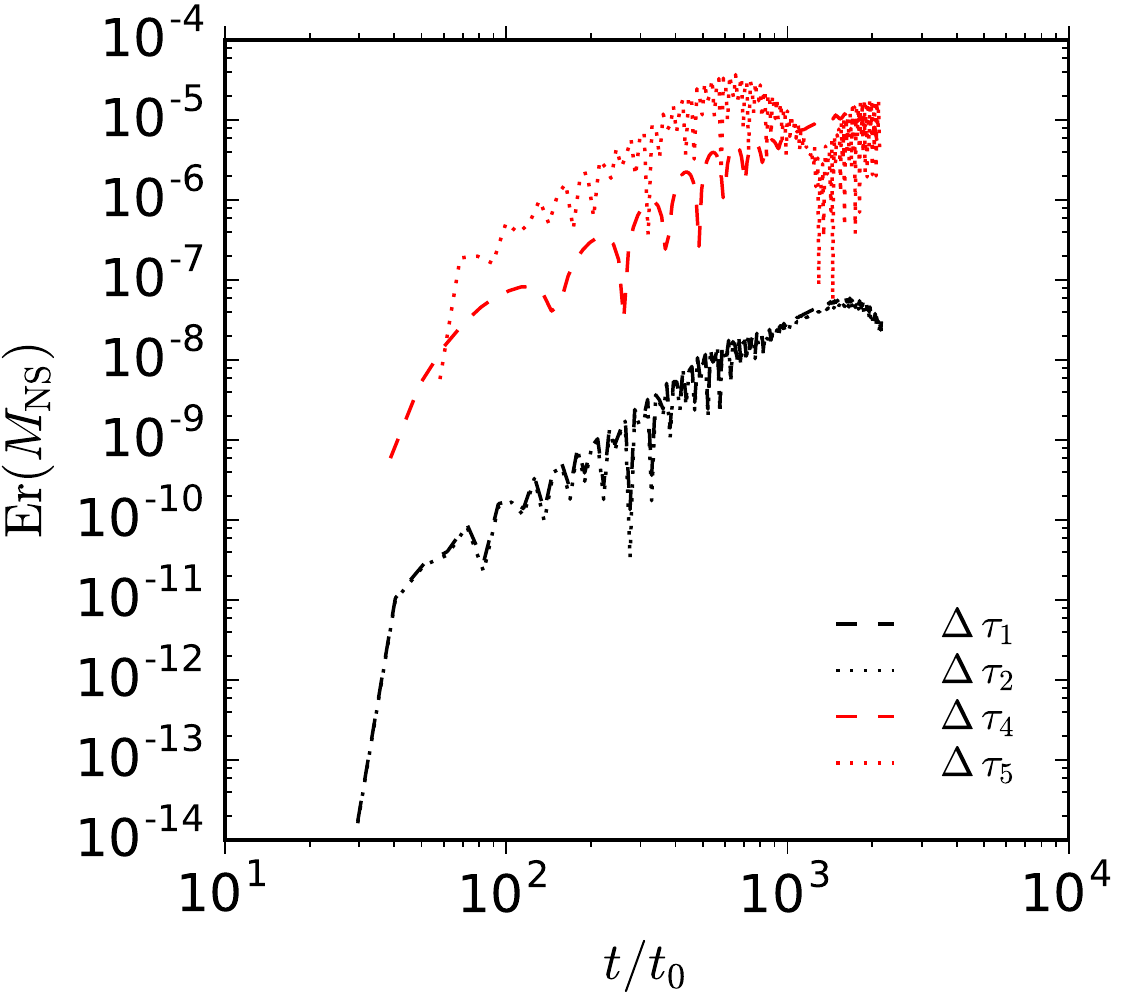}\\
\includegraphics[width=0.45\hsize,clip]{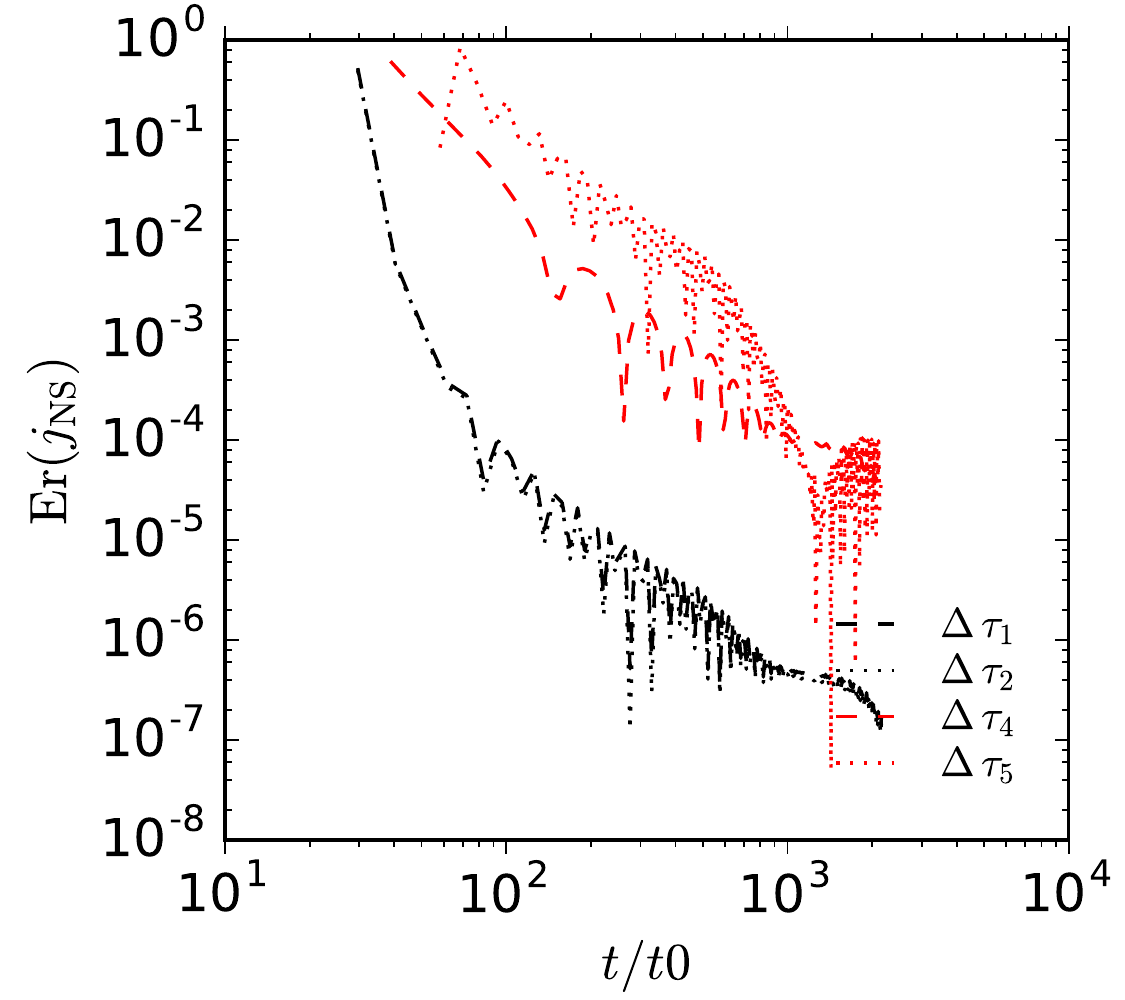}\includegraphics[width=0.45\hsize,clip]{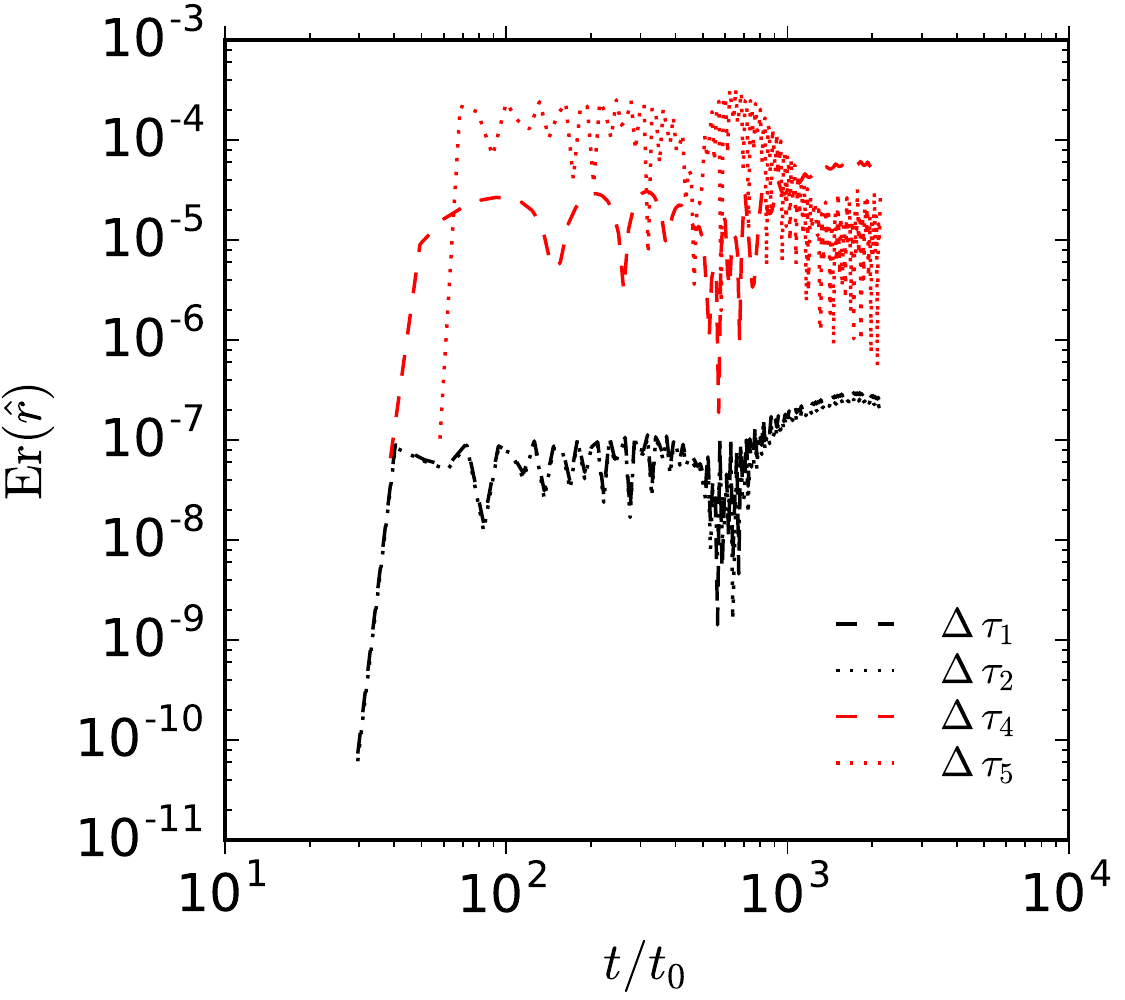}
\caption{Convergence test for the NS accreted mass $M_{\rm acc}(\tau)$, gravitational mass $M_{\rm NS}(\tau)$, angular momentum $J_{\rm NS}(\tau)$ and the gravitational capture radius measured from the supernova center normalized to the binary separation, i.e. $\hat{r} = 1-R_{\rm cap}(\tau)/a$. We select here five values of the stepsize: $\Delta \tau_i=0.0581,0.1453,0.5812,9.6874,29.062$ for $i=1,...,5$. The parameters of the binary system in this example are: the CO core of the $M_{\rm ZAMS}=20~M_\odot$ progenitor, an initial NS mass of $2~M_\odot$, and an orbital period $P=3$~h. The relative error increases for stepsizes higher than $\Delta \tau_{3}$ and decreases for stepsizes lower than it, which indicates convergence.}\label{fig:ctest}
\end{figure}

We use for the numerical integration the Adams method implemented in the Python library SciPy version 0.17.1. This requires that the user sets, besides the system of ordinary different equations, a minimum and a maximum integration stepsize. For the integration we set both to the same value, say $\Delta \tau$. To perform this test we select five values of the stepsize: $\Delta \tau_i \equiv\tau_{\rm acc,0}/N_i$ with $N_i = 1,20,50,200,500$ for $i=1,...,5$ and $\tau_{\rm acc,0}=t_{{\rm acc},0}/t_0$ is the dimensionless time at which the accretion process starts, i.e. the instant at which the first ejecta layer reaches the NS capture radius. We denote as $A_{\Delta \tau_{i}}$ the numerical value of the quantity $A$ computed with the stepsize $\Delta \tau_{i}$, and then compute the relative error with respect to the computed value using a reference stepsize, i.e.: ${\rm Er}(A)\equiv |A_{\Delta \tau_{\rm ref}}-A_{\Delta \tau_{i}}|/A_{\Delta \tau_{\rm ref}}$. Fig.~\ref{fig:ctest} shows the convergence test for a binary system with the following parameters: the CO core of the $M_{\rm ZAMS}=20~M_\odot$ progenitor, an initial NS mass of $2~M_\odot$, and an orbital period $P=3$~h. For these parameters we have $\tau_{\rm acc,0}\approx 29$ and so the stepsizes are: $\Delta \tau_i=0.0581,0.1453,0.5812,9.6874,29.062$, respectively for $i=1,...,5$. The stepsize of reference in this test is $\Delta \tau_{\rm ref} = \Delta \tau_{3}=0.5812$. We can see that, as expected, the relative error increases for stepsizes higher than $\Delta \tau_{3}$ and decreases for stepsizes lower than it, indicating convergence. All the results shown in this article are for $\Delta \tau_{3}$, which corresponds to a different numerical value for different binary systems, since the value of $\tau_{\rm acc,0}$ is specific to each system.

\bibliographystyle{apj}
\bibliography{references}

\end{document}